\documentclass{ws-rv961x669}
\usepackage{ws-rv-van}     
\usepackage{ws-rv-thm}     
\usepackage{subfigure}     
\usepackage{comment}
\makeindex
\usepackage{ulem}
\usepackage{graphicx}
\usepackage{float}
\usepackage{texvc}
\usepackage{amsmath,amssymb,amsfonts}

\newcommand{\Msun}{M$_{\odot}$}
\newcommand{\Msunyr}{M$_{\odot}$yr$^{-1}$}
\newcommand{\kms}{km\,s$^{-1}$}

\newcommand{\Msunh}{M$_{\odot}h^{-1}$}

\begin{document}

\chapter[Massive black holes in galactic nuclei - theory/simulations]{
V.    Massive black holes in galactic nuclei - Theory and Simulations\\ \label{title_ch5}}

\author[Tiziana Di Matteo, Daniel Angl\'es-Alc\'azar and Francesco Shankar] {Tiziana Di Matteo \footnote{McWilliams Center For Cosmology, Carnegie Mellon University,
5000 Forbes Avenue, Pittsburgh PA, 15213.}, Daniel Angl\'es-Alc\'azar \footnote{Department of Physics, University of Connecticut, 196 Auditorium Road, U-3046, Storrs, CT 06269-3046, USA. Center for Computational Astrophysics, Flatiron Institute, 162 Fifth Avenue, New York, NY 10010, USA.} and Francesco Shankar \footnote{School of Physics and Astronomy, University of Southampton, Highfield, Southampton SO17 1BJ, UK.}}


\begin{abstract}
Massive black holes are fundamental constituents of our cosmos, 
from the Big Bang to today.
Understanding their formation from cosmic dawn, 
their growth, and the emergence
of the first, rare  quasars in the early Universe remains one of our greatest
theoretical and observational challenges.
Hydrodynamic cosmological simulations
self-consistently combine the processes of structure formation at
cosmological scales with the physics of smaller, galaxy scales. They
capture our most realistic understanding of massive black holes and their
connection to galaxy formation and
have become the primary avenue for theoretical 
research in this field. 
The space-based gravitational wave interferometer, LISA, will open 
up new investigations into the dynamical processes involving massive black holes.
Multi-messenger astrophysics brings new exciting prospects for tracing the origin, growth
and merger history of massive black holes across cosmic ages.
\end{abstract}


\body


\section{Introduction}\label{intro_sec1}
In this chapter we will take a journey through our cosmic history to examine the role of black holes from the Big Bang to today.
Black holes are fundamental components of our Universe, and they play
a major role in our understanding of galaxy formation. As discussed in Chapter IX, black holes forming from
the collapse of the first density peaks and
associated processes at cosmic dawn (the first epoch of galaxy
formation) are likely to lead to a significant population of 'seed'
black holes that merge and grow. Here we will examine the emergence of the
first population of supermassive black holes, the first quasars, that
occurs within the first billion years of our cosmic history.
 As we link the formation of the first black holes to the first quasars,
we also study the formation of black holes in our standard paradigm of
structure formation. We will describe how we understand structure
formation within the context of cosmological simulations and semi-analytic models.
State-of-the art cosmological simulations include the formation and
growth of black holes and can make direct predictions for current and
future observations.  We will discuss the growth of black holes in the
centers of galaxies, and how mergers and gas accretion induced by large
inflows and mergers connect the physics of  small scales to 
larger cosmological scales. Cosmological simulations allow us to
study directly the co-evolution of and connection between black hole
growth and galaxy formation and the emergence of the fundamental
relations between central black holes and their host galaxy properties.

\section{Cosmological Simulations of galaxy formation and black holes}

\subsection{Brief Introduction, Motivation and Challenges}

Structure formation and evolution in cosmology encompasses the description of
the rich hierarchy of structures in our Universe, from individual galaxies and
groups to clusters of galaxies up to the largest scale filaments along which
smaller structures align. This so-called 'cosmic web' arises from the
gravitational growth of the initial matter inhomogeneities, seeded at the time
of inflation. The rate at which structure forms depends on the initial power
spectrum of the matter fluctuations, now well measured \cite{Spergel2006} 
and
on the expansion rate of the universe, which is regulated by its matter
content (the largest component of which is dark matter), radiation and dark
energy.  The standard cosmological model has been very successful at
predicting a wide range of phenomena, so that it has become worthwhile to
devote the largest computer resources to studying structure formation.

To do this, we need to develop computer simulations that cover a vast dynamic
range of spatial and time scales: we need to include the effect of
gravitational fields generated by superclusters of galaxies on the formation
of galaxies, which in turn harbor gas that cools and makes stars and is being
funneled into supermassive blackholes the size of the solar system.
Ultimately the study of structure formation should provide a true
understanding of how galaxy formation takes place in the universe, and so
allow us to use the many observations of galaxies and their clustering to gain
insights into the nature of the two greatest mysteries of modern physics, dark
matter and dark energy while also reproducing the formation and evolution of galaxies and their black holes across cosmic history.

There are two conflicting requirements that make the study of hierarchical
structure formation extremely challenging.  In order to have a statistically
significant representation of all structure in the Universe, the volume
studied needs to be large but the individual particle mass needs to be small to adequately
resolve the scale length of the structures which form and the appropriate
physics. This implies a need for an extremely large $N$, where $N$ is the
number of particles. Depending on the problem, a dynamic range of $10^{10}$ or
more can be necessary in principle.

The largest computer models of galaxy formation have traditionally involved
the properties of dark matter only but the part of the Universe astronomers
observe is made up of ordinary matter (gas, stars etc.).  In order to make
direct contact with observations and predictions from our theories we must
simulate the detailed hydrodynamics of the cosmic plasma. In addition, there
is strong observational evidence for a close connection between the formation
and evolution of galaxies and of their central supermassive black holes.
Cosmic structure formation is nonlinear, involves a large variety of
multi-scale physics and operates on large timescales making large scale
numerical simulations the primary means for its study.
Both approaches (dark matter only and full gas-dynamics, star formation and
black hole physics) can be used in concert to make progress. 

In hydrodynamic cosmological simulations, the complex non-linear interactions  of gravity, hydrodynamics, forming stars, and black holes are treated in
a large, representative volume of the universe. In this approach the
physics at these much smaller galaxy scales is hence self-consistently
coupled to large cosmological scales. These are therefore our most
powerful predictive calculations linking the part of the universe we
observe (stars, black holes etc..) to the underlying dark matter and
dark energy. They 
capture our most realistic understanding of black holes and their
connection to galaxy formation.

Over the last few years it has become possible, with newly developed
and more sophisticated codes, higher fidelity physical models as well
as large enough computational facilities, to simulate  statistically
significant volumes of the universe (down to $z=0$) with sufficient
detail to resolve the internal structure of individual galaxies and
follow the growth, mergers and evolution of black holes in their
centers. We will review these in this Chapter.
The prospect that we are in a position to use cosmology, i.e. the science 
of the Gigaparsec horizon, in our simulations to make predictions for the mass
distribution in the inner regions of galaxies and their central black holes is extraordinary.

\subsection{What we simulate, codes and physics}
To simulate structure formation in the Universe we need to account for
its full cosmic matter-energy content. Matter comes in two basic types:
ordinary {\em baryonic matter} (e.g.~atoms, stars, planets, galaxies)
which accounts for $15\%$ of the total matter content, and {\em dark
matter} which accounts for the remaining $85\%$. In addition, there is a
mysterious {\em dark energy} field which actually dominates the energy
density of the universe today, with a contribution of 75\%, while matter
constitutes only about 25\%.  

The simulations which include black holes, and that are the subject of this Chapter, are carried out in our standard $\Lambda$CDM cosmology.
They assume that the Universe has a component of its energy density driven by the cosmological constant $\Lambda$.
Dark energy, in the form of $\Lambda$ is then simply introduced in the initial conditions and to solve for the cosmological expansion via the  Friedmann equation. It is capable of providing the acceleration in the cosmic expansion compatible with our present observational constraints. 

Cosmological sumulation use a number of different
algorithms to self-consistently simulate the matter fluids (dark and baryonic) components according
to their appropriate physical laws.

\noindent{\bf Dark Matter:}
For dark matter, which in our standard cosmological models is thought to behave as a perfectly collisionless
fluid, the N-body method is used, where a finite set of particles
samples the underlying distribution function. As the only appreciable
interaction of dark matter is through gravity, the evolution of the
system obeys the Poisson-Vlasov equation.  For the computation of the
gravitational field, cosmological codes often use an FFT mesh solver on large-scales
coupled to a hierarchical multipole expansion of the gravitational field
based on a tree-algorithm~\cite{Barnes1986} on small scales, leading
to a uniformly high force resolution throughout the computational volume \citep{springel05}.

For the large scales,  many cosmological codes use
 a hierarchical multipole expansion (organized in a ``tree'') to
calculate gravitational forces. In this method, particles are
hierarchically grouped, multipole moments are calculated for each node,
and then the force on each particle is obtained by approximating the
exact force with a sum over multipoles. The list of multipoles to be
used is obtained with a so-called tree-walk, in which the allowed force
error can be tuned in a flexible way. A great strength of the tree
algorithm is the near insensitivity of its performance to clustering of
matter, and its ability to adapt to arbitrary geometries of the particle
distribution.
While the high spatial accuracy of tree algorithms is ideal for the
strongly clustered regime on small scales, there are actually faster
methods to obtain the gravitational fields on large scales. In
particular, the well-known particle-mesh (PM) approach based on Fourier
techniques is probably the fastest method to calculate the gravitational
field on a homogeneous mesh. The obvious limitation of this method is
however that the force resolution cannot be better than the size of one
mesh cell, and the latter cannot be made small enough to resolve all the
scales of interest in cosmological simulations.
Many codes offer a compromise between the two methods. The
gravitational field on large scales is calculated with a particle-mesh
(PM) algorithm, while the short-range forces are delivered by the
tree. Thanks to an explicit force-split in Fourier space, the matching
of the forces can be made very accurate. With this TreePM hybrid scheme,
the advantages of PM on large-scales are combined with the advantages of
the tree on small scales, such that a very accurate and fast
gravitational solver results. A significant speed-up relative to a plain
tree code results because the tree-walk can now be restricted to a small
region around the target particle as opposed to having to be carried out
for the full volume.
\begin{figure}
 \includegraphics[width=0.99\textwidth]{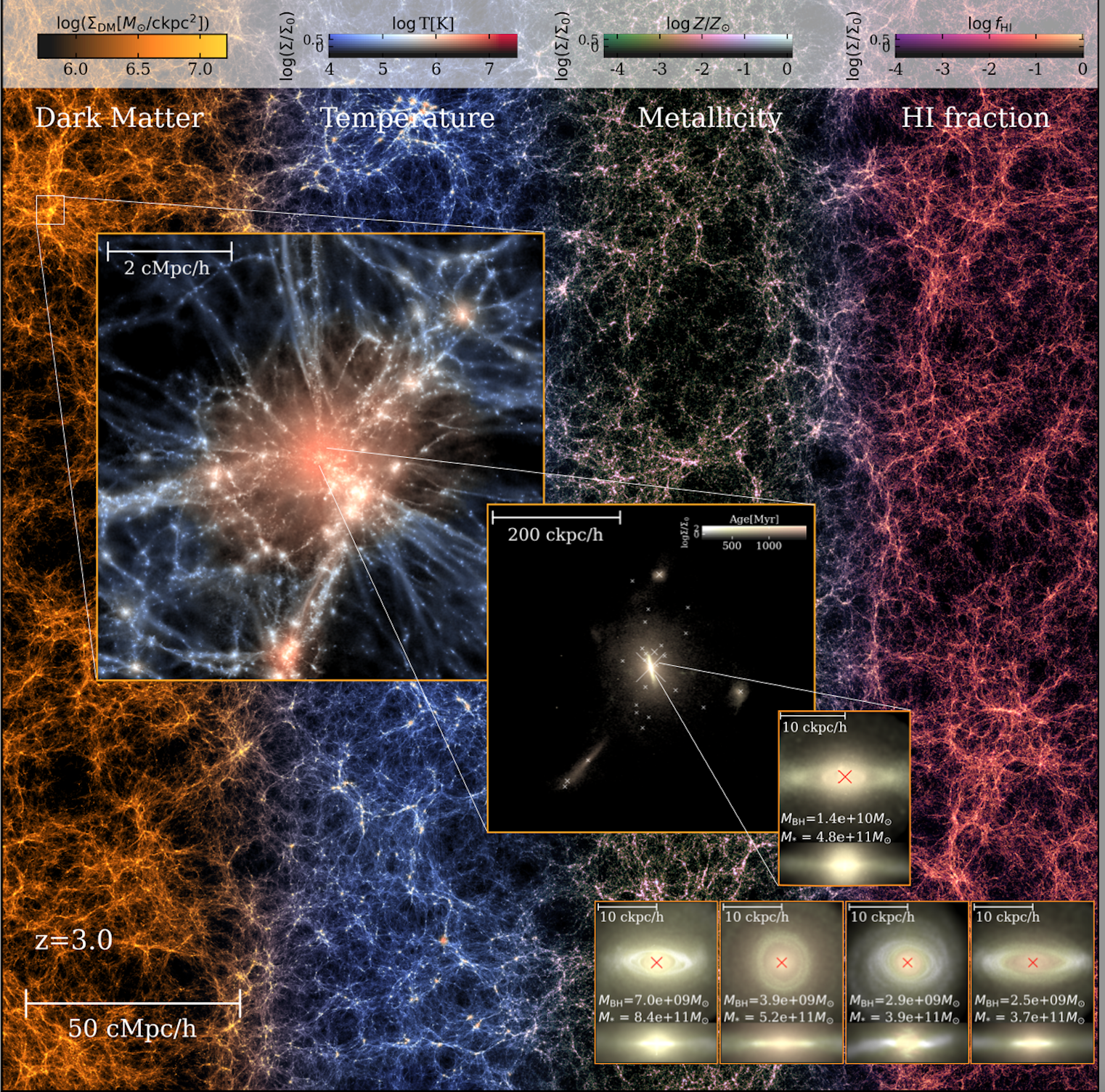}
  \caption{Illustration of state-of-the art hydrodynamical simulation at $z=3$. \textit{Background:} A $250 Mpc/h \times 250$ Mpc/h slice of the ASTRID simulation, from left to right the colour shows dark matter density (orange), gas temperature (blue), metallicity (purple) and neutral hydrogen fraction (red) respectively.
  \textit{First inset (Upper left):} Gas density field coloured by temperature, showing a $7$ Mpc/h  zoomed-in region centred on a massive halo with $M_{\rm h} = 3\times10^{13}$ \Msun.
  \textit{Second inset:} further zoom into a $500 $ kpc/h region, showing the stellar density field centred on an ultra-massive $10^{10}$ \Msun BH. The white crosses in the panel mark the positions of SMBHs in this region with the cross size scaled by the BH mass.
  \textit{Third inset:} the morphology of the host galaxy in face-on (upper panel) and edge-on (lower panel) views in a $20$ kpc/h region around the central SMBH. Colours show the stellar age with older stars being redder.
  The bottom insets show some randomly chosen galaxies hosting 10$^9$ BHs. Credit: Yueying Ni.}
  \label{fig:SimGallery}
\end{figure}

\noindent{\bf Baryonic Matter:}
Baryonic matter is evolved using a mass discretization of the Lagrangian
equations of gas dynamics. 
In cosmology and galaxy formation simulations, both Eulerian and
Lagrangian methods have been used to discretize the cosmic gas. Eulerian
methods offer the principal advantage of high accuracy for shock
capturing and low numerical viscosity.

In Lagrangian codes or Smooth Particle Hydrodynamics based codes, the baryonic matter is evolved using a mass
discretization of the Lagrangian equations of gas dynamics. The code
employs a particle-based approach to hydrodynamics, where fluid
properties at a given point are estimated by local kernel-averaging
over neighboring particles, and smoothed versions of the equations of
hydrodynamics are solved for the evolution of the fluid (SPH). 

With newer hydrodynamic algorithms such as AREPO \cite{Springel2010} and GIZMO \cite{Hopkins2015_Gizmo},
 an unstructured Voronoi tessellation of the simulation
volume allows for dynamic and adaptive spatial discretization, where a
set of mesh generating points are moved along with the gas flow. This
mesh is used to solve the equations of ideal hydrodynamics using a
second order, finite volume, directionally un-split Godunov-type
scheme, with an exact Riemann solver. The code has been thoroughly
tested and validated on a number of computational problems and small
scale cosmological simulations \cite{Springel2010},
  \cite{Bauer2011}, \cite{Sijacki2012}, \cite{Vogelsberger2012}, \cite{Torrey2012},
  \cite{Genel2013}, \cite{Nelson2013} demonstrating excellent shock capturing
properties, proper development of fluid instabilities, low numerical
diffusivity and Galilean invariance, making it thus well posed to
tackle the problem of galaxy formation. In recent years, these algorithms have enabled a computational approach to the full  problem of galaxy formation.

\subsubsection{Physical processes, galaxy formation}
The galaxy formation model in recent simulations is based on the
inclusion of: 
\\
(i) Gas cooling and photo-ionization: the cooling
function is calculated as a function of gas density, temperature,
metallicity, UV radiation field, and AGN radiation field.
\\
(ii) Star formation and ISM model: the simulations adopt a subgrid
model for the ISM, computing an effective equation of state assuming
a two-phase medium of cold clouds embedded in a tenuous, hot
phase. Star formation occurs stochastically and follows the Kennicutt-Schmidt law.
\\
(iii) Stellar evolution and feedback: stellar populations return mass
to the gas phase through stellar winds and supernovae. The simulations
also employ a kinetic stellar feedback scheme, which generates a wind
with velocity scaled to the local DM dispersion, and mass loading
inferred from the available SN energy for energy-driven winds.

In the next section we review in more detail the physical implementation
of the BH physics in our cosmological simulations, which is central to this chapter.

\section{Black Holes in Galaxy Formation Simulations}

A growing number of cosmological hydrodynamic simulations incorporate subgrid models for black hole seeding (\S\ref{sec:seeds}), dynamics (\S\ref{sec:dyn}), growth (\S\ref{sec:acc}), and impact of feedback (\S\ref{sec:feedback}), which we review in this section.  
Large volume cosmological simulations are a primary tool to model statistical populations of black holes and galaxies and their connection to large scale structure.
Examples of large volume simulations with black hole physics include:
Magneticum \citep{Hirschmann2014,Steinborn2015}, 
Horizon-AGN \citep{Dubois2014,Volonteri2016}, 
Eagle \citep{Schaye2015,Rosas-Guevara2015,Rosas-Guevara2016}, 
Illustris \citep{Genel2014,Vogelsberger2014,Sijacki2015}, 
MassiveBlack-II \citep{Khandai2015_MassiveBlack2,DeGraf2015},
BlueTides \citep{Feng2016,DiMatteo17, Huang2018,Ni2020_QSOobscuration}, 
Romulus \citep{Tremmel2017_Romulus,Ricarte2019,Sharma2020}, 
IllustrisTNG \citep{Weinberger2017,Pillepich2018,Habouzit2019,Terrazas2020}, 
SIMBA \citep{Dave2019_Simba,Thomas2019,Thomas2021,Borrow2020},
Astrid \citep{Chen2021_Astrid,Ni2021_astrid,Bird2022_Astrid}, and
CAMELS \citep{Villaescusa-Navarro2021_CAMELS,Villaescusa-Navarro2022_CAMELSpublic}.
Cosmological zoom-in simulations of smaller volumes are ideal to study the co-evolution of black holes and galaxies at higher resolution for individual systems (or reach the galaxy cluster regime) while maintaining a full realistic cosmological setting.  Examples of zoom-in simulations with black hole physics include:
Apostle \citep{Sawala2016}, 
Auriga \citep{Grand2017_Auriga},
NIHAO \citep{Wang2015_nihao,Blank2019},
Cluster-EAGLE \citep{Bahe2017,Barnes2017},
The Three Hundred \citep{Weiguang2018_The300,Weiguang2022_The300},
Choi et al. \citep{Choi2015_CosmoSim,Choi2018_AGNsizes,Choi2020_AGNmetals},
Costa et al. \citep{Costa2014_QSOenvironment,Costa2015,Costa2018b},
MARVELous Dwarfs and the DC Justice League \citep{Bellovary2019,Bellovary2021,Applebaum2021},
New-Horizon \citep{Volonteri2020,Dubois2021_newhorizon}, and
FIRE \citep{Angles-Alcazar2017_BHsOnFIRE,Hopkins2018_FIRE2methods,Angles-Alcazar2021,Catmabacak2022,Hopkins2022_FIRE3,Wellons2022}.

While we focus on simulations of black holes in a cosmological context, idealized models of galactic nuclei, isolated galaxies/halos, and galaxy mergers play a crucial role in our understanding of black hole seeding, growth \citep{}, dynamics, and feedback, informing the development of improved subgrid models for cosmological simulations.

\begin{figure}
\includegraphics[width=\textwidth]{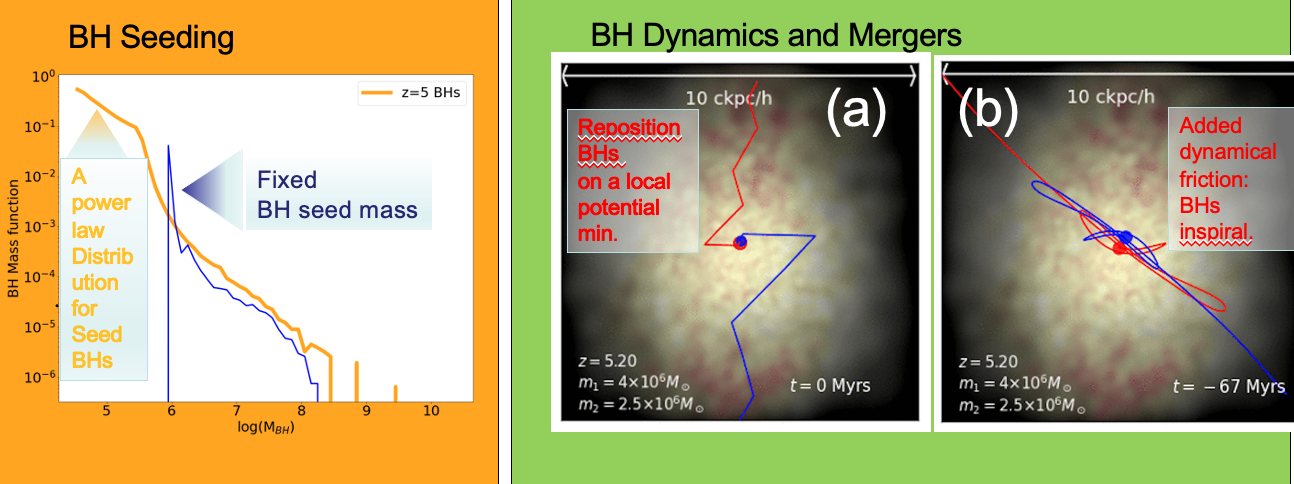} 
\caption{BH seeds and merger models in simulations: Illustration of BH mass function for BH Seeds in a power law distribution vs fixed BH mass (left).
BH repositioning on the local potential minimum, BH endowed inspiraling orbits with dynamical friction added in simulations.}
\label{fig:BHseeds}
\end{figure}

\subsection{Black hole seeds}\label{sec:seeds}

Several different scenarios exist for the formation of the initial black hole ``seeds'' that eventually grow to become massive black holes populating the centers of galaxies (see Chapters III and IX, and reviews by \citet{Volonteri2010} and \citet{Inayoshi2020_ARAA}).  Popular models include the formation of light seeds ($M_{\rm seed }\sim 10^{2}$\,\Msun) as remnants of population III stars \citep{Madau2001,Volonteri2003} and the formation of massive
seeds ($M_{\rm seed } \sim 10^{5}$\,\Msun) by direct collapse in pre-galactic haloes \citep{Begelman2006,LodatoNatarajan2006,LodatoNatarajan2007}.  Despite much recent work, major uncertainties remain on the formation time, initial mass, birth place, and overall number density of black hole seeds. Regardless of seed formation scenario, the relevant physical processes occur well below the resolution of cosmological simulations, which must therefore adopt simple subgrid models to introduce the initial seed black holes. 

Since the first cosmological simulations including black hole growth and feedback, a common approach to the seeding problem has been to simply assume that every halo above a given threshold dark matter (or stellar) mass hosts a central black hole \citep{DiMatteo2008,Booth2009,Angles-Alcazar2017_BHfeedback,Schaye2015,Steinborn2015,Feng2016,Weinberger2017,Ni2021_astrid}, without attempting to mimic the physics or outcome of any specific seed formation scenario.  In practice, halos are selected for seeding by regularly running a ``Friends-of-Friends'' (FoF) halo finder on-the-fly as the simulation proceeds, with typical linking length of $\sim$0.2 times the mean particle separation.  Halos that satisfy the seeding criteria are then assigned a seed black hole by converting the most bound or highest density gas element into a collisionless particles, if the FOF group does not already contain a black hole particle.  The mass threshold for seeding is usually resolution dependent and chosen such that halos are seeded as soon as they are resolved with a sufficient number of particles, but some simulations choose to seed black holes at later times in higher mass galaxies once they are expected to grow more efficiently \citep{Dave2019_Simba,Thomas2019}.

An alternative to halo-based models relying on FoF group finding is to create seed black holes based on local gas conditions \citep{Bellovary2010,Bellovary2011}.  When a gas element satisfies the star formation criteria in a given simulation, it can become a black hole seed (instead of a star particle) with a probability that can be adjusted to control the overall efficiency of seed formation.   In some simulations, this probability is weighted such that black hole seeds form preferentially at the lowest metallicities \citep{TaylorKobayashi2014,Habouzit2017,Tremmel2017_Romulus,Bellovary2019,MaLinhao2021,Hopkins2022_FIRE3,Wellons2022}, in qualitative agreement with various theoretical models of seed formation that rely on the presence of near pristine gas.   Additionally, more restrictive conditions require seeds to form preferentially in converging flows, at the highest surface densities and gravitational accelerations, and in local regions with no pre-existing black holes \citep{Habouzit2017,Hopkins2022_FIRE3,Wellons2022}.

Adopted black hole seed masses in simulations range from $M_{\rm seed }\sim 10^{2}$\,\Msun~\citep{Hopkins2022_FIRE3,Wellons2022} to $M_{\rm seed } \sim 10^{6}$\,\Msun~\citep{Weinberger2017,Tremmel2017_Romulus}, roughly covering the full range of theoretical expectations.  The seed mass is often below the resolution limit of simulations even for heavy seed formation scenarios, which requires (Lagrangian) simulations to track separately the ``physical'' mass of the black hole (starting at $M_{\rm seed}$) from the actual ``dynamical'' mass of the corresponding collisionless particle \citep{Springel2005_BHmodel}.  While constant mass is more common, some recent simulations incorporate more complex astrophysical scenarios by adopting a distribution of black hole seed masses.  Examples include a power-law distribution of heavy seeds:
\begin{equation}
\label{equation:power-law}
    P(M_{\rm seed}) = 
    \begin{cases}
    0 & M_{\rm seed} < M_{\rm seed,min} \\
    \mathcal{N} (M_{\rm seed})^{-n} & M_{\rm seed,min} \leq M_{\rm seed} \leq M_{\rm seed,max} \\
    0 & M_{\rm seed} > M_{\rm seed,max}
   \end{cases}
\end{equation}
where $\mathcal{N}$ is the normalization factor, $M_{\rm seed,min} = 3 \times 10^4$\,\Msunh~is the minimum seed mass, $M_{\rm seed,max} = 3 \times 10^5$\,\Msunh~is the maximum seed mass, and $n = -1$ defines the power-law distribution \citep{Ni2021_astrid}.  
Other simulations attempt to represent the spectrum of light black hole seeds that could form from population III remnants \citep{Habouzit2017}, based on theoretical expectations for the stellar initial mass function and the fate of stars of different masses \citep{Woosley2002,Hirano2014}.

Black holes in massive galaxies grow by many orders of magnitude and therefore their final mass at $z=0$ should be insensitive to the initial seed mass.  On the other hand, black holes in lower-mass galaxies are expected to grow significantly less and may retain memory of the initial conditions \citep{VolonteriNatarajan2009,vanWassenhove2010,Bonoli2016,Habouzit2017,Angles-Alcazar2017_BHsOnFIRE,Bellovary2019}.  In either case, the choice of seed mass may have strong implications depending on the accretion model (\S\ref{sec:acc}).  If the accretion rate is strongly dependent on black hole mass, then low mass seeds may never reach the conditions for efficient growth even in massive galaxies.  In contrast, if accretion is weakly-dependent on black hole mass, then black holes may converge to a similar mass regardless of the initial seed \citep{Angles-Alcazar2013,Angles-Alcazar2015,Angles-Alcazar2017_BHfeedback,Catmabacak2022}.  
The choice of black hole seed model can also lead to widely different black hole number densities and halo occupation fractions \citep{Volonteri2010,Habouzit2016,DeGraf2020}. \\

\subsection{Black hole dynamics and mergers}\label{sec:dyn}
\begin{figure*}
    \includegraphics[width=0.99\textwidth]{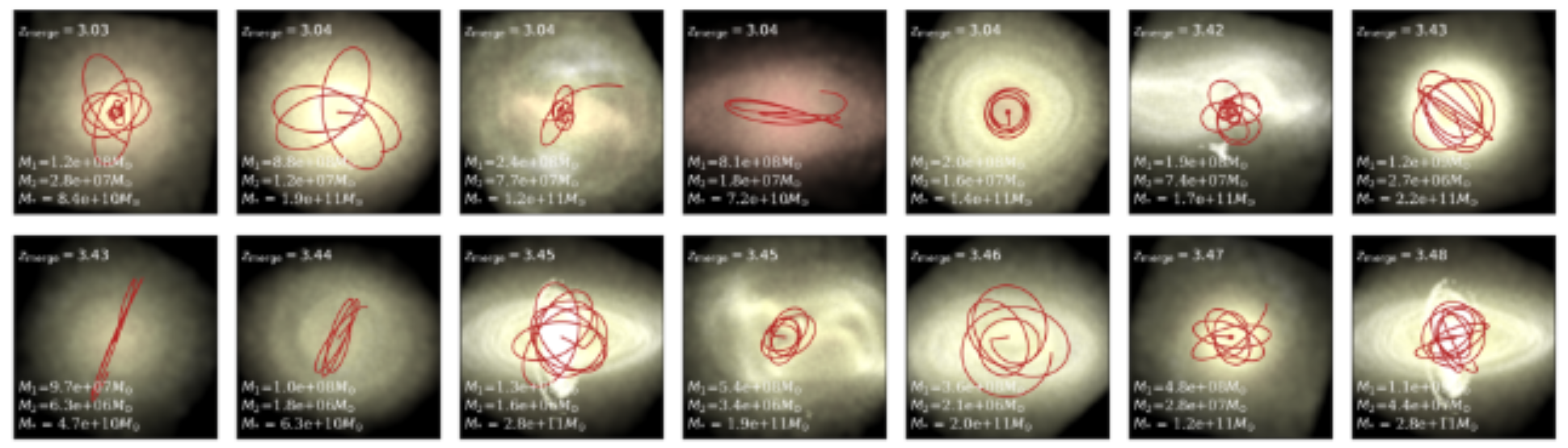}
    \caption{The last few orbits (starting from $\sim 80\,{\rm Myrs}$ before the merger) of a small sample binaries in the large cosmological simulation (Astrid) plotted on their host galaxies. The distance from left to right of each image is $8\,{\rm ckpc}/h$. The brightness corresponds to the stellar density, and the colors show the stellar age with older stars being redder. The red curves are the BH pairs' position relative to their center of mass. Some BH orbits circularize over time (e.g. third row, fifth column), although the majority of the orbits still remain eccentric when merging. Credit: Nianyi Chen.\citep{Chen2021_Astrid}}
    \label{fig:BHorbits}
\end{figure*}



After seed formation, the trajectory of black holes is governed by gravitational interactions with the gas, stellar, and dark matter components.  Any motion of the black hole relative to a collisionless background of stars and dark matter leads to dynamical friction, 
effectively acting as a ``drag force'' owing to the integrated effect of successive gravitational two-body encounters \citep{Chandrasekhar1943}.  
This process can help the black hole sink towards the gravitational potential minimum and remain near the center of the galaxy.
Similarly, the hydrodynamic disturbance generated by a black hole moving relative to the background gas distribution generates a wake slowing down the black hole \citep{Ostriker1999,Beckmann2018}, though feedback in the form of winds or radiation from the black hole can significantly reduce gas dynamical friction or even result in positive net acceleration \citep{Park2017,delValle2018,Gruzinov2020,Toyouchi2020,Bogdanovic2022_LRR}.

Dynamical friction is not properly resolved in cosmological simulations owing to limited mass and gravitational force resolution.  Resolving the gravitational radius of influence of a black hole with mass $M_{\rm BH}$, 
\begin{equation}
r_{\rm inf} \equiv \frac{G \, M_{\rm BH}}{\sigma^2} \approx 1\,{\rm pc} \left( \frac{M_{\rm BH}}{10^7\,{\rm M}_{\odot}} \right) \left( \frac{\sigma}{200\,{\rm km\,s}^{-1}} \right)^{-2},
\end{equation}
where $\sigma$ is the velocity dispersion of the surrounding gas, stellar, or dark matter component,
is only possible in idealized models \citep{Pfister2017} or cosmological simulations with strong hyper-refinement in the nuclear region \citep{Angles-Alcazar2021}. 
However, the typical spatial scale resolved in large-volume cosmological simulations can be orders of magnitude larger ($\sim$1\,kpc), with individual resolution elements representing the gas, stellar, and dark matter components that can be significantly more massive than the black hole particle.

In order to avoid spurious black hole trajectories owing to unresolved dynamical friction, a common approach in galaxy formation simulations is to artificially reposition the black hole to the location of the most bound particle within the neighboring gas (and/or stellar) distribution \citep{Springel2005_BHmodel,Booth2009,Sijacki2015,Feng2016,Angles-Alcazar2017_BHfeedback,Weinberger2017} or within the FoF group that contains the black hole \citep{Dave2019_Simba}.    
Black hole repositioning operates every time step, effectively pinning the black hole location to the center of the host galaxy.  In this scheme, close galaxy encounters and mergers can yield unphysical behavior with non-smooth trajectories of black hole particles, which is partially mitigated by only allowing black hole repositioning if the relative velocity of the most bound particle is lower than some fraction of the local sound speed or their mutual escape velocity.  Alternatively, more gradual repositioning techniques rely on displacing the black hole continuously by small increments in the direction of the stellar mass gradient or the location of the minimum potential \citep{Okamoto2008,Wurster2013,Bahe2021,Hopkins2022_FIRE3,Wellons2022}, resulting in smoother black hole trajectories.  As intended, black hole repositioning increases the density of the ambient gas and reduces the relative velocity of the black hole, increasing the accretion rate and the overall efficiency of black hole feedback in simulations \citep{Wurster2013,Tremmel2017_Romulus,Bahe2021}.  However, artificial repositioning precludes the study of black hole dynamics in galaxies and overestimates the rate of black hole merger events, requiring post-processing calculations to account for slower orbital decays \citep{Chen2022_dynfric}.

Boosting the dynamical mass of the black hole until its physical mass becomes significantly larger than the dark matter particles is an alternative to allow for non-trivial gravitational dynamics while avoiding stochastic trajectories of low-mass black holes \citep{Angles-Alcazar2017_BHsOnFIRE}, which can mimic the effect of black holes embedded in tightly bound stellar structures with large effective mass that can sink more efficiently to the galactic center \citep{Biernacki2017,MaLinhao2021}.
A more physically motivated approach to modeling black hole dynamics in cosmological simulations is to include a subgrid prescription for dynamical friction \citep{Tremmel2015_DynFriction,Tremmel2017_Romulus,Bellovary2019,Pfister2019,Ni2021_astrid,Chen2022_dynfric,MaLinhao2021}.  Most current models are based on the traditional \citet{Chandrasekhar1943} dynamical friction formula assuming a homogeneous, infinite, idealized background medium with isotropic velocity distribution, where the deceleration experienced by a black hole of mass M$_{\rm BH}$ is given by: 
\begin{align}
\mathbf{a}_{\rm DF} &= -4\pi\mathrm{G}^2\,\mathrm{M_{\rm BH}}\,\mathrm{m}_{\rm a}\,\mathrm{ln}\Lambda\frac{\mathbf{v}_{\rm BH}}{\mathrm{v}_{\rm BH}^3}\int_{0}^{\mathrm{v_{\rm BH}}} d\mathrm{v}_{\rm a}\mathrm{v}_{\rm a}^2 f(\textbf{v}_{\rm a}) \label{eq:dynfric1} \\
&\approx -4\pi\mathrm{G}^2\,\mathrm{M_{\rm BH}}\,\rho(<\mathrm{v}_{\rm BH})\,\mathrm{ln}\,\Lambda\,\frac{\mathbf{v}_{\rm BH}}{\mathrm{v}_{\rm BH}^3}, \label{eq:dynfric2}
\end{align}
where $\textbf{v}_{\rm BH}$ is the black hole velocity relative to the surrounding medium, $m_{\rm a}$ and $\textbf{v}_{\rm a}$ are the masses and velocities of the background particles, and the second expression replaces the integral over the velocity distribution $f(\textbf{v}_{\rm a})$ by the density $\rho(<\mathrm{v}_{\rm BH})$ of particles moving slower than the black hole \citep{Tremmel2015_DynFriction}.  The Coulomb logarithm $\mathrm{ln}\Lambda \approx \mathrm{ln}(b_{\rm max}/b_{\rm min})$ depends on the maximum ($b_{\rm max}$) and minimum ($b_{\rm min}$) impact parameters.  The spatial scale at which $\rho$, $\textbf{v}_{\rm BH}$, and $\Lambda$ are measured introduces some ambiguity in the calculation of dynamical friction.
In practice, the gravitational force resolution scale (or the size of the black hole ``interaction'' kernel) is often used to evaluate Equations~\ref{eq:dynfric1}-\ref{eq:dynfric2}, though more recent dynamical friction estimators attempt to generalize better to cases which violate these idealized conditions and remove the ambiguity in estimating ill-defined continuum quantities \citep{MaLinhao2021}.

Some simulations implement a drag force on the black hole from the surrounding gas distribution motivated by the analytical approximation of \citet{Ostriker1999}:
\begin{equation}
\label{eq:drag}
    \mathbf{F}_{\rm drag} = -4 \pi\rho \left( \frac{G M_{\rm BH}}{c_{\rm s}} \right)^2 \times \mathcal{I(M)}\frac{\bf{v}_{\rm BH}}{\mathrm{v}_{\rm BH}},
\end{equation}
where here $\rho$ is the gas density, $c_{\rm s}$ is the sound speed, and $\mathcal{I(M)}$ is a non-dimensional function that encapsulates the dependence on the Mach number $\mathcal{M} = \mathrm{v}_{\rm BH}/c_{\rm s}$ \citep{Dubois2015_SNa,Chen2022_dynfric}.  Some models simplify this expression to depend explicitly on the gas accretion rate $\dot{M}_{\rm BH}$, with deceleration given by $\mathbf{a}_{\rm drag} \approx - \mathbf{v}_{\rm BH} \, \dot{M}_{\rm BH}/M_{\rm BH}$ \citep{Biernacki2017,Ni2021_astrid}.  Dynamic friction from stars generally dominates over the gas drag \citep{Pfister2019,Chen2022_dynfric} but some simulations artificially boost the gas drag to help stabilize black hole trajectories when not including dynamic friction from collisionless particles \citep{Dubois2015_SNa,Dubois2021_newhorizon} or in gas-dominated systems at high redshift \citep{Trebitsch2020}.
In either case, despite the addition of subgrid dynamical friction, scattering through two-body interactions can still affect the black hole trajectory if its mass is similar to that of the background particles.  In practice, this implies that cosmological simulations often still need to either artificially boost the dynamical mass of the black hole or employ massive seeds ($\gtrsim10^6$\,\Msun) to avoid spurious dynamical heating \citep{Tremmel2015_DynFriction,Tremmel2017_Romulus,Pfister2019,Chen2022_dynfric}.

Despite the technical difficulties of modeling black hole dynamics in a galaxy evolution context, many independent studies indicate that efficient accretion requires black holes to remain tightly bound to the galaxy center where the highest average gas densities occur \citep{Angles-Alcazar2017_BHsOnFIRE,Trebitsch2020,Bahe2021,Catmabacak2022}. 
This can be achieved by dynamic friction in relatively massive galaxies with well-defined, dense, and stable central regions, but it may not be possible in turbulent, clumpy, high-redshift galaxies where frequent dynamical perturbations from mergers, bursty star formation, stellar feedback, and massive clumps can significantly increase the time-scale for orbital decay of even massive black hole seeds \citep{Fiacconi2013,Roskar2015,Angles-Alcazar2017_BHsOnFIRE,Tamburello2017,Tremmel2017_Romulus,Tamfal2018,Tremmel2018,Pfister2019,Ricarte2019,Volonteri2020,MaLinhao2021}.
This black hole ``sinking problem'' \citep{MaLinhao2021} can explain the observed off-center location of AGN in dwarf galaxies \citep{Reines2020,Mezcua2020}, attributed to wandering black holes \citep{Bellovary2019,Bellovary2021,Ricarte2021,Sharma2022}, but may represent a challenge for the early growth of the highest redshift QSOs and significantly affect the rate of black hole--black hole mergers.

Galaxy merger remnants will inevitably contain two or more massive black holes that may eventually merge, but cosmological simulations lack the resolution to follow this process in detail.  In simulations implementing artificial repositioning, any two black holes are allowed to merge instantaneously if they are located within their interaction kernel \citep{Sijacki2015,Feng2016,Weinberger2017}.  Given the $\sim$kpc scale resolution of large-volume simulations, black holes can merge during close fly-bys even at high relative velocity during the early stages of galaxy interactions, artificially enhancing the merger rate.  In some simulations with repositioning, merging black holes are also required to have a relative velocity lower than their mutual escape velocity \citep{Schaye2015,Angles-Alcazar2017_BHfeedback,Dave2019_Simba}.  This helps prevent black holes from merging during the initial stages of galaxy mergers, though the actual black hole velocities are ill-defined owing to the repositioning approach.  Simulations implementing subgrid dynamical friction, or modeling black holes massive enough that  dynamical friction is resolved, can follow the evolution of merging black holes for longer time and implement a more physically meaningful gravitational binding criteria for allowing black hole mergers at the resolution scale \citep{Angles-Alcazar2017_BHsOnFIRE,Tremmel2017_Romulus,Chen2022_dynfric}.   
Gravitational-wave recoils of merging black holes are usually neglected in cosmological simulations but may influence their dynamics, growth, and overall impact of feedback \citep{Blecha2011,Sijacki2011,Blecha2016}. \\

\subsection{Black hole accretion}\label{sec:acc}

\begin{figure*}
\begin{center}
\includegraphics[width=0.9\textwidth]{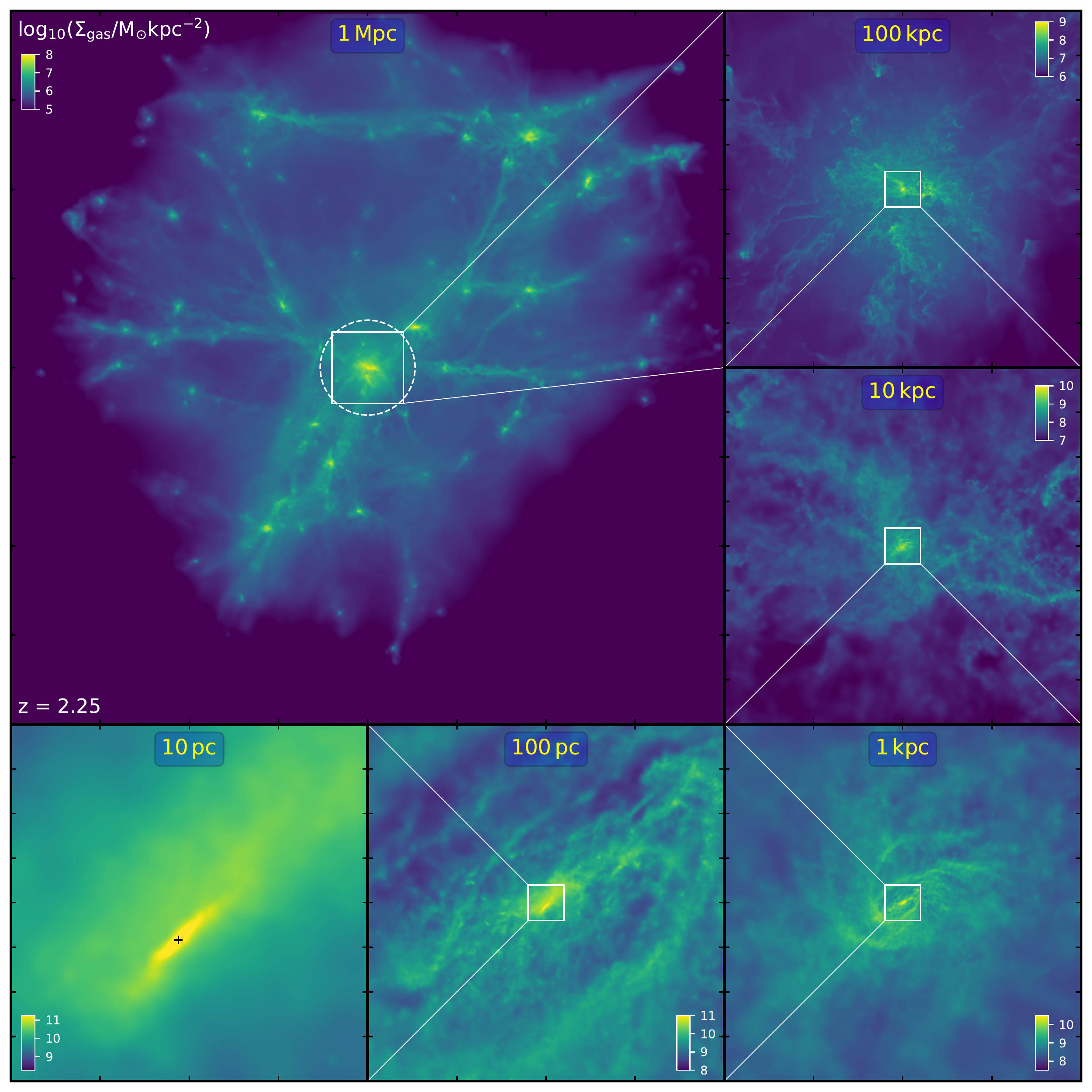}
\end{center}
\caption{
Distribution of gas across scales in a cosmological hyper-refinement simulation of a massive galaxy ($M_{\rm star} \sim 10^{10.5}$) at $z\sim2$. The top left panel shows a Mpc-scale region containing tens of galaxies connected by cosmic filaments. Subsequent panels zoom in progressively into the nuclear region of the most massive galaxy and down to the vicinity of the central massive black hole, resolving a pc-scale, rotationally supported disk of accreting gas. Strong gravitational torques from non-axisymmetric perturbations in the stellar potential drive a sub-pc gas inflow rate of a few \Msunyr, sufficient to power a luminous quasar.  Cosmological large-volume simulations cannot resolve the dominant mechanisms driving gas transport on scales $\lesssim$100\,pc--1\,kpc, requiring the implementation of sub-grid accretion prescriptions to parameterize the inflow of gas on unresolved scales.  Figure reproduced from \citet{Angles-Alcazar2021}.
}
\label{fig:qsomap} 
\end{figure*}

Cosmological large-volume simulations cannot resolve the dominant mechanisms driving gas transport from galaxy scales down to the black hole accretion disk, requiring the implementation of sub-grid accretion models to infer the inflow of gas on unresolved scales (Fig.~\ref{fig:qsomap}).
The Bondi accretion model \citep{Bondi1944,Bondi1952} is the most widely used prescription for black hole growth in galaxy formation simulations since its first implementation in idealized galaxy merger simulations \citep{Springel2005_BHmodel,DiMatteo2005}. For a black hole of mass $M_{\rm BH}$,
moving at velocity $v$ relative to a uniform distribution of gas with
density $\rho$ and sound speed $c_{\rm s}$, the Bondi rate is given by
\begin{equation}\label{eq:bondi}
\dot{M}_{\rm Bondi} = \alpha \, \frac{4\pi \, G^{2} \, M_{\rm BH}^{2} \, \rho}{(c_{\rm s}^{2}+v^{2})^{3/2}},
\end{equation}
where $G$ is the gravitational constant and $\alpha$ is a normalization factor often included to boost the accretion rate (see below).  The gas properties are measured locally around the black hole within an interaction {\it kernel} usually defined to contain the nearest fluid elements.  Depending on the resolution of the simulation, the black hole kernel represents a physical size typically ranging from $\sim$10--100\,pc in idealized and high-resolution cosmological zoom-in simulations to $\sim$1\,kpc in simulations of large cosmological volumes or very massive objects such as galaxy clusters.  The Bondi radius $r_{\rm B} \equiv G M_{\rm BH}/c_{\rm s}^2$ represents the gravitational black hole radius of influence relative to gas supported by thermal pressure and it is often not resolved in cosmological simulations.

The strong dependence $\dot{M}_{\rm Bondi} \propto M_{\rm BH}^{2}$ implies a transition between suppressed accretion rate at low $M_{\rm BH}$ and very fast growth at high $M_{\rm BH}$.  Assuming constant gas density and sound speed, Equation~\ref{eq:bondi} yields
\begin{equation}\label{eq:bondi_evol}
M_{\rm BH}(t) = \frac{M_{\rm seed}}{1 - t/t_{\rm B}},
\end{equation}
where 
\begin{equation}\label{eq:bondi_time}
t_{\rm B} = \frac{c_{\rm s}^3}{ \alpha \, 4\pi \, G^{2} \, \rho \, M_{\rm seed}}
\end{equation}
is the timescale for divergence in $M_{\rm BH}$ (assuming $v=0$) and $M_{\rm BH}(t=0) \equiv M_{\rm seed}$ is the initial mass of the black hole seed.  For typical kpc-scale conditions of star-forming gas in large volume simulations ($c_{\rm s} \sim 20$\,\kms and $\rho \sim 0.13\,m_{\rm p}\,{\rm cm}^{-3}$, with $m_{\rm p}$ the proton mass), where the multi-phase ISM is not resolved, we obtain
\begin{equation}\label{eq:bondi_time_phys}
t_{\rm B} \approx 100\,{\rm Gyr} \,  \left (\frac{1}{\alpha} \right ) \,   \left (\frac{10^5\,{\rm M}_{\odot}}{M_{\rm seed}} \right ),
\end{equation}  
implying that a black hole with $M_{\rm seed}=10^5$\,\Msun~would never reach efficient growth in standard Bondi accretion ($\alpha = 1$).  In order to mitigate this problem, a boost factor of order $\alpha \sim 100$ is often adopted, which partially compensates the inability of cosmological simulations to resolve the Bondi radius and the multiphase structure of gas \citep{Springel2005_BHmodel}.  Alternative implementations include a density-dependent boost factor $\alpha \propto \rho^{2}$ \citep{Booth2009}, or evaluating Bondi accretion for the inferred cold gas phase (increasing $\rho$ and decreasing $c_{\rm s}$)  in the context of subgrid ISM models with relatively high mean gas temperatures owing to stellar feedback \citep{Pelupessy2007}.  These implementation choices together with $M_{\rm seed}$ can thus have significant effects on early black hole growth.

In the opposite regime, Bondi accretion reaches supra-exponential growth on very short timescales for massive black holes, with $t_{\rm B} \approx 100$\,Myr for $M_{\rm BH}=10^8$\,\Msun.  In practice, the divergence in $M_{\rm BH}$ at $t = t_{\rm B}$ (Equation~\ref{eq:bondi_evol}) is avoided by limiting the accretion rate to the Eddington limit, which represents the maximum growth rate that can be achieved through spherically symmetric accretion in the presence of radiation pressure:
\begin{equation}\label{eq:edd}
\dot{M}_{\rm Edd} = \frac{4\pi \, G \, m_{\rm p} \,M_{\rm BH}}{\epsilon_{\rm r}\,\sigma_{\rm T}\,c},
\end{equation}
where $\epsilon_{\rm r}$ is the radiative efficiency, $\sigma_{\rm T}$ is the Thomson scattering cross-section, and $c$ is the speed of light.   Some models allow black holes to exceed $\dot{M}_{\rm Edd}$ by factors of a few, which is consistent with detailed radiation hydrodynamic simulations of non-spherical accretion flows indicating that super-Eddington feeding is indeed possible \citep{Jiang2014,Jiang2019}, provided that the inflow rate from larger scales is sufficiently high \citep{Lupi2016,Angles-Alcazar2021}.
Accretion at the Eddington rate yields exponential growth with $e$-folding time given by the Salpeter time,
\begin{equation}\label{eq:salpeter_time}
t_{\rm S} \equiv \frac{\epsilon_{\rm r}\,\sigma_{\rm T}\,c}{4\pi \, G \, m_{\rm p}} \approx 45\,{\rm Myr},
\end{equation}
where we assume $\epsilon_{\rm r} = 0.1$ \citep{YuTremaine2002}.  Continuous Eddington growth would thus quickly produce over-massive black holes, implying that Eddington-limited Bondi accretion requires strong self-regulation by AGN feedback to reproduce observations such as the black hole--galaxy scaling relations \citep{Kormendy2013,Graham2016}.

Equation~\ref{eq:bondi} neglects radiative cooling, the angular momentum of inflowing gas, the gravitational influence of the gas and stellar components, and gas consumption by star formation among other key physical processes affecting gas transport in galaxy discs.  High-resolution simulations of galactic nuclei show that Bondi accretion may indeed fail to reproduce gas inflow rates by orders of magnitude under a variety of conditions \citep{Hopkins2010_MultiScale,Hopkins2011_Analytic,Hobbs2012,Gaspari2013,Hopkins2016_NuclearSims,Negri2017,Beckmann2018,Angles-Alcazar2021}, demonstrating the challenge of developing accurate predictors of gas inflow rates across multiple spatial scales. 

Despite these limitations, models based on Bondi accretion have been very successful at reproducing global galaxy and black hole observables when coupled with suitable AGN feedback prescriptions \citep{DiMatteo2008,Schaye2015,Sijacki2015,Steinborn2015,Feng2016,Rosas-Guevara2016,Habouzit2021_Mbh,Ni2021_astrid}.
These include variations of the Bondi model that attempt to account for the gas angular momentum.
Considering a reduction of the Bondi radius owing to a decreased effective gravitational potential due to gas rotational support yields a suppression of Bondi accretion by a factor $\propto (c_{\rm s}/v_{\rm \phi})^4$ when $v_{\rm \phi} > c_{\rm s}$, where $v_{\rm \phi}$ represents the gas rotational velocity at the resolution scale \citep{Tremmel2017_Romulus}. Considering instead the characteristic inflow time of Bondi accretion compared to the timescale of viscous accretion from the circularization radius yields a suppression of Bondi accretion by a factor $\propto (c_{\rm s}/v_{\rm \phi})^3$ \citep{Rosas-Guevara2015}.  In either case, Bondi-based accretion prescriptions retain the strong $M_{\rm BH}^{2}$ dependence.

Adaptive refinement techniques can achieve significantly higher dynamic range than standard cosmological simulations by splitting fluid elements dynamically (thereby increasing resolution) as gas approaches the central black hole \citep{Beckmann2019,Angles-Alcazar2021}, albeit at the expense of significant increase in computation cost (Fig.~\ref{fig:qsomap}).  Cosmological hyper-refinement simulations of gas-rich, quasar-host galaxies show that the inflowing gas across 1\,pc--10\,kpc scales is primarily cool, with rotational support dominating over turbulence and thermal pressure \citep{Angles-Alcazar2021}. Under a range of conditions, these simulations show that gravitational torques from multi-scale stellar non-axisymmetries dominate angular momentum transport over gas self-torquing and pressure gradients, with the gas inflow rate on sub-pc scales weakly dependent on $M_{\rm BH}$.  This is consistent with earlier idealized simulations of galactic nuclei \citep{Hopkins2010_MultiScale,Hopkins2011_Analytic} and points to a departure from the assumptions in Bondi-based accretion models.

Under the assumption that gravitational torques from non-axisymmetric perturbations in the stellar component (driven by galaxy interactions and instabilities in self-gravitating nuclear gas disks) induce strong gas orbit crossing and shocks that dissipate energy and angular momentum, the inflow rate is given by 
\begin{equation}\label{eq:torque_base}
\dot{M}_{\rm inflow}(R) \sim \frac{ |a|\, M_{\rm gas}}{t_{\rm dyn}},
\end{equation}
where $M_{\rm gas}$ is the gas mass within $R$, $t_{\rm dyn} \equiv (R^{3}/GM_{\rm enc})^{1/2}$ is the dynamical time, $M_{\rm enc}$ is the total enclosed mass within $R$ (including the gas and stellar components in addition to $M_{\rm BH}$), and $|a|$ is the fractional amplitude of the non-axisymmetric perturbation to the potential \citep{Hopkins2011_Analytic}.  Equation~\ref{eq:torque_base} and additional considerations about the dominant perturbation modes on different scales and their amplitude, the dependence of global gravitational instability on bulge-to-disk ratio, and the local balance between inflow and star formation represent the basis of an alternative accretion rate estimator based on gravitational torques:
\begin{equation}\label{eq:torque} 
\begin{split}
\dot{M}_{\rm Torque} \; \approx \; \alpha_{\rm T}  \, f_{\rm d}^{5/2} \times 
\left ( \frac{M_{\rm BH}}{10^{8}\,{\rm M_{\odot}}} \right )^{1/6} 
\left ( \frac{M_{\rm enc}(R)}{10^{9}\,{\rm M_{\odot}}} \right ) \\
\times \left ( \frac{R}{100\,{\rm pc}} \right )^{-3/2}  \left (1 +
\frac{f_{0}}{f_{\rm gas}} \right )^{-1} \, {\rm M_{\odot}\,yr^{-1}},
\end{split}
\end{equation}  
where $f_{\rm d} \equiv M_{\rm d}/M_{\rm enc}$ is the disk mass fraction (including both stars and gas), $f_{\rm gas}$ is the gas mass fraction in the disk component, $f_{0} \approx 0.31 \, f_{\rm d}^{2} \, (M_{\rm d}/10^{9}{\rm M_{\odot}})^{-1/3}$, and all quantities are evaluated within a distance $R$ corresponding the size of the black hole kernel \citep{Hopkins2011_Analytic}.  The normalization factor $\alpha_{\rm T} \approx 1$--10 parametrizes the dependence of sub-pc inflow rates on the assumed Schmidt-Kennicutt law of star formation on unresolved scales and the galaxy stellar density profile.

Cosmological simulations implementing gravitational torque-driven accretion show important qualitative differences compared to Bondi-based accretion models \citep{Angles-Alcazar2013,Angles-Alcazar2015,Angles-Alcazar2017_BHfeedback,Angles-Alcazar2017_BHsOnFIRE,Catmabacak2020,Wellons2022}.  The inflow rate driven by gravitational instabilities and resulting torques is nearly independent of $M_{\rm BH}$.  This implies that black holes do not show the characteristic behavior seen in Equation~\ref{eq:bondi_evol} for Bondi-based models, where a marked transition occurs from suppressed black hole growth at low $M_{\rm BH}$ to supra-exponential growth at high $M_{\rm BH}$.  Instead, gravitational torque-driven inflow predicts efficient growth of under-massive black holes (without the need for an $\alpha$ accretion boost factor) and does not require strong self-regulation of massive black holes by galaxy-scale AGN feedback.  Explicitly including the physics of gravitational torques between the stellar and gas components therefore yields qualitatively different results to that of modifications of Bondi accretion that attempt to incorporate angular momentum transport.

The gravitational torque model assumes conditions relevant for black hole fueling in self-gravitating gas disks with a dominant stellar component.  These conditions are less likely to apply in the gas dominated regime at early times, where other processes such as scattering of dense gas clouds or turbulent transport may be required \citep{Bournaud2011,Hobbs2011,Gabor2013}, or the gas poor regime at late times in massive elliptical galaxies where Bondi accretion of hot, pressure supported gas with low angular momentum may be a better representation.  
Ideally, cosmological simulations should implement black hole accretion prescriptions able to incorporate the dominant gas transport mechanisms operating on different regimes as the host galaxy evolves.  A first attempt at implementing a hybrid accretion model considered separately the cold ($T<10^5$\,K) and hot ($T>10^5$\,K) gas components within the black hole kernel, modeling the accretion of cold, rotationally supported gas following the gravitational torque model and the accretion of hot, pressure supported gas following the Bondi parameterization \citep{Dave2019_Simba,Thomas2019,Thomas2021}.

In addition to modeling the inflow of gas across galaxy scales, simulations should also consider the transport of gas within the black hole accretion disk itself. Some models incorporate an intermediate accretion disk reservoir from which the black hole grows at a rate motivated by accretion disk theory \citep{Power2011,Fiacconi2018,Talbot2021,Hopkins2022_FIRE3,Wellons2022}.  In some cases, simulations track the evolution of black hole spin, which can modify the radiative efficiency of the accretion disk and the efficiency of accretion-driven winds and jets \citep{Dubois2014_spin,Bustamante2019,Talbot2021}. \\

\subsection{Black hole feedback}\label{sec:feedback}      
            

Radiatively efficient accretion disks convert a significant fraction of the rest mass energy of accreted material into radiation, with the total bolometric luminosity given by
\begin{equation}\label{eq:Lbol}
L_{\rm bol} = \epsilon_{\rm r} \, \dot{M}_{\rm BH}\,c^2,
\end{equation}
where the radiative efficiency depends on the spin of the black hole and can range from $\epsilon_{\rm r} \sim 0.05$--0.35 \citep{Bardeen72}.
The total integrated power radiated by the central massive black hole is thus expected to be orders of magnitude larger than the binding energy of the host galaxy,
\begin{equation}\label{eq:bh_energy}
\frac{ \epsilon_{\rm r} \,M_{\rm BH}\,c^2}{M_{\rm star}\,\sigma^2} \sim 100 \times \left ( \frac{\epsilon_{\rm r}}{0.1} \right ) \left ( \frac{300\,{\rm km\,s}^{-1}}{\sigma} \right )^2,
\end{equation}
where $\sigma$ is the galaxy stellar velocity dispersion and we have assumed a typical black hole mass to galaxy stellar mass ratio $M_{\rm BH}/M_{\rm star} \sim 10^{-3}$.  Radiation can affect the thermodynamic state of gas thorough Compton, photoionization, and photo-electric heating and driving winds through radiation pressure on free electrons and dust or via coupling to spectral lines.  Accretion disks can also drive mechanical winds and jets that can extract a larger fraction of the accreted rest mass energy $\dot{M}_{\rm BH}\,c^2$ compared to radiation \citep{Fabian2012}. Even if only a small fraction of $L_{\rm bol}$ couples to the surrounding gas through some of these processes, Equation~\ref{eq:bh_energy} implies that AGN feedback can have a significant impact in galaxy evolution \citep{Silk1998,DiMatteo2005,Murray2005}.

Observations of AGN feedback in action include fast nuclear outflows \citep{Tombesi2013,Nardini2015}, galaxy-scale winds \citep{Greene2012_QSOoutflow,Cicone2014,Zakamska2014,Wylezalek2020}, radio-emitting jets \citep{Fabian2012,Hlavacek-Larrondo2012}, and ionized QSO proximity zones \citep{Trainor2013,Eilers2017}.  Regardless of the specific form of feedback, the energy is originated in the central engine on scales comparable to the Schwarzschild radius,
\begin{equation}\label{eq:Rs}
R_{\rm s} \equiv \frac{ 2 G M_{\rm BH}}{c^2} \approx 10^{-6}\,{\rm pc} \times \left ( \frac{M_{\rm BH}}{10^{7}\,{\rm M}_{\odot}} \right ),
\end{equation}
but can affect the properties of gas out to Mpc scales.  Modeling the generation and impact of feedback across more than ten orders of magnitude in spatial scales is computationally unfeasible.  The implementation of AGN feedback in galaxy formation simulations is thus schematic by necessity, owing to limited resolution but also to uncertainties in observational constraints and the lack of a full theoretical understanding of the different feedback channels.

A popular AGN feedback model since its first implementation in idealized galaxy merger simulations assumes that a fraction $\epsilon_{\rm f}$ of the bolometric luminosity couples to the surrounding gas in the form of thermal energy \citep{Springel2005_BHmodel,DiMatteo2005}.  The input energy $\dot{E} = \epsilon_{\rm f}\,L_{\rm bol}$ is deposited isotropically into the nearest gas resolution elements, typically within the black hole kernel used to evaluate the accretion prescription (\S\ref{sec:acc}).  This thermal feedback model represents the basis for many AGN feedback implementations in large-volume cosmological simulations \citep{DiMatteo2008,Dubois2014,Hirschmann2014,Schaye2015,Feng2016,Tremmel2017_Romulus,Weinberger2017,Ni2021_astrid}.  The feedback efficiency $\epsilon_{\rm f}$ is a free parameter that represents the expected impact of AGN feedback, without necessarily corresponding to a specific physical mechanism, and it is usually calibrated to reproduce the observed black hole--galaxy scaling relations.  The overall effect of thermal feedback depends on the resolution of the simulation, the accretion prescription, and other implementation details, but typical values in the range $\epsilon_{\rm f}\sim 0.05$--0.15 have been shown to yield efficient self-regulation of black hole growth while generating strong outflows that help reduce the star formation rate in galaxies.  

Reproducing the observed decline in the abundance of galaxies at the high mass end \citep{Baldry2012,Bernardi2017} and the bimodality in galaxy colors with a growing population of red galaxies at low redshift \citep{Baldry2006,Brammer2009} requires efficient quenching of star formation.
Rapid cooling in metal-rich gas can decrease the efficiency of thermal feedback by radiating away a fraction of the injected energy, which may represent a challenge to fully quenching massive galaxies.  Some simulations artificially accumulate the feedback energy produced by each black hole until it is enough to heat the neighboring gas elements to a minimum temperature $T \sim 10^9$\,K where cooling is inefficient, reducing cooling losses \citep{Schaye2015}. Other approaches resort to temporarily inhibiting cooling of gas that has received a thermal feedback injection \citep{Tremmel2017_Romulus}.  
Alternatively, simulations incorporating mechanical energy deposition instead of thermal feedback benefit from the fact that the injected momentum cannot be radiated away, reducing cooling losses and potentially increasing the impact of feedback \citep{Choi2015_CosmoSim}.  

The momentum injection rate owing to radiation pressure on dust can be approximated as
\begin{equation}\label{eq:rad_dust}
\dot{P}_{\rm rad} = \tau\, \frac{L_{\rm bol}}{c},
\end{equation}
where $\tau > 1$ corresponds to the total FIR optical depth in the nuclear region assuming absorption of UV radiation
by dust and multi-scattering re-radiation in the infrared.  Idealized galaxy merger simulations suggest that radiation pressure on dust assuming $\tau = 10$ can efficiently self-regulate black hole growth \citep{Debuhr2010,Debuhr2012}, and more detailed radiation hydrodynamic simulations show that radiation pressure-driven feedback may be an important ingredient in regulating star formation in compact starbursts in the initial obscured QSO phase, producing galactic winds qualitatively different to that of thermal feedback models \citep{Costa2018a,Costa2018b}.  The increased computational cost of radiation hydrodynamics has limited its use in galaxy formation simulations, but approximate radiative transport methods allow more efficient implementation of radiative feedback in cosmological simulations \citep{Hopkins2016_NuclearSims,Hopkins2022_FIRE3}.

The momentum and kinetic energy imparted by mechanical winds can be modeled in simulations by prescribing the mass loading factor $\eta_{\rm m} \equiv \dot{M}_{\rm out} / \dot{M}_{\rm BH}$ (parameterizing the mass outflow rate relative to the black hole accretion rate) and the corresponding outflow velocity $v_{\rm out}$.  With these choices, the ``momentum loading" and the kinetic energy efficiency relative to the bolometric luminosity are given by
\begin{equation}\label{eq:facc}
\eta_{\rm p}  \, \equiv \, \frac{\dot{P}_{\rm wind}}{L_{\rm bol}/c}  \, = \, \frac{\eta_{\rm m}}{\epsilon_{\rm r}} \, \left ( \frac{v_{\rm out}}{c} \right ),
\end{equation}
\begin{equation}\label{eq:ekin}
 \epsilon_{\rm k} \, \equiv \, \frac{\dot{E}_{\rm wind}}{L_{\rm bol}} \, = \, \frac{\eta_{\rm m}}{2\,\epsilon_{\rm m}} \left ( \frac{v_{\rm out}}{c} \right )^{2}.
\end{equation} 
In some implementations of mechanical winds in Lagrangian hydrodynamics, gas resolution elements within the black hole kernel receive a velocity kick corresponding to $v_{\rm out}$ with a probability set to reproduce the prescribed mass outflow rate on average \citep{Choi2012_BHmodel,Choi2015_CosmoSim,Angles-Alcazar2017_BHfeedback}.  Alternatively, new gas elements can be introduced into the simulation to represent the wind at higher mass resolution \citep{Richings2018_MolecularOutflow,Torrey2020,Richings2021,Hopkins2022_FIRE3,Wellons2022}.
 In the case of grid-based simulations, mass and momentum are injected into neighbouring gas cells according to the prescribed values of $\eta_{\rm m}$ and $v_{\rm out}$ \citep{Kim2011}.  Isotropic winds are generally implemented by assigning velocity kicks directed radially outward from the black hole, but equatorial winds \citep{Hopkins2016_NuclearSims} and collimated winds \citep{Dave2019_Simba} have also been implemented in simulations.

The appropriate mass loading and wind velocity depend on the resolution of the simulation, which sets the physical scale at injection and therefore the type of winds that are represented.  High resolution simulations (usually idealized) can attempt to model nuclear winds similar to those observed in broad absorption line QSOs \citep{Gibson2009} and in absorption against X-ray emission from AGN \citep{Tombesi2013,Nardini2015,Gofford2015}, with mildly relativistic velocity ($v_{\rm out} \sim 30,000$\,\kms) and mass outflow rate comparable to the black hole accretion rate ($\eta_{\rm m} \sim 1$) \citep{Hopkins2016_NuclearSims,Richings2018_MolecularOutflow,Torrey2020,Richings2021,Costa2020}.  These winds appear to generate on accretion-disk scales ($<$100\,$R_{\rm s}$) and are broadly consistent with predictions from accretion disk simulations, corresponding to a kinetic energy efficiency $\epsilon_{\rm k} \sim 0.05$ similar to the feedback coupling efficiency used in simulations implementing thermal feedback.  Analytic models suggest that inefficient cooling of high-velocity shocked winds entraining ISM gas results in energy-conserving outflows where the momentum flux can be boosted by $\eta_{\rm p} \sim 20$ \citep{Faucher-Giguere2012_WindModel,Zubovas2012}, which may explain the highly mass-loaded ($\eta_{\rm m} \sim1,000$) and slower ($v_{\rm out} \sim 1,000$\,\kms) outflows observed on kpc scales \citep{Greene2012_QSOoutflow,Cicone2014,Zakamska2014}.  Cosmological large-volume simulations cannot resolve this process but can instead implement mechanical AGN winds with the boosted momentum flux and velocity expected on kpc scales \citep{Angles-Alcazar2017_BHfeedback,Dave2019_Simba}.

Some simulations implement two different modes of AGN feedback motivated by observations of two distinct populations of AGN, with (1) ``quasar-mode'' (or ``radiative-mode'') AGN associated with radiatively efficient accretion at high Eddington ratio ($\lambda_{\rm Edd}\equiv \dot{M}_{\rm BH}/\dot{M}_{\rm Edd} \gtrsim 0.01$) in less massive black holes growing in star-forming galaxies, and (2) ``jet-mode'' (or ``radio-mode'') AGN associated with radiatively inefficient accretion at low Eddington ratio ($\lambda_{\rm Edd} \lesssim 0.01$) in more massive black holes in early type-galaxies \citep{Heckman2014}.
Quasar-mode feedback is associated with intense radiation and galaxy-scale outflows with high momentum flux in luminous AGN while jet-mode feedback is primarily driven by highly collimated jets of relativistic particles which inflate hot, X-ray emitting bubbles in galaxy groups and clusters.
The role and impact of each AGN feedback mode is still not fully understood and their outcome in simulations depends on their specific implementation, but quasar-mode feedback is believed to be important in regulating black hole growth and central densities in massive galaxies while jet-mode feedback is believed to be most relevant to prevent cooling of hot gas in massive halos and maintain massive galaxies quenched. In practice, the distinction of feedback modes in simulations allows for different forms of energy injection and efficiencies operating in different regimes, which provides more flexibility to accomplish quenching in massive galaxies, groups, and clusters, while not significantly affecting lower mass galaxies.

The first two-mode AGN feedback model in cosmological hydrodynamic simulations was implemented as thermal feedback injection either in the vicinity of the black hole for the quasar mode or at two locations outside of the galaxy and in opposite sides to represent the expansion of hot bubbles driven by jet-mode feedback \citep{Sijacki2007}.  A similar model was implemented into the Illustris large volume simulation \citep{Genel2014,Vogelsberger2014}, with the thermal coupling efficiency increasing from $\epsilon_{\rm f} = 0.05 \rightarrow 0.35$ when black holes transition from quasar to radio mode feedback, successfully quenching galaxies but over-evacuating gas from high mass halos \citep{Genel2014}.  A more recent implementation of radio-mode feedback in IllustrisTNG replaced thermal coupling by the injection of mechanical winds with spherical symmetry and higher coupling efficiency ($\epsilon_{\rm k} \sim 1$), producing a red galaxy sequence and simultaneously more realistic thermodynamic profiles in large haloes \citep{Weinberger2017}.  
Other implementations of radio-mode feedback in large-volume simulations attempt to represent relativistic jets, though resolving their propagation and structure remains a challenge even in idealized simulations \citep{Bourne2017,Weinberger2017_jet,Su2021_Jets}. 
The Horizon-AGN simulation \citep{Dubois2014} implements quasar-mode feedback as ``standard'' thermal coupling and a transition to mechanical bipolar outflows with velocity $\sim$10,000\,\kms, corresponding to an increase in feedback efficiency $\epsilon_{\rm f} = 0.15 \rightarrow 1$.  In contrast, the SIMBA simulation \citep{Dave2019_Simba} implements collimated outflows with constant momentum flux ($\eta_{\rm p} = 20$) in both feedback modes \citep{Angles-Alcazar2017_BHfeedback}, with a mass-dependent outflow velocity increasing from $v_{\rm out} \sim$ 1,000 $\rightarrow$ 8,000\,\kms~when transitioning from quasar mode to full-speed jets. 
Implementations of two-mode AGN feedback in cosmological hydrodynamic simulations usually rely on a threshold in Eddington ratio ($\lambda_{\rm Edd} \sim 0.01$) above and below which black holes operate in the quasar and jet feedback mode, motivated by observations and accretion disk theory.  However, the success of some models requires additional criteria such as a minimum black hole mass required to transition to radio-mode feedback  \citep{Dave2019_Simba} or a mass-dependent threshold in Eddington ratio such that higher mass systems can more easily transition into the powerful radio mode while low-mass galaxies remain unaffected \citep{Weinberger2017}. 

The strong degeneracy between model parameters controlling black hole seeding, dynamics, accretion, and feedback remains a challenge for building fully-predictive black hole prescriptions in cosmological simulations. 
This has been emphasized in a large suite of hundreds of cosmological zoom-in simulations from the FIRE project \citep{Hopkins2018_FIRE2methods,Hopkins2022_FIRE3} varying independently the numerical schemes and efficiency parameters for black hole accretion and mechanical, radiative, and cosmic ray feedback \citep{Wellons2022}.  Several plausible combinations of models satisfy observational constraints (e.g., stellar mass--halo mass relation and black hole--galaxy scaling relations) but it is non-trivial to reproduce scaling relations across the full mass range (from dwarfs to massive galaxies) without relying on artificial numerical implementations.  Many model variations produce qualitatively incorrect results regardless of parameter choices, which may help discriminate models \citep{Wellons2022}. Interestingly, cosmic rays accelerated by massive black holes are emerging as a plausible important feedback channel \citep{Su2021_Jets,Wellons2022}, currently ignored in most large-volume cosmological simulations. \\

\section{Analytic evolutionary models of Supermassive Black Holes in a cosmological context}

\subsection{Semi-analytic models}

Semi-analytic models (SAMs) have been conceived as an alternative methodology to numerical simulations to model galaxies in a cosmological context from first principles \citep{Cole00,Benson03,Granato2004,Baugh06,Bower06,Cattaneo06,Croton2006,DeLucia2006,Menci06,Monaco07,Somerville99,Somerville2008,Guo11,Benson12,Hirschmann12,Henri13,Croton16,Lacey16,Lagos18,Cattaneo20,Fontanot20}. SAMs start from dark matter N-body simulations or analytic merger trees. A number of physically and/or observationally motivated analytic recipes are then set up in SAMs to control the cooling, star formation and ejection of baryons in/from the potential wells of the host dark matter haloes through cosmic time. Each of these recipes is characterized by one or more adjustable parameters, which are then fitted to best fit a variety of independent data sets on the galaxy population at different epochs and environments. SAMs represent a valuable and flexible tool to more rapidly test against observational data the effectiveness of some theoretical models and/or to bracket the values of some relevant physical parameters. 

Although SAMs had been originally designed to mostly focus on galaxy evolution, a lot of attention has been devoted in the last decades to also include in SAMs the formation and evolution of SMBHs \citep{Granato2004,Bower06,Croton2006,Monaco07,Marulli2008,Menci08,Bonoli2009,Benson12,Fanidakis12,Hirschmann12,Fontanot20}. Similarly to hydrodynamic simulations, SAMs start by seeding galaxies with black holes of varying mass, from light seeds of a few tens to hundred solar masses (Pop III remnants), to massive seeds of a few thousands solar masses (direct collapse from supermassive stars). In more recent times, a third avenue has been proposed to form SMBH seeds, which lies somewhat in between the previous two models, and it is based on the accretion of stellar mass black holes that fall onto the centre via dynamical friction in the dense environments of starforming galaxies. Given the initial high star formation rates of the host galaxies, calculations have shown that central black holes could grow rapidly and efficiently towards the formation of seed black holes comparable to those from direct collapse.    

Seed black holes are then allowed to grow, similarly to galaxies, via gas accretion and mergers with other (intermediate and supermassive) wandering black holes brought in by past mergers with other galaxies. Gas accretion onto SMBHs has been often associated to galaxy mergers, during which the gas accretion rate onto the SMBH has been approximated by the relation \citep{Granato2004,Fontanot20}
\begin{equation}
    \dot{M}_{\rm BH, Q} = f_K f(V_{\rm vir}) m_{\rm ratio} M_{\rm cold}\, , 
    \label{EqQSOmode}
\end{equation}
where $M_{\rm cold}$ is the mass of available cold gas, $f_K$ is an efficiency parameter, usually chosen in a way to match the local scaling relations between SMBH mass and host galaxy stellar/bulge mass, $m_{\rm ratio}$ is the mass ratio of the merging galaxies, and $f(V_{\rm vir})$ is a function of the virial velocity of the host halo, usually parameterised as \citep{KauffHaehnelt}
\begin{equation}
 f(V_{\rm vir})=\frac{1}{1+\left(\frac{V_{\rm vir}}{280\, {\rm km/s}}\right)^{-2}}
 \label{EqVvir}
\end{equation}
to take into account the possibility that in lower-mass galaxies/haloes, with shallower potential wells, gas is more easily expelled and thus less available for accretion onto the SMBH. The accretion formula in Equation~\ref{EqQSOmode} is usually referred to as ``quasar-mode'', as it easily generates large accretion rates onto the SMBH and thus high luminosities. 

The quasar-mode accretion has not always been strictly associated to galaxy mergers in SAMs, but also to a more general formation epoch of the host halo. In this approach, each virialization/formation of a host dark matter halo, identified by its early phase of fast collapse, promotes infall and cooling of baryons. Baryonic cooling is then rapidly followed by bursts of star formation which can induce, via different physical processes such as photon radiation drag, the formation of a central gas reservoir of low angular momentum, which can in turn feed the central SMBH \citep{Umemura1993,Granato2004,Monaco07}
\begin{equation}
    \dot{M}_{\rm low J}=f_{low J}\Psi,
    \label{EqlowJ}
\end{equation}
where $\Psi$ is the galactic star formation rate and $f_{low J}$ a free parameter, again usually tuned to reproduce the scaling relations between SMBHs and galaxies or the AGN luminosity functions. Equation~\ref{EqlowJ} has been often adopted as an intermediate step to better mimic the gradual loss of angular momentum in the gas component from the larger to the smaller scales around the SMBH. Equation~\ref{EqlowJ} has been used in merger models.

The actual accretion onto the central SMBH from the reservoir of low angular momentum gas has then been modelled on the viscous accretion timescale \citep{Granato2004,Lapi06}
\begin{equation}
\dot{M}_{\rm BH}=f_{\rm BH}\frac{\sigma_B^3}{G}\left(\frac{M_{\rm res}}{M_{\rm BH}}\right)^{3/2} \left(1+\frac{M_{\rm BH}}{M_{\rm res}}\right)^{1/2}\, ,
\label{EqViscAcc}
\end{equation}
where $M_{\rm BH}$, $M_{\rm res}$, and $\sigma_B$ are, respectively, the mass of the central SMBH, the mass of the gas reservoir, and the stellar velocity dispersion of bulge. We note that in Equation~\ref{EqViscAcc}, and in fact in all analytic recipes of SMBH accretion, the rate of gas accretion onto the central SMBHs is usually capped at a chosen multiple of the SMBH Eddington accretion rate. 

The accretion onto the central SMBH has also been sometimes modelled in SAMs following a ``light-curve'', i.e., the varying accretion rate suggested by hydrodynamic simulations and analytic models. Such light-curve models are typically characterized by a fast (super-)Eddington accretion phase, which lasts until the SMBH reaches a peak mass marking the self-regulation limit between accretion and feedback, followed by a power-law decline as a function of time of the type \citep{Granato2004,Hopkins06LF,Lapi06,Malbon07,Marulli2008,Bonoli2009,Fontanot20}
\begin{equation}
\dot{M}_{\rm BH}=\frac{\dot{M}_{\rm Edd}}{1+\left(\frac{t}{t_{\rm Edd}}\right)^2}\, . 
\label{Eqdescphase}
\end{equation}
It has been shown several times in the literature that replacing instantaneous accretion with a gradual redistribution of gas accretion modulated by a physically motivated light curve, provides a closer match to different AGN observables, most notably the shape and evolution of the AGN bolometric luminosity function, and the large-scale AGN clustering \citep{Lidz06}.

\begin{figure}[t!]
\includegraphics[width=\textwidth]{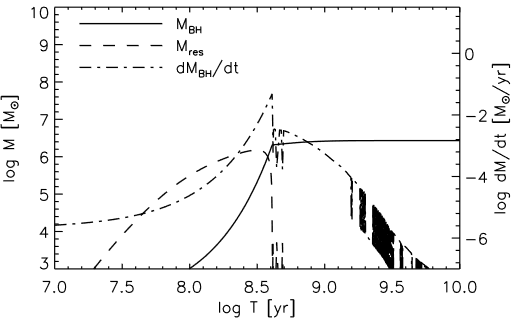}
\caption{Dependence on galactic age of the SMBH mass, reservoir mass, and SMBH accretion rate \citep{Lapi06}).}
\label{fig:LapiLC}
\end{figure}

We show in Figure~\ref{fig:LapiLC} a light curve that is not parameterized but it is generated by the balance between accretion and feedback in a self-regulated mode. It is straightforward to distinguish the initial, Eddington-limited growth of the SMBH from an initial seed, followed by a descending phase regulated by the availability of gas in the reservoir and the surrounding interstellar medium. The transition between the self-regulated mode to the starvation mode is what sets the scaling relation between SMBH mass and host galaxy stellar velocity dispersion \citep{Granato2004}.

While in SAMs the accretion of cold gas controls the growth rate of SMBHs in the quasar-mode regime, as given in Equation~\ref{EqQSOmode}, the accretion of hot gas onto a central SMBH occurs from a static hot
halo around the SMBH host galaxy, and it is expressed as \citep{Fontanot20}
\begin{equation}
\dot{M}_{\rm BH, R}=k_{\rm R}\left(\frac{M_{\rm BH}}{10^8\, M_\odot}\right)\left(\frac{f_{\rm hot}}{0.1}\right)\left(\frac{V_{\rm vir}}{200\, {\rm km\, s^{-1}}}\right)^3\, ,
\end{equation}
where $f_{\rm hot}$ is the fraction of the total halo mass in the form of hot gas, $V_{\rm vir}$ is the virial velocity of the host halo, which is proportional to the total mass of hot gas in the halo, and $k_{\rm R}$ is an adjustable parameter, tuned to reproduce the local scaling relations between SMBHs and their host galaxies.
The radio-mode AGN feedback mostly acts at stalling or heating up the late re-accretion of cold gas in the host galaxy (the so-called ``cooling flow''), along with limiting the growth of the central SMBH, which always occurs at very sub-Eddington regimes. 

As in hydrodynamic simulations, explicit inclusion of AGN feedback in the quasar-mode regime have been included in some SAMs. Inspired by analytic arguments \citep{SilkRees1998}, the first examples of quasar-mode feedback in SAMs assumed that at any given time of episodic accretion, the central SMBH back reacts by heating the interstellar medium and removing a fraction of gas from the cold phase $\frac{M_{\rm cold}}{M_{\rm gas}}$ at a rate \citep{Granato2004}
\begin{equation}
    \dot{M}_{\rm cold}^{\rm QSO} \propto \frac{L_K}{\sigma^2}\frac{M_{\rm cold}}{M_{\rm gas}}\, ,
    \label{eq:QSOmodeGranato}
\end{equation}
where $L_K$ is the kinetic luminosity of the quasar and $\sigma$ the stellar velocity dispersion of the host galaxy. It is interesting to note that SAMs inclusive of only quasar-mode feedback are still capable of reproducing the same local galaxy stellar mass function as in models inclusive of explicit radio-mode feedback, as long as the quasar-mode feedback is powerful enough to ejecting cold and infalling gas beyond the virial radius \citep{Granato2004}.


\subsection{Semi-empirical models}

Along the past decades, several groups have attempted to probe the growth SMBHs, or at least to set some general constraints on their evolutionary patterns and average properties, by adopting a number of ``data-driven'' approaches, without necessarily adopting ab-initio cosmological models. One of the very first of these attempts is the so-called So{\l}tan ``argument'' \citep{Soltan82}. The aim of this seminal work was to yield an average estimate of the radiative efficiency $\epsilon_{\rm r}$ of SMBHs (and thus of their spin) via the relation 
\begin{equation}
    \rho_{\rm BH}=\frac{1-\epsilon_{\rm r}}{\epsilon_{\rm r} c^2}\Psi\, ,
    \label{Soltan}
\end{equation}
which is the equivalent of Equation~\ref{eq:Lbol} but at ``population'' level. 
In Equation~\ref{Soltan}, $\rho_{\rm BH}$ is the relic local mass density of SMBHs, estimated from galaxy number densities converted to SMBH number counts via any of the local scaling relations between SMBH mass and host galaxy property, such as stellar/bulge mass \citep{HaringRix}, stellar velocity dispersion \citep{GrahamSigma}, or even light profile \citep{GrahamLight} or host dark matter haloes \citep{Bogdan}. The $\Psi$ term in Equation~\ref{Soltan} represents instead the integrated AGN emissivity across redshifts and (bolometric) luminosities. Along the years, the application of Equation~\ref{Soltan} has yielded a wide range of values $\epsilon_{\rm r}\sim 0.06-0.30$ \citep{Marconi04,Shankar13,Ueda14,Shankar20,Ananna20}, mainly due to the still noticeable systematics in bolometric corrections \citep{Duras20}, SMBH scaling relations \citep{Shankar16,Shankar20}, and AGN obscured fractions \citep{Ueda14,Ananna20}.

\begin{figure}[t!]
\includegraphics[width=\textwidth]{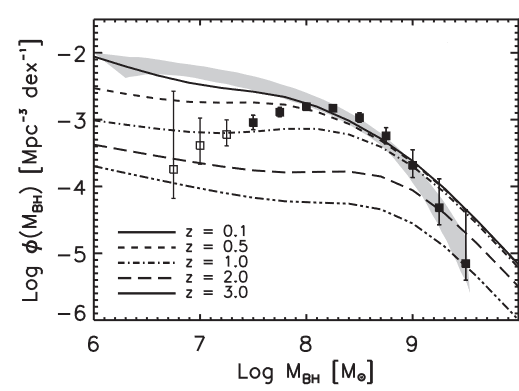}
\caption{The SMBH mass function at different redshifts, as labelled, as predicted from the continuity equation models presented in Equation~\ref{eq:ContEq}, assuming a mean radiative efficiency of $\epsilon_{\rm r}=0.06$, and compared to local measurements of the SMBH mass function (grey area \citep{Shankar09} and data points \citet{Vika09}).}
\label{fig:SMBHMFfromCE}
\end{figure}

Continuity equation models \citep{SmallBlandford,Marconi04,Shankar09,Aversa15,TucciVolonteri} can be considered as a ``differential'' generalization of So{\l}tan’s argument. Given values of the radiative efficiency and the Eddington ratio distribution, the bolometric AGN emissivity at any time can be linked to the average growth rate of SMBHs of the corresponding mass, via a continuity equation describing the average mass ``flow'' of SMBHs of any given mass, and thus capable of predicting the time evolution of the global SMBH mass function $n(M_{\rm BH},t)$. This continuity equation is usually written as 
\begin{equation}
    \frac{\partial n(M_{\rm BH},t)}{\partial t}=-\frac{\partial \left[\langle \dot{M}_{\rm BH} \rangle n(M_{\rm BH},t)\right]}{\partial M_{\rm BH}}\, ,
    \label{eq:ContEq}
\end{equation}
where $\langle \dot{M}_{\rm BH} \rangle$ is the average accretion rate averaged over the full population of black holes of mass $M_{\rm BH}$ at time $t$. Figure~\ref{fig:SMBHMFfromCE} shows an example of the SMBH mass function at different redshifts, as labelled, as predicted from Equation~\ref{eq:ContEq}, giving in input the bolometric AGN luminosity function derived from X-rays \citep{Ueda14}, and adopting an average value of the radiative efficiency $\epsilon_{\rm r}=0.06$. It is interesting to show that, without any specific fine-tuning, the continuity equation model is able to naturally generate a SMBH mass function that well aligns with the number density of local SMBHs at all masses. Allowing for a significant fraction of (dry) SMBH mergers in Equation~\ref{eq:ContEq} tends to overproduce the high-mass end of the SMBH mass function with respect to local data \citep{Shankar13,Aversa15}. However, given the still unclear systematics in the SMBH scaling relations adopted to infer the local SMBH mass function \citep{Shankar16,ShankarNat}, and the uncertainties on the radiative efficiency and its dependence on SMBH mass \citep{DavisLaor}, no firm conclusions can be drawn yet on the importance of SMBH mergers in shaping the SMBH mass function along cosmic time. Continuity equation models have been further utilised to impose constraints on the mean Eddington ratio distribution characterizing SMBH growth in time. Several studies have suggested that mean Eddington ratio should be below unity \citep{Cao10} and progressively decreasing towards low redshifts \citep{Aversa15,Shankar13} to accommodate the decreasing emissivity in the AGN luminosity function \citep{Ueda14} and the local values of the observed fractions of AGN in galaxies \citep{Shankar13}. 

\begin{figure}[t!]
\includegraphics[width=\textwidth]{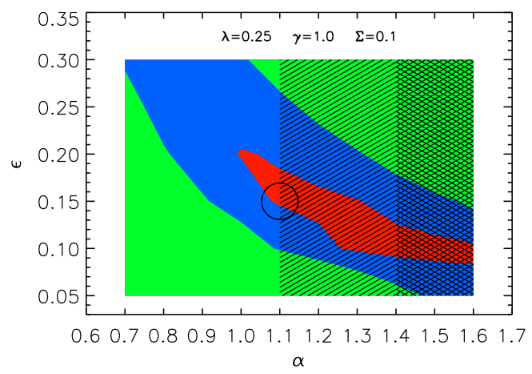}
\caption{The $\chi^2$ per degree of freedom as a function of the radiative efficiency and slope $\alpha$, the normalization of the $M_{\rm BH}\propto V_{\rm vir}^{\alpha}$ relation, with other parameters fixed at the
values listed on top of the panel. The blue and red areas define the regions where the $\chi^2$ for the luminosity is below 3 and 1.5, respectively.}
\label{fig:SMBHMFfromCEHalo}
\end{figure}

In a more general attempt to probe the coevolution of SMBHs and their host galaxies in a transparent, data-driven approach, several groups have included SMBHs in either semi-empirical models of galaxy evolution \citep{Shankar10,Hopkins_SEM,Yang17,Shankar20,Trinity}, or by assigning galaxies and SMBHs at fixed redshift (ignoring evolutionary links) to derive some basic properties on their number densities, clustering, and scaling relations \citep{Georgakakis19,ShankarNat,AirdCoil,Allevato21,Viitanen21,Carraro22}. Let's start by reviewing the latter type of approach, which are particularly relevant to the
creation of active and normal galaxy mock catalogs \citep{ConroyWhite}, a vital component of the planning of imminent extragalactic surveys such as Euclid
\citep{Euclid} and the Vera C. Rubin Observatory Legacy Survey of Space and Time (LSST) \citep{LSST}. In these more basic models, galaxies are first assigned via ``abundance matching techniques'', based on the number equivalence between galaxy and dark matter number density counts, on top of N-body dark matter simulations \citep{Kravtsov04,Vale04,Shankar06,Moster10,Moster13,Behroozi13,Kravtsov18}. SMBHs are then assigned to host galaxy/haloes by either assuming a direct scaling with the host dark matter haloes \citep{Shankar10}, or via a relation with their host galaxies \citep{ShankarNat,Allevato21,Trinity}. An example of this procedure is shown in Figure~\ref{fig:SMBHMFfromCEHalo}, in which an underlying scaling is assumed between SMBH mass and halo virial velocity of the type $M_{\rm BH}\propto V_{\rm vir}^{\alpha}$, through which the halo mass function is converted into a SMBH mass function and then into an AGN luminosity function via a mean Eddington ratio \citep{Shankar10}. The plot reports the constraints on the radiative efficiency $\epsilon_{\rm r}$ and slope $\alpha$ of the SMBH-halo relation derived from the matching with high-redshift AGN number counts. In other instances, the SMBH mass is bypassed and instead galaxies are directly assigned an X-ray AGN luminosity as extracted from a $P(\lambda)$ Eddington ratio distribution as determined from X-ray and optical AGN surveys, which usually well approximated by a Schechter function \citep{Georgakakis19,AirdCoil}.  

Full-fledged semi-empirical models (SEMs) are based on a cutting-edge ``data-driven'' methodology that avoids the modelling of galaxy and SMBH growth and assembly within dark matter haloes from first principles, as in the more traditional modelling approaches discussed in the previous Section. The main input parameter in SEMs is a monotonic relation between host galaxy stellar mass and its host halo mass (SMHM relation hereafter), which is derived from the equivalence between the cumulative number densities of galaxies and dark matter haloes. Applying these ``abundance matching'' techniques, one can assign galaxies to dark matter haloes in an N-body simulation at any specified redshift. Central galaxies along the main progenitor branches of their dark matter merger trees can be reinitialised 
via a time-dependent SMHM relation, and can also gradually transform their morphologies from discs to bulges via ``in-situ'' processes such as disc instabilities, or ``ex-situ'' processes such as mergers with other galaxies. Satellite galaxies are those associated via the SMHM relation to each dark matter branch merging to the main progenitor \citep{Hopkins09,Zavala12,Shankar14}. The SMHM relation is thus capable to convert a dark matter halo assembly history into a galaxy merger history. By using an epoch-dependent SMHM relation, it can be shown that one can predict the mean assembly and merger histories of galaxies \citep{Hopkins_SEM,Cattaneo11,BuchanShankar,Grylls19}, along with their star formation histories, which are ultimately computed from the difference between the total growth and the contribution from mergers \citep{Grylls20SFH}. It is clear that the predicted merger and star formation histories will strongly depend on the input SMHM relation \citep{Grylls20Mergers,Oleary}. Alternatively, star formation histories can be fitted as a function of halo properties and integrated forward in time to derive, when coupled to the contribution from mergers, the full assembly histories of galaxies \citep{Moster18,Behroozi19}. 

In the context of a SEM, the growth and assembly histories of black holes can be derived in multiple ways. One of the most straightforward one is to adopt a monotonic mapping between galaxy properties and the mass of their central black hole, as informed by direct observations at different redshifts. For example, one could assume a constant or slowly-varying scaling between the black hole mass and the host galaxy stellar mass or stellar velocity dispersion \citep{Trinity}, and predict the rate at which black holes would grow in mass and merge following the assembly histories of their host galaxies and dark matter haloes \citep{Yang17,Shankar20}. Other groups have instead included some physically-motivated light curves of SMBHs into the merger histories of their host dark matter haloes/galaxies, to infer properties on, e.g., AGN luminosity functions \citep{WyitheLoeb,Hopkins05,Hopkins06LF,Shen09,ShankarIAU}, AGN clustering \citep{Haiman01,MartiniWeinberg,Lidz06,Shen09}, SMBH merger rates and implied gravitational waves \citep{Sesana16}, or more general galaxy properties \citep{Hopkins06UnifiedModel}.

\section{Model predictions across cosmic time}

\subsection{Large Volume Simulations: The first, rare quasars at $z> 6$}
\begin{figure}[t!]
\includegraphics[width=\textwidth]{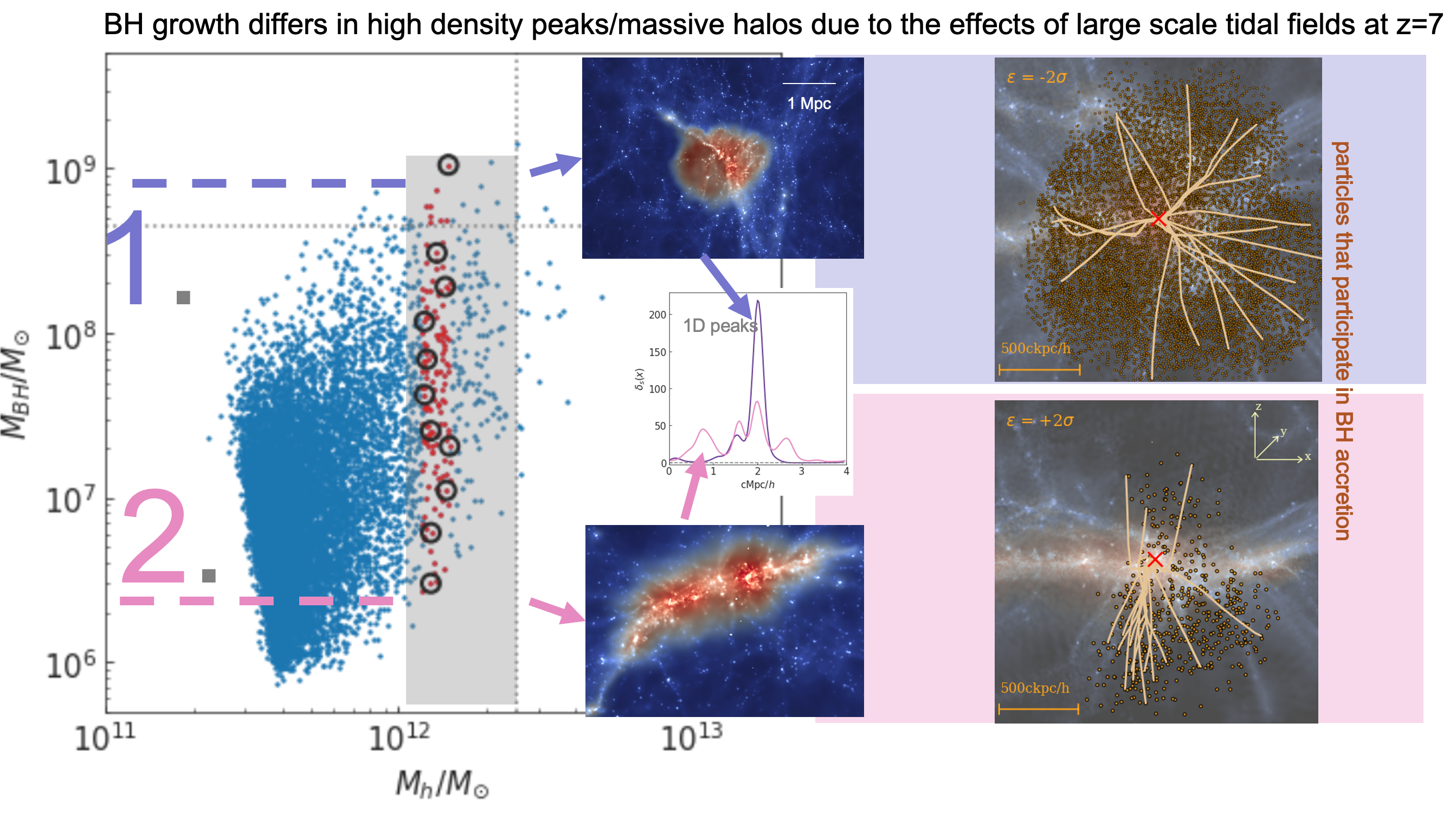}
\caption{The BH masses as a function of halo mass in the BT simulation and examples of low and high tidal field environments for $5-
sigma$ peaks at $z=7$. The gray shaded area highlight the BH masses in 5-$sigma@$ peaks ($10^{12} M_{\odot}$) halos. The BH mass in galaxies within these halos can range from $ 10^6 $ \Msun  (example 2) to $ 10^9$ \Msun (example 1).
In the center the large scale environments of 1 and 2. On the left panel we show all the particles that end up contributing to BH accretion by $z=7$. In region 1, a region of low tidal field,  particles 
have accreted quasi-radially in region 2, of large tidal fields the only particles that accrete flow perpendincular to the main large scale density.}
\label{fig:firstqso}
\end{figure}
\begin{figure}
    \centering
    \includegraphics[width=1.0\textwidth]{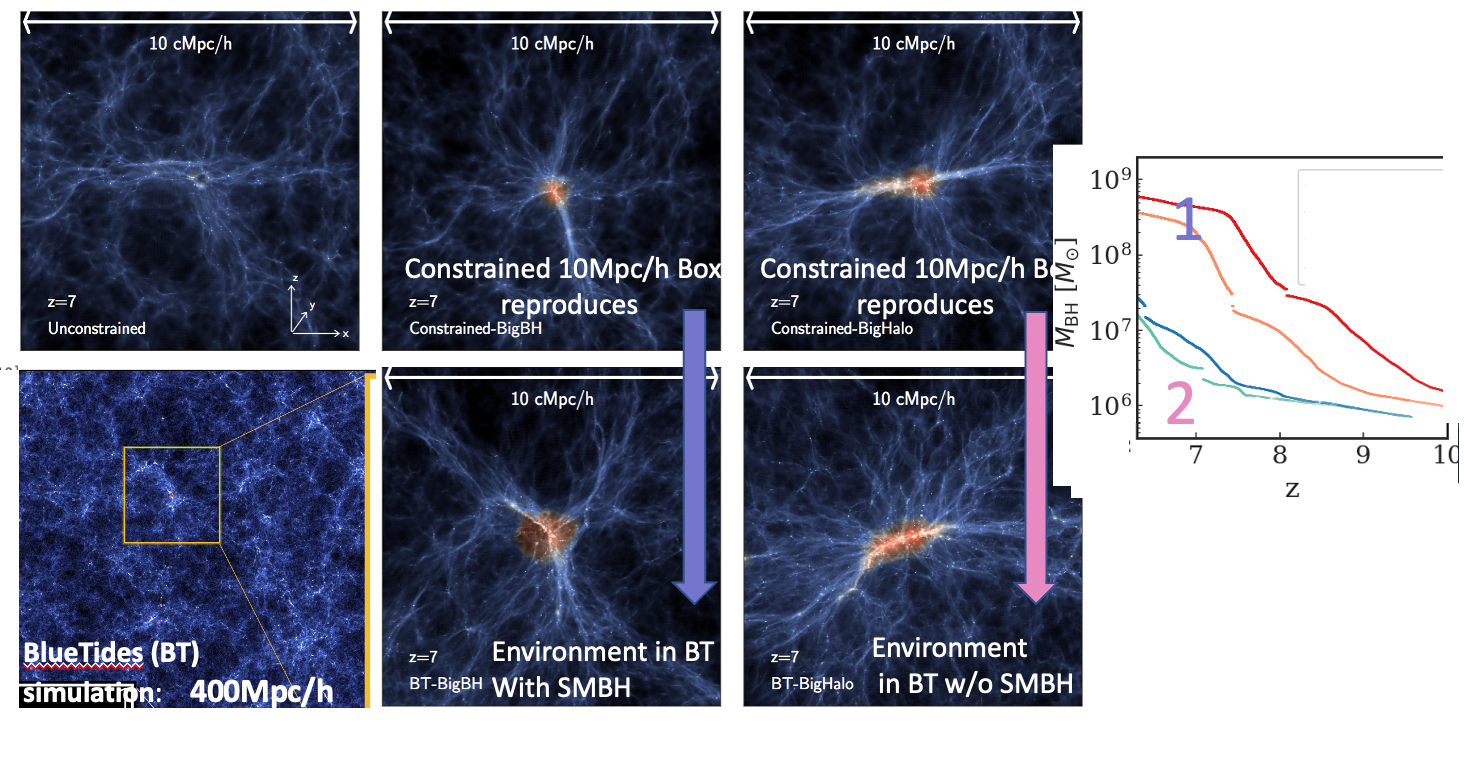}
    \caption{{\small Illustration of the gas density field around the SMBH at $z=7$. 
    The top middle and top right panels are the simulations with constrained IC of 1 and 2 and using the peak parameter sets extracted from IC of \textsc{BlueTides} 400Mpc/h simulation (in the corresponding bottom panels).  The right panel shows the actual BH history in these two highly biased regions in constrained runs which match the exact results in BT.}}
    \label{fig:constrainedsims}
\end{figure}

How the first supermassive black holes formed in the early Universe and
evolved into the bright quasars, among the most 
luminous sources in the cosmos, is still an open question hotly debated by the community.
In order to address it, one must simulate
a large volume of the Universe, sufficient to include a statistical
sample of bright quasars and their associated large scale structures,
as well as having enough mass resolution to successfully model
the physics and gas inflows close to black holes (the scales of an actual back hole accretion disk are still impossible to resolve in a fully cosmological simulations).

In Chap III, the authors
focus more on the
formation of the black hole seed
populations and the
physics of pristine
gas on very small
scales, which are typically not captured in large cosmological volume we discuss here. We are therefore
focusing our discussion on what we can learn about the
infall of gas
on larger scales and
subsequent growth of the central black hole seed.

In the current and coming decade, a new generation of astronomical instruments,
all in the billion dollar class will start making observations of the
Universe during the period of the first stars and quasars, and opening
up the last frontier in astronomy and cosmology. Those that are
specifically targeting this epoch as their highest priority include
the Square Kilometer Array radio telescope, the NASA James Webb Space
Telescope,  and several huge ground-based
telescopes, such as the Thirty Meter Telescope, the European Extremely
Large Telescope, and the Giant Segmented Meter Telescope, each of
which have collecting areas an order of magnitude larger than the
current largest telecopes. The scientific community has obviously
decided that research targeting the epoch of the first quasars and black hole
formation and growth  matters
enormously. These  observation and experiments will gain a lot of value
if we have theories to test our models.

To decide which physical problem is best suited to which scale
of computation one should consider both the
space density of the relevant objects, which governs the size of the
computational volume, and the mass resolution needed to 
follow the relevant small scale physical processes. In the case
of the first quasars  a volume
of 1 Gpc/h is needed in order to have a statistical sample of
at least 10 objects of the type seen at high redshifts by the Sloan Digital
Sky Survey. The minimum mass resolution can be gauged using some smaller volume
simulations that can be used to study 
convergence, i.e. carrying out cosmological
simulations of black hole formation with different particle masses.

At high redshifts the requirements are quite stringent, as galaxies are extremely compact, with sizes close to a few kpc in scale at best, and the structure and inflow within
each galaxy needs to be resolved. This pushes the resolution requirements to only a few hundred parsec.
With Peta and upcoming Exascale computational facilities this has just become
possible: a qualitative advance, running,  arguably, the first
complete simulation (at least in terms of the hydrodynamics and
gravitational physics) of the creation of the first galaxies and quasars
in simulations at Gpc scales. This is one of the primary aims of the BlueTides simulation,
which we use to discuss 
the emergence of the first quasars.
BlueTides is the largest simulation yet run with full physics (hydrodynamics, 
star formation, black holes),
 and is targeted at the early Universe of galaxies and quasars.
 The simulation can be compared
to cutting edge observations from the Hubble Space telescope, 
finding good agreement with the properties of observed galaxies 
when the Universe was only 10 percent of its current age ($z \sim 6$). 

{\bf The first quasars: high overdensity and low tidal field environment}
With the recent high-resolution, large volume simulations we have been able to investigate directly the property of the environments/density field that causes the growth of the most
massive black holes in the early Universe, and found that tidal fields
in the large-scale (megaparsec and more) environments of black hole
hosting galaxies play a critical role. The necessary growth (black hole mass in simulations at $z=6-7$ is consistent with
the detection of highly luminous quasars at $z> 6$) implies
sustained, critical accretion of material to grow and power
them. Given a black hole seed scenario, it is important to understand which
conditions/environments in the early Universe allow the fastest black hole
growth. Large scale hydrodynamical cosmological simulations of
structure formation allow us to explore the conditions conducive to
the growth of the earliest supermassive black holes. We use the
cosmological hydrodynamic simulation BlueTides, which incorporates a
variety of baryon physics in a $(400 Mpc/h)^3$ volume with 0.7 trillion
particles to follow the earliest phases of black hole critical
growth. 

At $z>6$ the most massive black holes (a handful) approach
masses of $10^9$ Msun with the most massive 
being found in an extremely compact spheroid-dominated host
galaxy. Examining the large-scale environment of hosts, we find that
the initial tidal field is more important than overdensity in setting
the conditions for early BH growth (Figure~\ref{fig:constrainedsims}). However necessary it is not sufficient.
It is only in those high-density (4-5$\sigma $ peaks of the density field) regions, in which large scale tidal fields are very suppressed that the BHs are able to
accretes copious amount of 'cold' gas
at a sufficiently fast rate. Only in low tidal field regions the gas falls along thin, radial
filaments straight into the center forming the most compact galaxies
and most massive black holes at earliest times. Equally high density region (massive halos) in high tidal
field regions instead easily acquire large coherent angular momentum, providing most of the accretion perpedicular to the main large scale filament which significantly suppressed the early BH and galaxy growth ( while influencing
the formation of the first population of massive compact disks). 
This can be seen clearly in Figure~\ref{fig:firstqso} 
where the environment of the most massive black hole is visualized alongside a massive disk galaxy \cite{DiMatteo2017} \cite{Tenneti2019}.

Recently effective techniques of constrained Gaussian realisations have been used and validated it in detail by comparing halo assembly histories (halo, stars, gas, and black hole growth) against large volume cosmological hydrodynamic simulations environments (Bluetides and ASTRID, see Figure~\ref{fig:firstqso} and Figure~\ref{fig:constrainedsims} \citep{Ni2022,huang20}.

The CR technique provides 
an efficient and powerful tool for generating the desired large-scale structures in cosmological simulations.  It can be used to effectively simulate a particular region of interest, using a small volume, with a given large scale feature and learn about the impact of large scale structure on the smaller galaxy scales which regulate star and BH formation history.
The technique works by generating ICs with imposed and precise constraints on the large-scale features of the primordial density field  
such as the height and shape of the density peaks,  as well as the peculiar velocity and tidal field at the site of the peak,  
which fully represent the environments in that region of a large volume, statistically representative simulation. 
The theory of constrained random fields was first set forth by \cite{Bertschinger1987,Binney1991} and extended by \cite{vandeWeygaert1996} which is what our implementation is based on.

This technique is similar to, another common method  of 'zoom-in' simulation, which is designed to completely reproduce a "user-selected" region (for example, a specific halo) from a low-resolution large volume simulation by tracing it back to the IC. 
The CR simulation technique reproduces a statistical representation of that halo at the prescribed resolution (this will not be the same exact, extracted-halo as in a 'zoom-in') and it has far higher computational efficiency that zoom-in simulations. 
The demand in computational resources is very modest
(factors on 1000 less then traditional zooms). One is able to carry out a systematic exploration to construct a sizeable suite of halo environments (with minimal computational effort) and run at high-resolution and with full physics.
This method is already fully validated, and working and we have performed sets of 20-30 constrained simulations using the environments from BT in \citep{Ni2022}. More work is upcoming using this technique to investigate different seed BH models and other different accretion prescriptions.

From the analytic side of cosmological models, several possibilities have been put forward to explain the growth of the first SMBHs \citep{Latif}, which all require either super-Eddington accretion onto relatively small, stellar-like BHs \citep{Madau14,Volonteri15,Lupi16,Regan19}, and/or accretion onto relatively massive seed BHs of the order of $10^3-10^5\, {M_{\odot}}$ \citep{DiMatteo2012,DiMatteo17,MayerBonoli}. 
In more recent times, it has been highlighted that the migration and merging of stellar compact remnants (neutron stars and stellar-mass BHs) via gaseous dynamical friction toward the central high
density regions of highly starforming (proto)galaxies, could potentially build up central BH masses of the order of $10^4-10^6\, {M_{\odot}}$, within a few $10^7$ yr, effectively providing heavy
seeds before standard disk (Eddington-like) accretion takes over \citep{Boco20}. 
(Proto)stars can also migrate and merge quickly to yield a massive black hole seed via a supermassive star \citep{Tagawa2020}.
It has been shown via semi-analytic models simultaneously incorporating both light and heavy seeds BHs, that a gap can be generated in both the low end of the predicted BH mass and luminosity functions at $4 \le z \le 6$, mainly induced by the light BH seeds, which cannot ``catch up'' with the more massive counterparts unless super-Eddington accretion and/or enhanced accretion during mergers are invoked \citep{Trinca22}.

\subsection{Black Hole--Galaxy Scaling Relations}

In the local universe, the discovery of close relationships between the masses of SMBHs and several properties of their bulges such as the stellar mass (${M_{\rm BH}}-M_{\star}$ relation) and the velocity dispersion (${M_{\rm BH}}-\sigma$ relation) have revolutionized our view of massive black holes, linking their growth to that of their host galaxy \citep{Kormendy2013,McConnell2013,Reines2015,Graham2016}.  
To understand the evolution of these relations at higher redshifts (mostly up to $z \sim 2-4$), observational studies rely on galaxies with active galactic nuclei for which SMBH mass estimates use the virial method (see Chapter VI).  Some of these studies find an evolution in which black hole growth precedes galaxy growth while other studies imply little or no evolution \citep{Trakhtenbrot2010,Bongiorno2014,Schulze2014,Shen2015,Sun2015,Willott2015}.  Systematic uncertainties in high redshift measurements are still large and come from both the method for black hole mass estimation and from measuring host galaxy properties.

A popular way to interpret these relationships is by assuming that SMBHs regulate their own growth and that of their hosts by coupling some (small) fraction of their energy output to their surrounding gas \citep{Silk1998,King2003,DiMatteo2005,Murray2005,Hopkins2007_BHplane}.  In this scenario, black holes grow until AGN feedback is able to unbind a significant fraction of gas from the host galaxy, shutting down their own growth, inhibiting star formation, and driving the black hole--galaxy correlations.  Other scenarios interpret the scaling relations as a consequence of mass averaging by hierarchical merging \citep{Peng2007,Hirschmann2010,Jahnke2011} or the result of a common gas supply for star formation and BH growth, regulated by gravitational torques \citep{Kauffmann2009,Chen2013,Angles-Alcazar2013,Angles-Alcazar2015,Angles-Alcazar2017_BHfeedback}. The scaling relations of black hole mass and global properties of the host galaxies form a way to understand the importance and the effects of AGN fueling and feedback.

\begin{figure*}
\begin{flushleft}
\includegraphics[width=\textwidth]{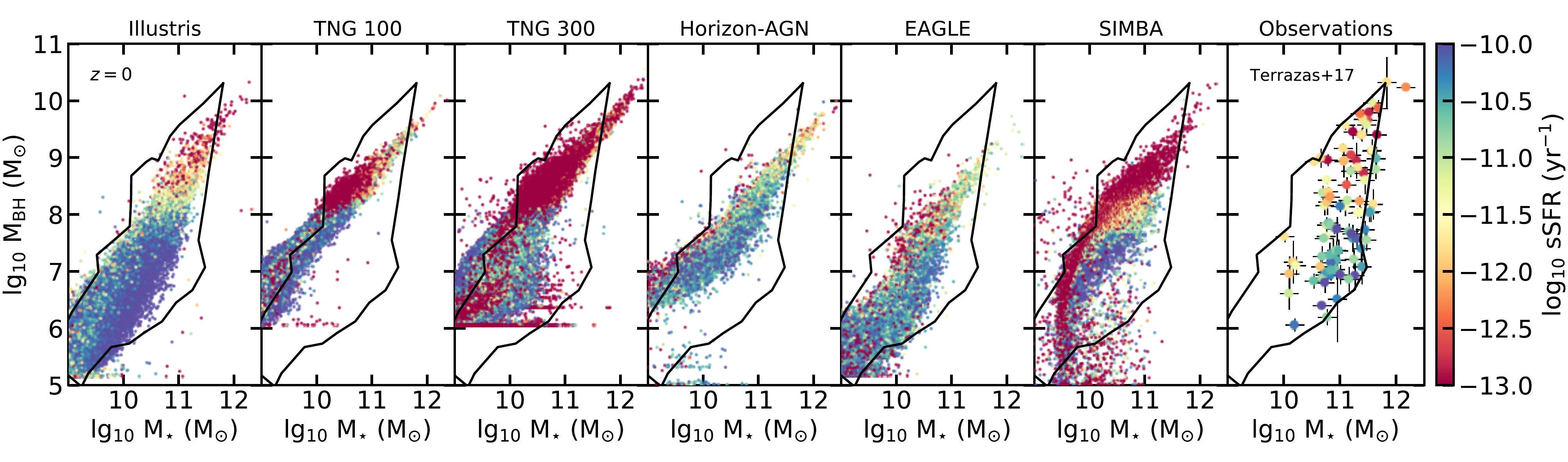}
\includegraphics[width=0.80\textwidth]{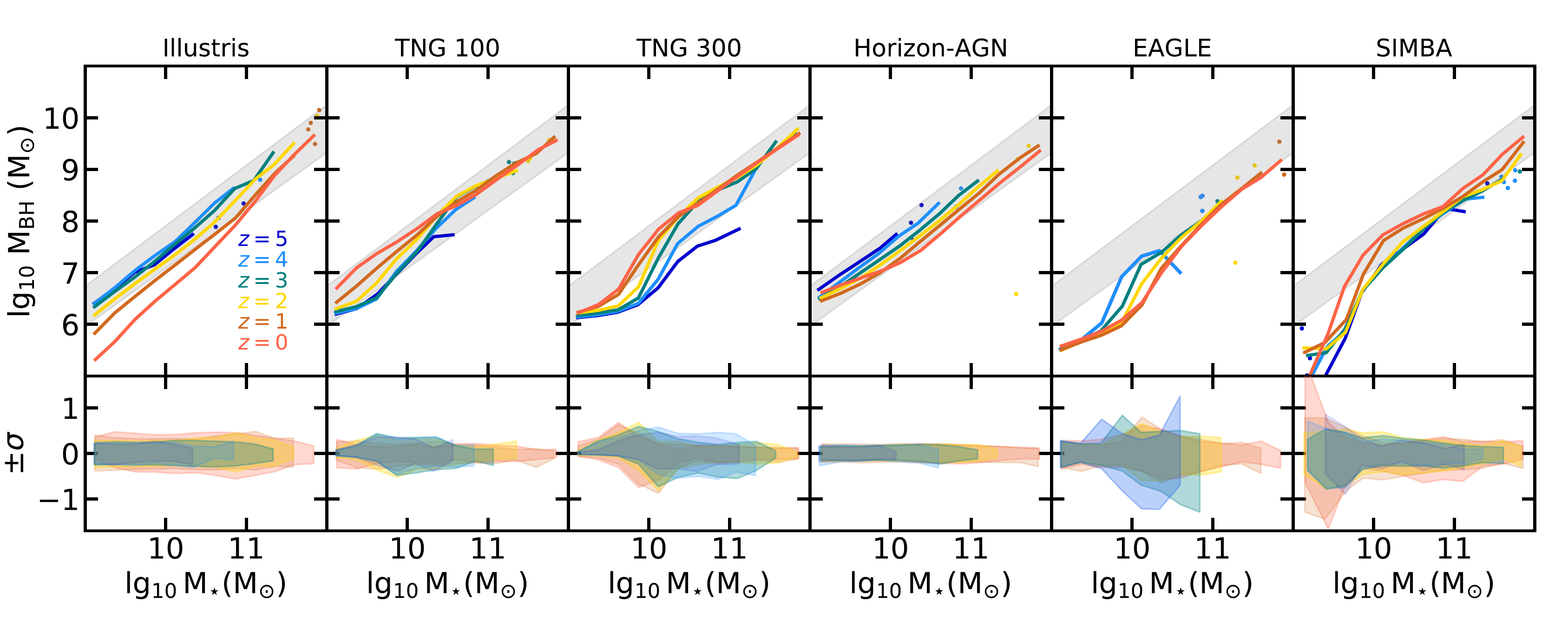}
\end{flushleft}
\caption{
Correlations between black hole mass and host galaxy stellar mass predicted by several large-volume cosmological simulations (figures reproduced from \citet{Habouzit2021_Mbh}). 
The top panels show individual black holes color-coded by the specific star formation rate (sSFR) of their host galaxy.  The black line indicates the region of the diagram occupied by the observational sample of \citet{Reines2015}, and the rightmost panel shows the observed star-forming and quiescent galaxies of \citet{Terrazas2017}.  Simulations generally agree with the observed $M_{\rm BH}$--$M_{\star}$ relation at $z=0$ and the observed anti-correlation of black hole mass and sSFR at fixed stellar mass (indicative of the negative impact of black hole feedback).  However, the detailed shape, normalization, and sSFR connection in the predicted scaling relation depends on the specific subgrid black hole model.
The bottom panels show the redshift evolution of the median $M_{\rm BH}$--$M_{\star}$ relation for the same simulations (and the 15-85th percentile of the distributions).  The grey shaded area indicates the range of variation between several best-fit observed relations \citep{Haring2004,Kormendy2013,McConnell2013}.  Some simulations show a systematic trend for higher black hole to galaxy mass ratio at higher redshift while others show the opposite trend, but current models are generally consistent with no strong evolution in the $M_{\rm BH}$--$M_{\star}$ relation.
}
\label{fig:msigma} 
\end{figure*}

Many cosmological hydrodynamic simulations of structure formation (e.g. Horizon-AGN \citep{Dubois2014,Volonteri2016}, MassiveBlack-II \citep{Khandai2015_MassiveBlack2,DeGraf2015}, Illustris \citep{Vogelsberger2014,Sijacki2015}, IllustrisTNG \citep{Weinberger2017,Li2020_TNGbhs,Terrazas2020}, EAGLE \citep{Crain2015,Rosas-Guevara2016}, and SIMBA \citep{Dave2019_Simba,Thomas2019}) have been used to predict the black hole--galaxy scaling relations for representative populations of black holes and compare them to observational constraints at different redshifts.
We show in Fig.~\ref{fig:msigma} examples of the black hole mass--stellar mass relation derived from a sub-set of recent simulations, as presented in \citet{Habouzit2021_Mbh}.  The detailed shape, normalization, and scatter depends on the specific subgrid black hole model but most simulations agree with the observed $M_{\rm BH}$--$M_{\star}$ relation at $z=0$, partially reflecting the careful tuning of free parameters to match observations.  The scatter in black hole mass at fixed stellar mass generally correlates with the specific star formation rate (sSFR) of their host galaxy, in agreement with observations \citep{Terrazas2017}.  The origin of the scatter is still not well understood, but a generic prediction of simulations is that overmassive black holes preferentially live in galaxies with lower sSFR due to the negative impact of AGN feedback.
Some simulations show a systematic trend with redshift where the normalization of the $M_{\rm BH}$--$M_{\star}$ relation increases at higher redshift while other simulations show the opposite trend, but current models are generally consistent with no strong evolution in the scaling relations.

Despite the overall agreement of cosmological simulations, the main physical driver of the scaling relations is still not fully understood.  Simulations implementing Bondi-like subgrid accretion \citep{DiMatteo2008,Sijacki2015,Volonteri2016} support the scenario of AGN feedback self-regulation, where the AGN feedback efficiency controls the normalization of the scaling relations. However, simulations implementing subgrid accretion driven by gravitational torques support a scenario where the scaling relations are the result of a common gas supply for star formation and black hole growth, where the black hole accretion efficiency (rather than feedback) controls the normalization of the scaling relations \citep{Angles-Alcazar2013,Angles-Alcazar2015,Angles-Alcazar2017_BHfeedback,Angles-Alcazar2017_BHsOnFIRE,Dave2019_Simba,Thomas2019}.  
Cosmological simulations also often differ in their predictions at the low mass end of the scaling relations, where black hole growth is very sensitive to the specific numerical implementations of black hole seeding and dynamics.  Recent simulations indicate that stellar feedback can also significantly suppress early black hole growth by efficiently evacuating gas from galactic nuclei  \citep{Dubois2015_SNa,Angles-Alcazar2017_BHsOnFIRE,Bower2017,Habouzit2017,Prieto2017,McAlpine2018,Trebitsch2018,Lupi2019,Lapiner2021,Catmabacak2022,Tillman2022}, which may explain observed undermassive black holes in low mass galaxies \citep{Graham2013,Savorgnan2016}, but the details depend on resolution and interstellar medium physics in the simulations.

Interestingly, simulations  now have enough sophistication that we can mock up
 observations to the point that we can
 test  the
  simulated population for possible selection biases in all types of observed relations\cite{DeGraf2015}\cite{DeGraf2012}.
  Indeed, we typically find that for samples selected  on the basis of $M_{\rm{BH}}$ or
  $M_*$ (and of size similar to those observed) the slopes can be steeper
  than for  randomly selected samples. Such sample selection also
  biases toward finding stronger evolution with redshift than for a
  random sample as they tend to pick objects at the high-end of the
  relation (consistent with stronger evolution). This is relevant for some of the claims of evolution in the 
  measurements \citep{Trakhtenbrot2010,Bongiorno2014}. It is an interesting direction that simulations
  can indeed be used to pin down possible observational biases from actual physical evolution.

\subsubsection{Semi-empirical models}
\begin{figure}
    \centering
   \includegraphics[width=0.9\textwidth]{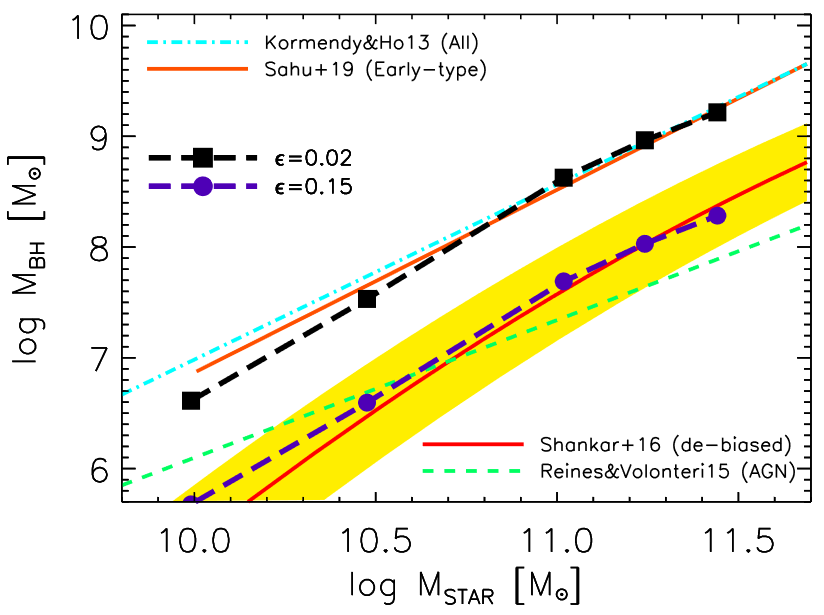}
    \caption{Correlations between central black hole mass and host galaxy total stellar mass in the local Universe \citep{Kormendy2013,Sahu19,Shankar16,ReinesVolonteri}. 
    The solid red line with its scatter (yellow region) is the de-biased $M_{\rm BH}-M_{\star}$ relation \citep{Shankar16}. The green
dashed line is the fit to the local AGN \citep{ReinesVolonteri}. Also included are the predicted average black hole mass as a function of host stellar
mass at $z = 0.1$ for two different values of the radiative efficiency, as labelled. Values of $\epsilon \sim 0.02$ are required (black long-dashed with filled squares) to
match the normalization of the raw black hole $M_{\rm BH}-M_{\star}$ relation for local dynamically measured quiescent black holes.}
    \label{fig:MbhMstar}
\end{figure}

In a semi-empirical approach, SMBHs are often included using a SMBH-galaxy scaling relation \citep{Trinity}, thus the latter cannot be defined as a true prediction of the model, though it is possible to still test, for example, how much SMBH mergers can influence the shape and or scatter of a given input SMBH mass-galaxy mass relation \citep{Hirschmann10}. In many instances, SMBHs have been included by adopting empirical or theoretically motivated relations between the SMBH accretion rate and the host galaxy stellar mass or star formation rate \citep{Yang17,Shankar20,Carraro22}, or by directly including the SMBH light curve during, e.g., galaxy mergers \citep{Hopkins06UnifiedModel,Shen09,ShankarIAU}. In most cases, when integrating forward in time the cumulative accretion rates, the relic SMBH masses appear to be linearly correlated with their host galaxy stellar mass, where the normalization of the relation is mostly controlled, as expected, by the chosen input radiative efficiency $\epsilon_{\rm r}$.   
Recent work \citep{Yang17,Carraro20} showed that the ratio between SMBH accretion rate, as traced by the X-ray AGN luminosity, and the SFR in the host galaxy, does not evolve with redshift, although it depends on stellar mass. When combining the mean black hole accretion rate as a function of stellar mass and redshift with empirical models of galaxy stellar mass growth \citep{Moster18} and integrating over time, the resulting $M_{\rm BH}-M_{\star}$ relation appears to be nearly independent of redshift \citep{Yang17,Carraro20,Shankar20,Carraro22}, indicating that stellar and black hole masses grow, on average, at similar rates, a conclusion supported by several independent studies on the scaling relations of AGN at different redshifts and luminosities \citep{Shen15,Suh20,Li21}. An example of this procedure is shown in Figure~\ref{fig:MbhMstar}, where the resulting $M_{\rm BH}-M_{\star}$ relation, obtained from direct integration of the SMBH accretion rate along each stellar mass accretion track, is plotted for two different values of the mean radiative efficiency, as labelled, and compared with different determinations of the $M_{\rm BH}-M_{\star}$ from local SMBHs. Similar approaches have also been applied to the $M_{\rm BH}-\sigma$ relation, showing evidence \citep{ShankarMbhSigma,Marsden22} for a negligible evolution of the latter up to, at least, $z\sim 2$.

\subsection{The Black Hole mass function and AGN luminosity functions}

Even though the QLF has been studied for over 30 years, we still do not
understand the fundamental physical parameters that regulate its shape
and evolution. Ultimately the evolution of the quasar luminosity
function (QLF) is one of the basic cosmological measures providing
insight into structure formation and its relation to black hole growth.
The QLF is typically described by two power-law components: flatter and
steeper at the faint and bright end respectively, and with a break luminosity
that evolves with redshift (luminosity density evolution). The bright-end
slope also appears to evolve becoming flatter at the highest redshifts
($z=5-6$), although the most recent
measurements are hinting that this may
not be the case \cite{Giallongo2015}\cite{Fan2019}.
\begin{figure}
\hbox{
\includegraphics[width=0.6\textwidth]{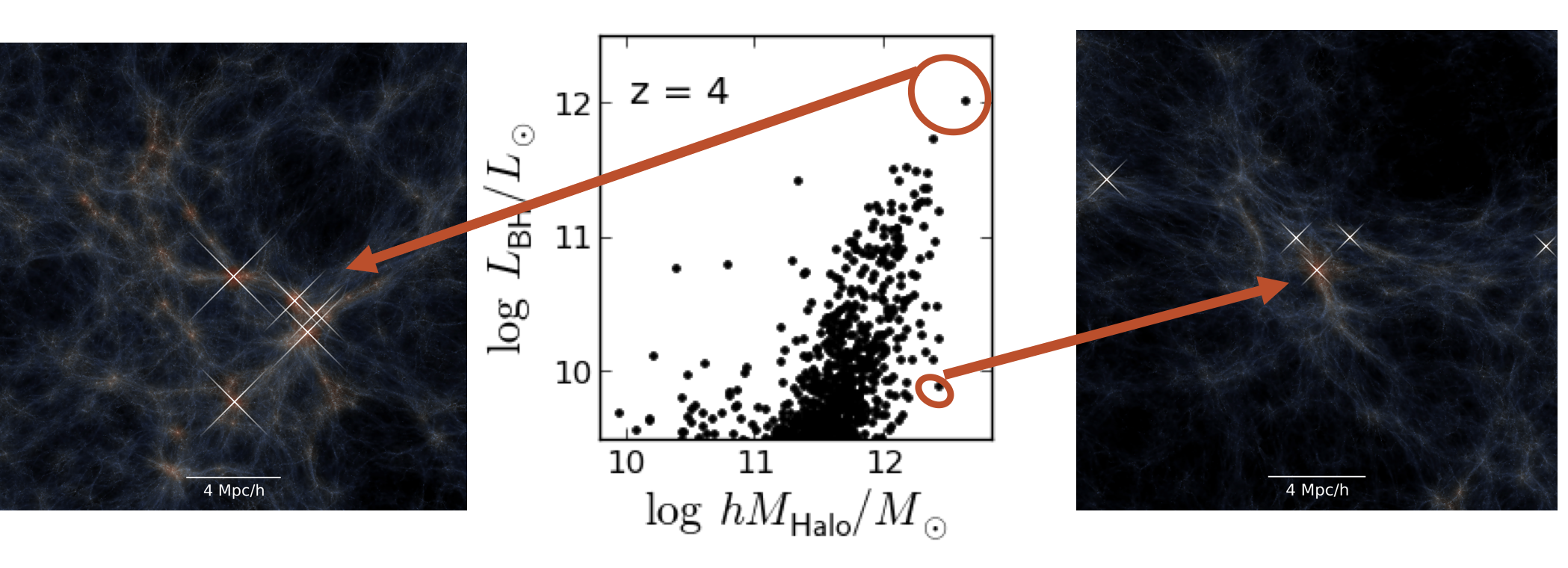}
\includegraphics[width=0.3\textwidth]
{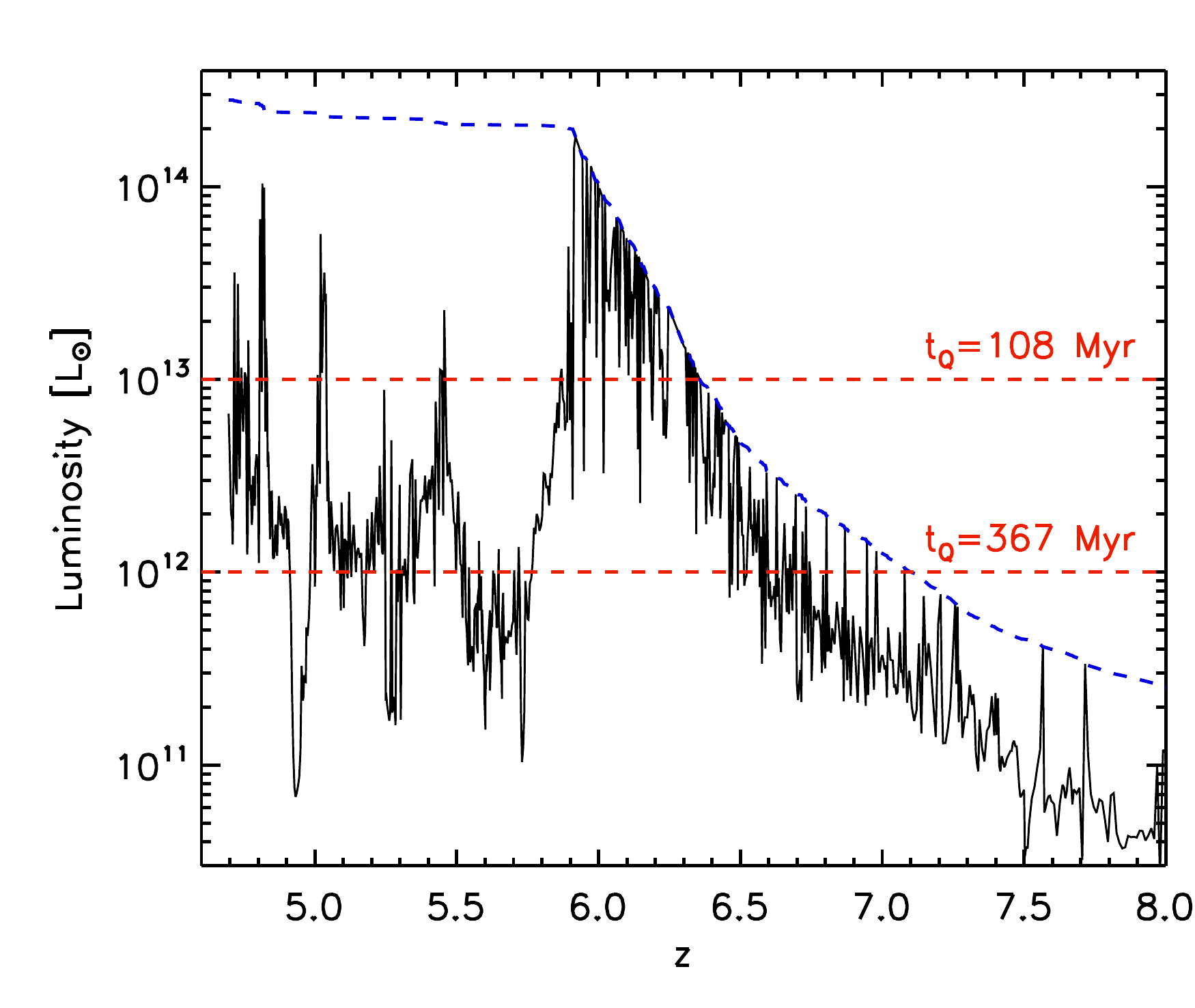}
}
\caption{ Left: An example
environment of quasars in simulations. Two galaxies are shown, along with their location in the scatter plot of BH mass vs
halo mass. Both halos have a mass $\sim 10^{12}$ \Msun,
but have experienced very different gas inflow and hence black hole growth.
   Right: An example of a bolometric lightcurve (over a selected redshift range to show details) from the bright
    quasar in the left panels. The dashed blue line
    shows the Eddington luminosity and the red quasar lifetimes at
    corresponding luminosities.}
\label{fig:z4env}
\end{figure}

Theoretical investigation of the QLF has been done using semi-analytical
models or halo models. 
Since, by construction, these models do not self-consistently follow black hole
growth, the quasar lightcurves (and luminosities) have to be calculated via
imposed prescriptions and a number of parameters are introduced for quasar
triggering, quasar lifetimes etc. So while these models offer more
flexibility for testing a variety of reasonable prescriptions and have
produced promising results it is still ideal to complement these approaches
with detailed hydrodynamic simulations. 
For example, in Figure~\ref{fig:z4env}
we showed a particular a view of two regions
that contain two massive BHs but with very different accretion histories, black hole masses and luminosities. In particular, these frames show the underlying distribution of
gas (color coded by temperature, red is hot and blue cold) with stars (in
white) and black holes indicated by the diffraction spikes whose size is scaled
by the QSO luminosity.  Many things are evident from this
picture. For example there is a clear distribution of quasar luminosities
related to large scale properties, 
and effects of BH feedback (clearly seen as
hot gas around the BHs).
The same mass halos, can be found in different large scale environments hosting dramatically different AGNs.
(see also Fig. \ref{fig:z4env}).
We see clearly that the effects 
modeled in  the simulations also directly provide for each BH  a
  detailed prediction of its full lightcurve across cosmic history
with high time resolution. This is directly predicted by the
interaction of the cosmological gas supply and resulting  BH
feedback. An example of a lightcurve and the predictions of the
associated luminosity functions derived from such a population is
shown in Fig~\ref{fig:z4env} (in black).

If we relate this to 
the QLFs, while the bright end QLF may
inform us about feedback, the faint-end where BHs are already mostly
self-regulated, should inform us about gas supply.
 In our simulations we typically find that although the low (high) luminosity ranges of the faint-end QLF are
dominated by low (high) mass black holes, a wide range of black hole masses
still contributes to any given luminosity range. The faint-end of the QLF can indeed be formed by quasars
  radiating well below their peak luminosities, rather than by quasars with
  low peak luminosities. 
This is consistent with the
complex lightcurves of black holes,
which show that any given black hole can 
undergo significant changes in its luminosity and hence (while its mass always grows) it can occupy different parts of the LF.
The complex light curve, and the resulting effects on the LFs are a result of the detailed hydrodynamics
followed in the simulations.

\begin{figure*}
\includegraphics[width=\textwidth]{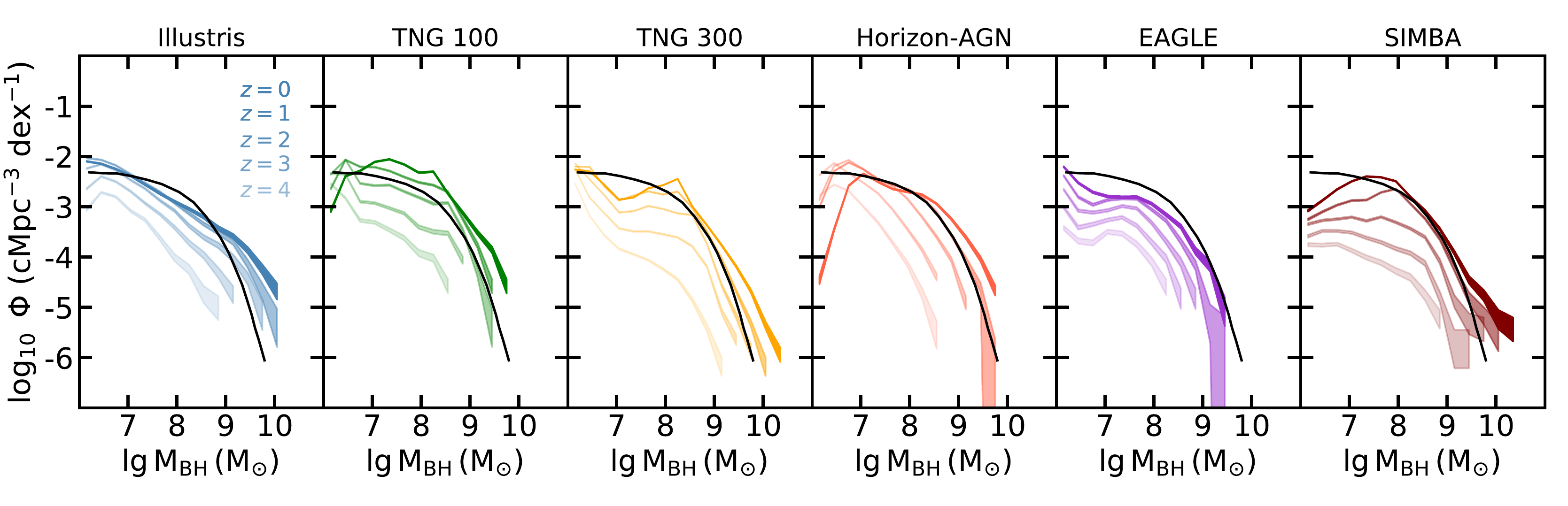}
\includegraphics[width=\textwidth]{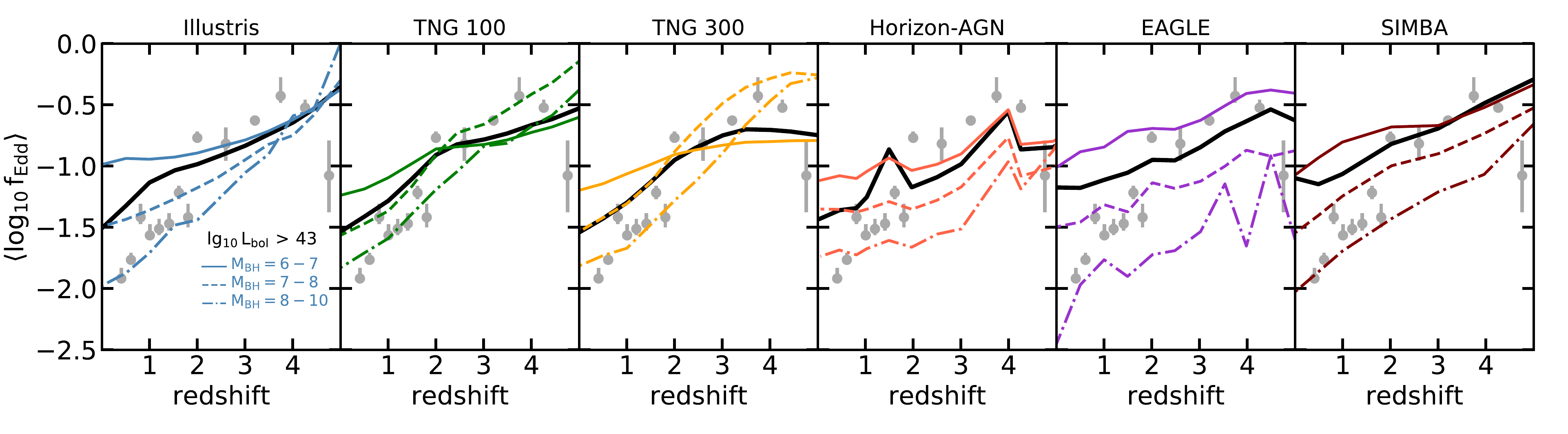}
\includegraphics[width=\textwidth]{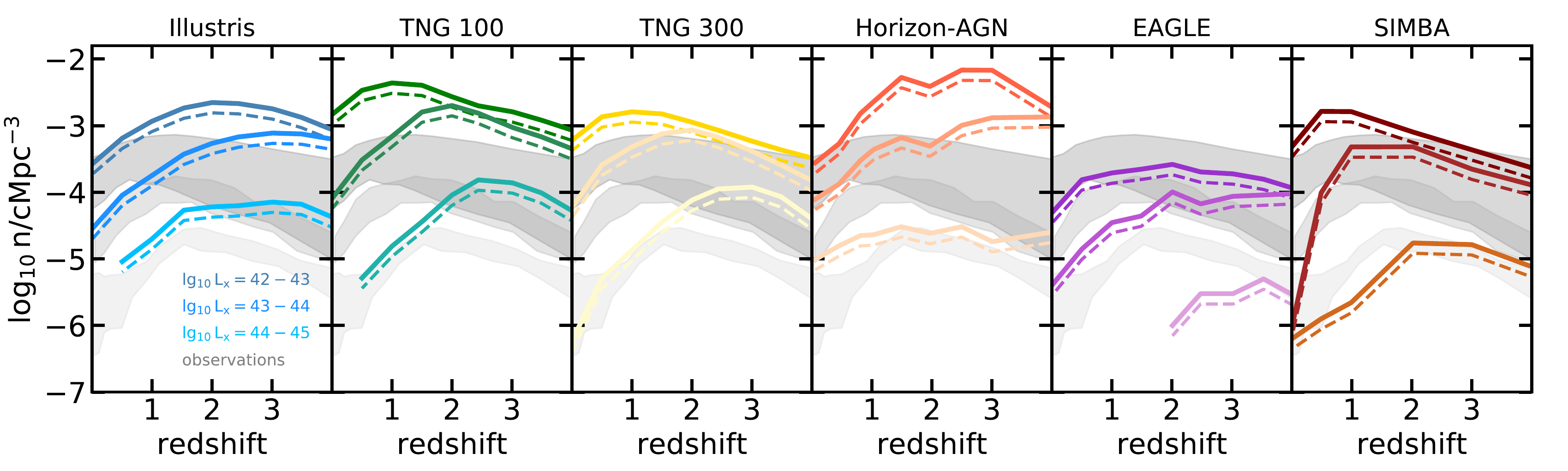}
\caption{Redshift evolution of the black hole mass function (top), the median Eddington ratio (middle), and the comoving number density of AGN of different hard X-ray luminosities (bottom) as predicted by several large-volume cosmological simulations (figures reproduced from \citet{Habouzit2021_Mbh,Habouzit2022_AGN}).
Simulations are compared against the black hole mass function at $z=0.1$ from the population synthesis model of \citet{MerloniHeinz2008} (black line; top panel), the observational constraints on Eddington ratios from \citet{ShenKelly2012} (grey points; middel panel), and the observed hard X-ray luminosity functions of \citet{Ueda2014, Aird2015,Buchner2015} (dark-to-light grey bands corresponding to increasing luminosity ranges; bottom panel).
Simulations qualitatively agree on the overall shape and build up of the black hole mass function, the decreasing median Eddington ratios at lower redshifts, and the early increase of AGN number density followed by a decline at low redshift, roughly following observational constraints.
}
\label{fig:LF} 
\end{figure*}

Fig.~\ref{fig:LF} shows examples of the black hole mass function, the typical Eddington ratio of AGN as a function of redshift, and the evolution in the comoving number density of AGN derived from a sub-set of recent simulations and presented in \citet{Habouzit2021_Mbh,Habouzit2022_AGN}.  As in Fig.~\ref{fig:msigma}, here we compare predictions from Horizon-AGN \citep{Dubois2014,Volonteri2016}, Illustris \citep{Vogelsberger2014,Sijacki2015}, IllustrisTNG \citep{Weinberger2017,Li2020_TNGbhs,Terrazas2020}, EAGLE \citep{Schaye2015,Rosas-Guevara2016}, and SIMBA \citep{Dave2019_Simba,Thomas2019}.
Cosmological hydrodynamic simulations generally agree in the overall build up of the black hole mass function over time and are in good agreement with constraints from population synthesis models.  This reflects the overall agreement in the predicted black hole--galaxy scaling relations and the evolution of the stellar mass function, which are primary observational targets to match in simulations and often used to constrain the subgrid parameters that control the efficiency of stellar and black hole feedback. 

While detailed black hole accretion histories depend on each specific model, most simulations agree between them and with observations on the overall redshift dependence of the median Eddington ratio of AGN.  Active black holes typically accrete gas at high Eddington ratios at early times ($z\sim4$) and decrease their specific growth rates at lower redshift, mimicking the overall decline in the specific star formation rate of galaxies \citep{kollmeier2006,Kauffmann2009,Shankar2013,Angles-Alcazar2015,Madau2014,Aird2018}.
Cosmological simulations are also used to predict the evolution in the number density of AGN at different luminosities, as illustrated in the bottom panels of Fig.~\ref{fig:LF}.  Most simulations agree in the overall shape of the number density evolution (roughly in agreement with observational constraints), with increasing number density of AGN (of any luminosity) at early times and decreasing number density at lower redshifts.  However, the amplitude and the redshift at which the maximum number density is reached can vary significantly from simulation to simulation \citep{Habouzit2022_AGN}.  Some models reproduce the observed ``downsizing'' effect, with brighter AGN reaching their peak number density at earlier times and fainter AGN becoming more abundant at later times
 \citep{Ueda2014,Aird2015}.

Continuity equation models have provided significant contribution in unveiling the evolution of the SMBH mass function, its duty cycle, and overall Eddington ratio distribution \citep{Marconi04,McLureDunlop04,Merloni04,Shankar04,Kollmeier06,Cao08,Silverman08,Aversa15}. The main results from these studies can be summarised as follows. The overall SMBH mass function growth via gas accretion is sufficient to fully reproduce the local SMBH mass function extracted from SMBH-host galaxy scaling relations. The impact of SMBH-SMBH mergers has the natural effect of increasing the 
the high-mass end of the SMBH mass function, which can then be reconciled with the high-mass end of the local
SMBH mass function via an increased radiative efficiency at high SMBH mass. The observed AGN fractions at low redshift requires a characteristic Eddington ratio $\lambda_c$ that declines at late times, and matching observed
Eddington ratio distributions requires a $P(\lambda)$ that broadens at low redshift. To reproduce the
observed increase of AGN fraction with black hole or galaxy mass, the $\lambda_c$ that
decreases with increasing SMBH mass, reducing the AGN luminosity associated with the
most massive SMBHs.

\subsection{Quasar Clustering}

\begin{figure}[t!]
\centering
\includegraphics[width=\textwidth]{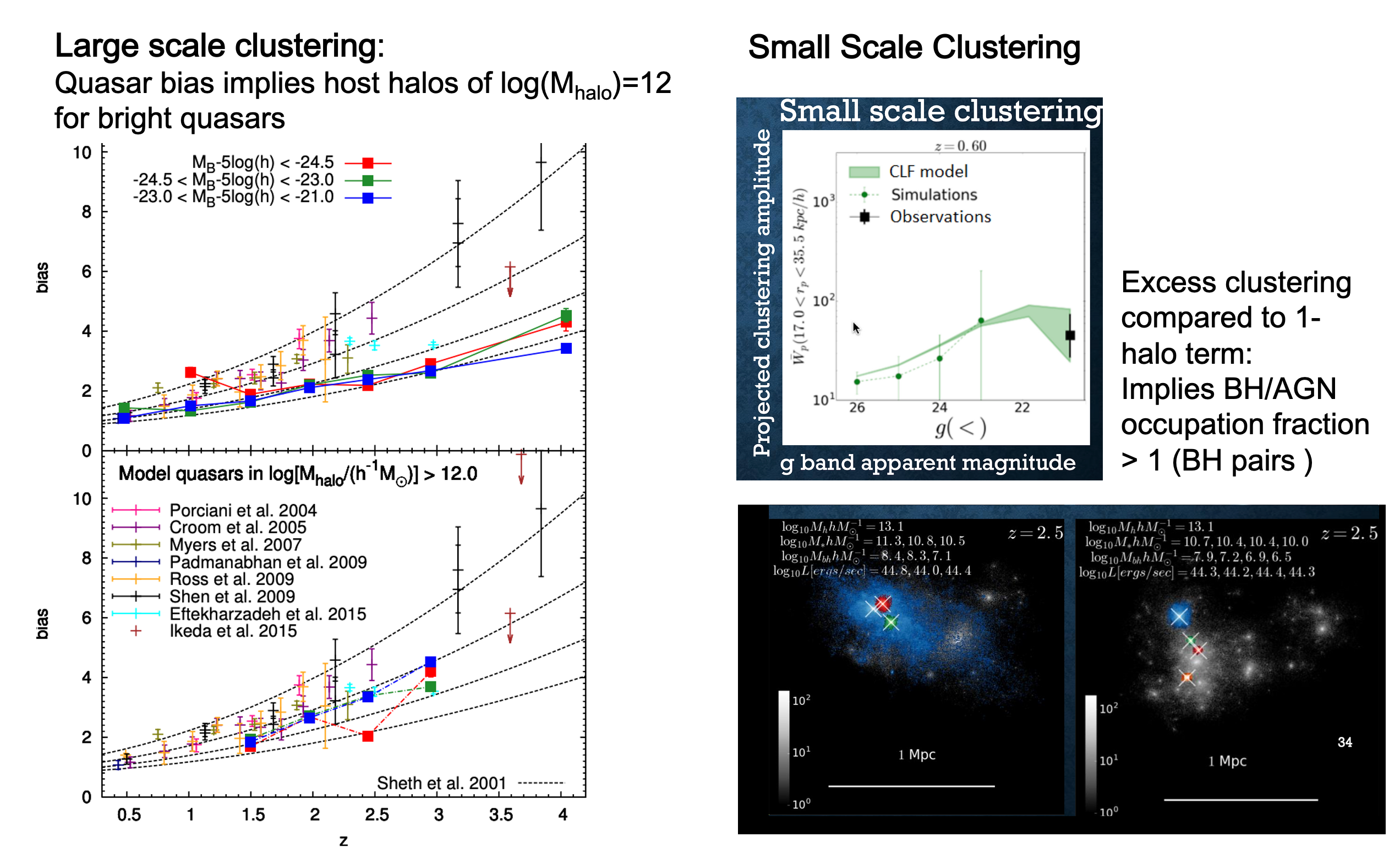}
\caption{Left: From \citep{DeGraf2011} Redshift evolution of the bias of bright
 intermediate and faint quasars (from \citep{oogi2016}).
The dashed
lines are halo bias factor evolution for fixed halo mass of
$log[M_{\rm halo}/(h^{-1}\,M_{\odot}$)] = 11.5, 12.0, 12.5 and 13.0 from bottom
to top, respectively, using the Sheth et al. (2001) fitting formula
with the Planck cosmology. Observational results are also plotted (plus signs and error bars). Bottom: the same as the top
panel, but for only model quasars which are hosted DM haloes
($ Mhalo \gtrsim 10^12 h^{-1} M_{\odot}$). Credit Oogi et al. 2016.
Right: From \citep{Bhowmick2018} Small scale clustering measurements and predictions from simulations.
$W_p$ is the volume-averaged projected correlation function averaged over $17.0 < r_p < 36.6$ kpc/h for quasars brighter than
a given magnitude threshold, which we denote by ‘g(<)’. The dashed lines correspond to predictions from the simulations.
he black squares correspond
to the observational constraints at g = 20.85.  Below, examples of systems of quasar pairs and triplets that give rise to the enhanced clustering.} 
\label{fig:clustering}
\end{figure}

Clustering measurements provide the means to better understand the relation
between quasars, their hosts and the underlying dark matter distribution, as
well as to allow estimates of quasar lifetimes
once coupled with QLF
constraints, e.g.;\cite{HaimanHui2001}, \cite{MartiniWeinberg2001}. With clustering we
learn about the co-evolution of quasars, mergers. For example, strong clustering would suggest
that quasars should reside in massive halos. If so, they should be rare and in
order to reproduce the quasar luminosity density, they must have long
lifetimes. Conversely, low spatial correlations would suggest more common
quasars, and thus shorter quasar lifetimes.

\noindent{\bf Large Scale Clustering}\\
The large scale quasar clustering properties
are quantified with the quasar bias, which is defined as
the square root of the ratio of the two-point correlation
function of the quasars to that of the dark matter. 
Comparing it with the bias of DM
halos predicted by, for example, Sheth et al. (2001), the typical DM
halo mass of the quasars is derived (see Figure~\ref{fig:SMBHMFfromCEHalo} for one recent example). The clustering properties
have been reported using some large-scale surveys.
Simulations have been used to compare to
constraints on clustering showing that quasar hosts at high-$z$
are consistent with the level of the observed quasar clustering bias for $10^{12}-10^{13}$\Msun halos. This is an
important starting point which gives us some confidence to further
pursue lower redshifts---towards the peak epoch of quasar
activity and overlap  with the and other upcoming  quasar surveys

Because of the detailed information and lightcurves and QLF
we have from the simulations  we can
directly derive host halo masses for a given luminosity range and
translate that into  predicted quasar clustering as a function of $z$
or luminosity.   from the ongoing BOSS and
upcoming eBOSS analyses and pushing large volume
simulations all the way to $z=2$, clustering as a function of luminosity
should be able to discriminate between different
models, and directly constrain duty cycles and the effects of
gas inflows and feedback in regulating quasar active phases.

\noindent{\bf Small Scale Clustering}\\
In addition to large scale behavior, the possibility of excess quasar
clustering on small scales has arisen in several recent studies.  While
some observed quasar pairs are believed to be the result physically distinct
quasar binaries (double nuclei), which would suggest quasars cluster much more
strongly on small scales than extrapolation of large scale clustering would
imply, indication perhaps  direct evidence for connection between galaxy mergers and quasar
activity  Several recent studies have
managed to probe even smaller scales, where they do indeed find (some level)
of excess.

Small scale clustering measurements for AGNs/quasars have
been also been significant interest over the last two decades as they
may constrain signatures of the physical processes that trigger AGN activity, such as galaxy mergers and the related efficiency of BH mergers resulting from presence  AGN pairs, triplets etc in the center of galaxies. Over the last 20 years the small-scale clustering of quasars, mainly from the
SDSS and 2dF-QSO surveys, at scales ranging from $\sim$ 10 kpc
to $\sim 1$ Mpc. 
Cosmological hydrodynamic simulations are valuable
tools to study AGN clustering (see Figure~\ref{fig:clustering}\citep{Bhowmick2018}). 
Large volume hydrodynamic simulations are
invaluable tools to study properties of AGN and quasar populations. 
Simulations currently indicate that the excess small scale clustering is due to quasars  pairs likely correspond to extremely luminous quasars in satellite galaxies which are triggered by galaxy mergers \citep{Bhowmick2017, Bhowmick2018}. Multiple sequence of such galaxy mergers can also lead to formation quasar triples and quadruples. Several detections of such exotic systems have been made in the recent past \citep{Farina2013, Djorgovski2007}.

\subsubsection{Semi-empirical models}
\begin{figure}
    \centering
   \includegraphics[width=0.9\textwidth]{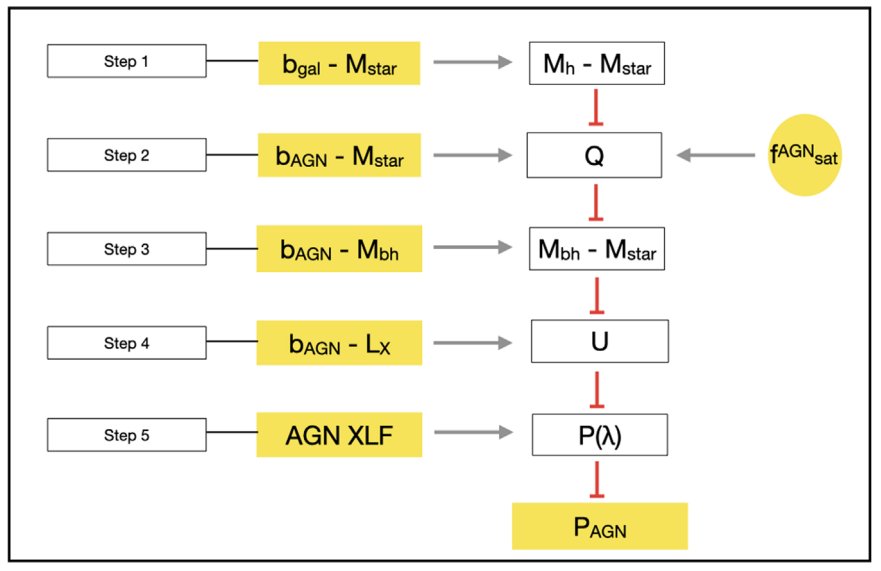}
    \caption{Sketch of how to build realistic AGN mocks. The dependence of each observable on one or a few input model parameters (open black boxes) is shown as red lines. From the comparison of observationally derived relations and the AGN mock catalog predictions, we can constrain (gray arrows) the input parameters. Additional observables, such as the fraction of satellite AGNs (filled yellow circle), can help in breaking the degeneracies among the input model parameters.}
    \label{fig:ViolaList}
\end{figure}

From the semi-empirical approach, significant insights have become available in the last years in relating the clustering of AGN and their hosts, to the 
physical properties of their central SMBHs \citep{ShankarNat,AirdCoil,Allevato21}.
It has been shown that model validation is often affected by degeneracies when comparing theoretical predictions and observational data, with the same AGN number densities and spatial distributions being reproduced by radically different models \citep{Allevato21}. However, recent work has also shown that, although multiple parameters are responsible for shaping SMBH demography through time, they all play different roles in generating different observables. For example, the stellar mass-halo mass relation can be
constrained by the large-scale clustering as a function of stellar mass, because the spatial distribution of AGN in relatively narrow bins of stellar mass is largely independent of, e.g., the level and frequency of AGN activity in the host galaxies, at least in the limit in which AGN
hosts are a random subsample of all galaxies of similar stellar
mass. On the other hand, the AGN large-scale bias as a function of SMBH mass can
be used to constrain the normalization and shape of the scaling relation between SMBH mass and host galaxy stellar mass \citep{ShankarNat}. Observational constraints on the AGN duty cycle can then be
derived from the comparison of the model predictions with the
measured AGN large-scale bias as a function of AGN luminosity. Finally, the combination of the AGN luminosity function and of the specific accretion rate distribution \citep{Georgakakis19} allow to constrain the input Eddington ratio distribution and duty cycle. Additional
observables can be considered, such as the average correlation between X-ray luminosity and host galaxy stellar mass/star formation rate in sample of active galaxies to constrain the mean Eddington ratios in AGN \citep{Carraro22}. Figure~\ref{fig:ViolaList} provides the different steps that one needs to follow to create a robust and realistic mock catalog of AGNs avoiding the risk of strong degeneracies, showing that the starting point is the large-scale clustering (bias $b$) as a function of galaxy stellar mass, SMBH mass, and AGN luminosity, with additional, complementary constraints offered by the AGN luminosity function, the relative fraction $f^{\rm AGN}_{\rm SAT}$ of satellite AGN in groups and clusters, and the specific accretion rate distribution $P_{\rm AGN}$. 

\subsection{AGN feedback and cosmology}


AGN feedback is key for reproducing the global evolution of the stellar
mass function and galaxy luminosity functions \citep{SomervilleDave2015}.
Simulations are also able to firmly predict the quasar bias across scales. 
The newest surveys coupled with these predictions
will allow us to determine how quasars probe large scale structure from very
small scales to Cosmic Microwave Background scales, which will be
important for planning future surveys.

Current and forthcoming cosmological experiments such as Dark Energy Spectroscopic Instrument \citep{DESI2016}, Dark Energy Survey \citep{DES2018}, Rubin Observatory's Legacy Survey of Space and Time (LSST) \citep{Ivezic2019_LSST}, the Roman Space Telescope \citep{Spergel2013_WFIRST}, and Euclid \citep{Laureijs2011_Euclid} rely on observed galaxy properties to constrain the properties of dark matter and dark energy with increasing accuracy. The fundamental challenge is that galaxy formation involves a complicated blend of different physical processes that is non-linearly coupled on a wide range of scales, leading to extremely complex dynamics. 
The required percent level accuracy to extract cosmological information from future surveys can only be reached through a much better understanding of galaxy formation in direct cosmological hydrodynamic simulations.
Massive black holes are one of the most important pieces of baryonic physics that we need to understand for constraining dark energy with upcoming weak lensing surveys.
AGN feedback can potentially spread baryons over multi-Mpc scales \citep{Borrow2020}, but the effects of AGN feedback on cosmological observables depend heavily on calibration from simulations.

\begin{figure}
\centering
\includegraphics[width=0.8\textwidth]{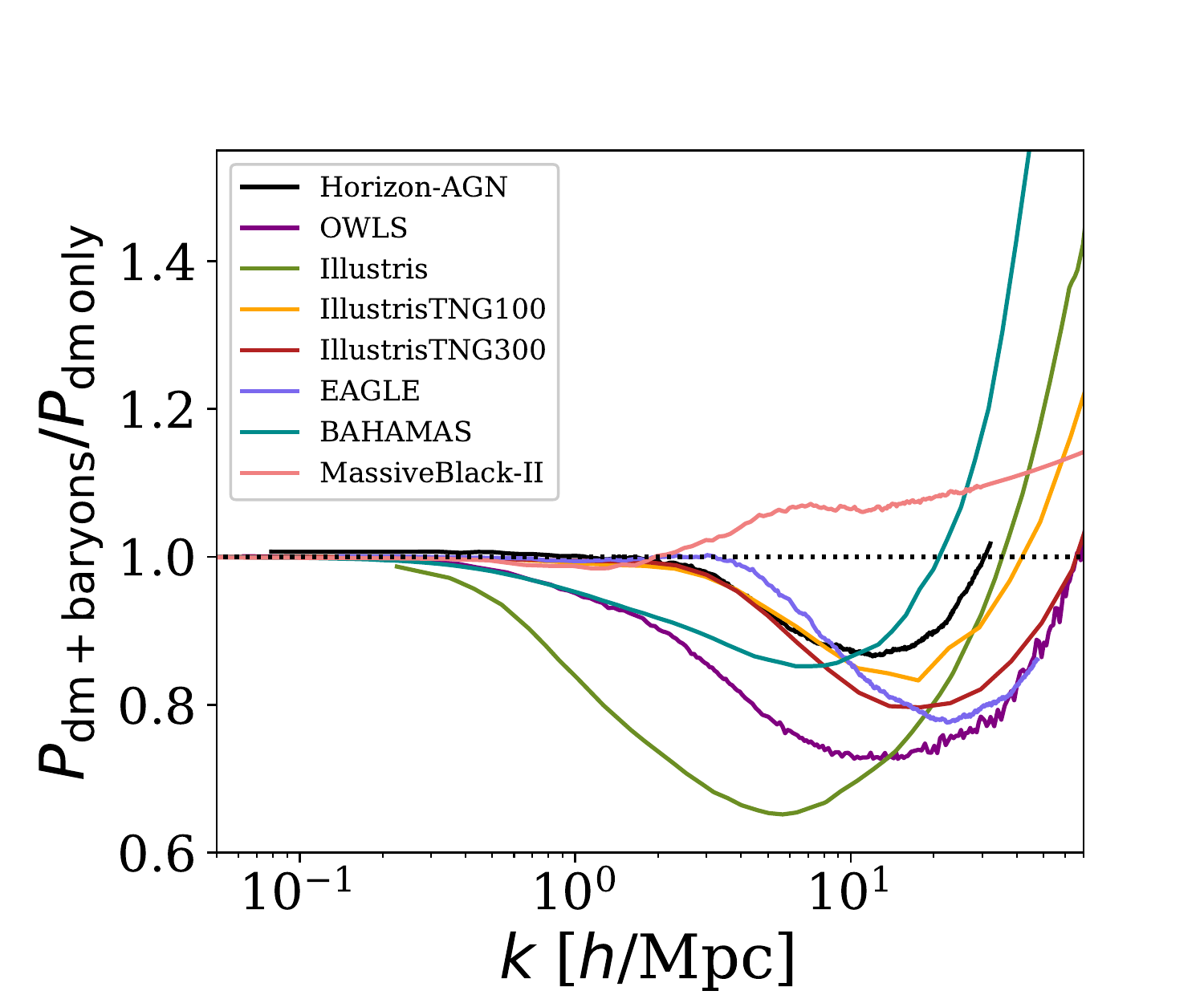}
\caption{Ratio of total matter power spectrum at $z=0$ in different cosmological hydrodynamic simulations to that of the corresponding dark matter-only simulations (figure reproduced from \citet{Chisari2019_review}).  The inclusion of baryonic physics increases the power on small scales relative to dark matter-only simulations owing to gas cooling and star formation, while generally suppressing power on large scales owing to the redistribution of gas and dark matter due to feedback processes.
}
\label{fig:Pk} 
\end{figure}


As an example, Fig.~\ref{fig:Pk} illustrates the importance of galaxy formation modeling in cosmology by comparing the impact of baryonic physics in the total matter power spectrum at $z=0$ as predicted by different cosmological hydrodynamic simulations \citep{Chisari2019_review}.
Predicted power spectra differ substantially from expectations based on dark matter-only simulations on a range of scales probed by cosmological surveys ($k \gtrsim 0.1\,h\,{\rm Mpc}^{-1}$).
Most simulations agree qualitatively on the overall impact of baryonic physics. Gas cooling and star formation increase the power on small scales relative to dark matter-only simulations while feedback processes suppress power on larger scales by ejecting gas out of halos, with the overall redistribution of baryons affecting also the dark matter component via back-reaction effects \citep{vanDaalen2011,Semboloni2011,Tenneti2015,Chisari2018,Springel2018,Chisari2019_review,vanDaalen2020,Villaescusa-Navarro2021_CAMELS}.  However, there is significant uncertainty in theoretical predictions, as indicated by the large quantitative differences between simulations \citep{Chisari2019_review,vanDaalen2020,Villaescusa-Navarro2021_CAMELS}.  Some models predict a suppression of power of $\gtrsim 30$\%~at wave numbers $k \sim10\,h\,$Mpc$^{-1}$ while others predict $<$10\%~suppression (or even enhancement of power) on similar scales. Effects on the power spectrum of this size must be accurately
taken into account in order to achieve precise constraints on dark energy through missions such as Euclid or Rubin-LSST. 
As Fig~\ref{fig:Pk} shows, the changes in the dark matter power spectrum induced by AGN feedback are of the same order of magnitude as those due to a different cosmological models \citep{Springel2018}.

Comprehensive studies of the impact of quasar-mode and radio-mode feedback in full cosmological semi-analytic models \citep{Granato2004,Croton2006,Menci06,Monaco07,Fanidakis12,Hirschmann12,Croton16,Fontanot20} have been performed using a variety of recipes for both accretion and feedback inspired by both hydrodynamic simulations \citep{Dubois13} and analytic models \citep{Faucher-Giguere2012_WindModel,Menci19}. First off, the inclusion of a light curve in SAMs to modulate accretion onto the central SMBH \citep{Hopkins06LF,Lapi06,Marulli2008,Bonoli2009}, provides a more extended, delayed triggering of the SMBH, and an improved match to the AGN luminosity function, especially at the faint end, and possibly to the large-scale clustering as a function of AGN luminosity. In the latest renditions of SAMs, the accretion onto the central SMBH is not only related to galaxy mergers, but to any event, e.g., disc instabilities or galaxy interactions \citep{Menci14,Fontanot20} that can generate loss of angular momentum in the gas and contribute to the reservoir of low angular momentum feeding the central SMBH. The main prominent consequence of the inclusion of quasar-mode or radio-mode AGN feedback in SAMs has been to reduce the number densities of massive galaxies \citep{Granato2004,Croton2006}. However, the inclusion of more refined AGN feedback recipes in SAMs in more recent times has revealed a plethora of additional interesting features that could be observationally testable \citep{Fiore17}. In some instances, the star formation rates tend to decrease at fixed stellar mass, and the amount of ejected material instead increases, whilst maintaining a low level of star formation in the most massive galaxies at all times, although in tandem with the effect of stellar feedback.

\subsection{Massive Black hole Binaries, Mergers and Gravitational Waves}
While black holes grow predominantly
via accretion, a second mode of black hole growth is through mergers which occur when dark
matter halos merge into a single halo, such that their black holes fall toward the
center of the new halo, eventually merging with one another.  
Mergers of
massive black holes are then a natural consequence of our current
hierarchical structure formation paradigm.
In much of what has been discussed in
this chapter and in our current understanding, massive black holes form and reside at the
centers of galaxies and hence they grow and merge closely intertwined
with their host galaxies.
The presence of luminous quasars observed
within the first billion years of the Universe highlights that the black hole seeds for
the massive black hole population were assembled at the cosmic dawn,
concurrently with the time of the formation of the first galaxies. In
our standard $\Lambda$CDM cosmology cosmic
structure formation occurs hierarchically by the continuous merging of
smaller structures and accretion of surrounding matter. SMBHs growth
and evolution is expected to follow a similar process in which black
hole seeds grow both though accretion and mergers with other BHs.

The upcoming Laser Interferometer Space Antenna (LISA) \citep{LISA2017arXiv170200786A} mission will be sensitive to low-frequency ($10^{-4}-10^{-1}$Hz) gravitational waves from the coalescence of MBHs with masses $10^4-10^7 M_\odot$ up to $z\sim 20$. 
 At lower frequencies, Pulsar Timing Arrays (PTAs) are already collecting data and the
Square Kilometer Array (SKA) in the next decade will be a major leap forward in
sensitivity. 
While MBH binaries are the primary sources for PTAs and LISA, these two experiments probe different stages of MBH evolution. 
PTAs are most sensitive to the early inspiral
(orbital periods of years or longer) of nearby ($z <1$) massive ($M_{\rm BH} \gtrsim 10^8\,M_\odot$) sources \citep{Mingarelli2017}. 
In contrast, LISA is sensitive to the inspiral, merger, and ringdown of MBH binaries at a wide range of redshifts \citep{Amaro-Seoane2012} and from smaller sources ($M_{\rm BH} \in [10^4M_\odot,10^7 M_\odot]$). 

In the last decade major efforts have been made to predict the event
rate of GWs in the frequency band of LISA \cite{Amaro2012},
  \cite{Amaro2013}. These predictions range from a few to a few hundred
events per year, depending on the assumptions underpinning the
calculation of the SMBHs coalescence rate. Early works derived the
SMBH coalescence rate from observational constraints such as the
observed quasar luminosity function, whilst more
recent studies have utilised semi-analytical galaxy formation models
and/or hybrid models that combine cosmological N-body simulations with
semi-analytical recipes for the SMBH dynamics \citep{WyitheLoeb2003,Enoki2004,Koushiappas2006, Micic2007, Sesana2009, Klein2016}.

In contrast to semi-analytic models, hydrodynamical simulations follow
the dynamics of the cosmic gas by direct numerical integration of the
equations of hydrodynamics, capturing non-linear processes that cannot
be described by simple mathematical approximations. Hence a more
complete and consistent picture of the evolution of SMBHs and their
host galaxies can be obtained.

Predicting SMBH mergers inevitably involves following a
variety of complex physical processes that cover many orders of magnitude in
physical scale. Black hole mergers occur at sub-parsec scales when
two galaxies within large dark matter halos are driven together by
large scale gravitational forces that drive the formation of the
cosmic web at $>$ Mpc cosmological scales. After the galaxy
merger, the central SMBHs are brought near the center of the main halo
due to dynamical friction against the dark matter, background stars,
and gas. Eventually the final SMBH merger occurs via the emission of
GWs. For reviews of SMBH dynamics in galaxy mergers we refer to \cite{ColpiDotti2011}\cite{Mayer2013}, \cite{Colpi2014}. The dynamical evolution of the
SMBH binary is expected to happen fast (coalescence timescale
10-100 Myrs) in  gas rich environments, thanks to the
efficient dissipation of angular momentum and energy from the binary.
Conversely, three-body interactions slow things down in gas poor
systems (leading to coalescence timescales $\sim$ Gyrs).

Newly developed, large volume hydrodynamic cosmological simulations
self-consistently combine the processes of structure formation at
cosmological scales with the physics of smaller, galaxy scales and thus
capture our understanding of black holes and their
connection to galaxy formation on many relevant scales. They
thus provide a good tool for predicting
for MBHB merger rates.
These simulations directly associate MBH binaries with their host galaxies, and they are carried out in large enough cosmological volumes to provide the statistical power to make merger rate predictions across cosmic time which are crucial for the upcoming observations. 

In order to accurately predict when MBH mergers occur in these simulations, one must account for the orbital decay and binary hardening timescales in a wide dynamical range. 
During galaxy mergers, the central MBHs start at large separation in the remnant galaxy (as much as a few tens of kpc). 
These MBHs then gradually lose their orbital energy and sink to the center of the remnant galaxy due to the dynamical friction exerted by the gas, stars, and dark matter around them \cite{Chandrasekhar1943}\cite{Ostriker1999}. 
When their separation is $\lesssim 1$ parsec, a MBH binary forms and other energy-loss channels begin to dominate, such as scattering with stars \citep{Quinlan1996, Berczik2006, Sesana2007b,Berentzen2009, Khan2011, Khan2013, Vasiliev2015}, gas drag from the circumbinary disk \citep{Haiman2009}, or, if relevant, three-body scattering with a third black hole \citep{Bonetti2018}.

Among these processes, only the dynamical friction decay affects the dynamics at orbital separation above the resolution of large-volume cosmological simulations. 
However, so far there is limited attempt to directly model dynamical friction (at small scales, close to the resolution) in the large-volume cosmological simulations mentioned above. 
In most cosmological simulations, once MBHs are within a given halo, they are simply repositioned to the minimum potential position of the host galaxy at each time step. 
For these simulations, (although sometimes the effects of subgrid dynamical friction are treated in post-processing), many spurious mergers occur during fly-by encounters. 
Among simulations that do include subgrid modeling of DF on-the-fly, \citep{Dubois2014} only includes the friction from gas but not stars, while \citep{Tremmel2017} and \citep{Hirschmann2014} model the dynamical friction from stars and dark matter particles. 
Most recently, \citep{Mannerkoski2021} uses a hybrid model to track the MBH dynamics during galaxy mergers on small scales, while including on-the-fly dynamical friction and stellar scattering computations.

Recent simulations directly incorporates additional dynamical friction modeling, 
\citep{Chen2021} for the MBH dynamics down to the resolution limit \citep{Hirschmann2014},\citep{Tremmel2015}.
With more physical modeling of the MBH dynamics, we can follow the in-simulation mergers for a more extended period of time over hundreds of Myrs, and almost completely prevent mergers during fly-by encounters. 
Moreover, for the first time we can aim to measure the orbital evolution and eccentricities of MBH pairs on sub-kpc scales. 
Such information should be important both for estimating the binary hardening timescales and for predicting the GW signals from the MBH mergers. 

The launch of LISA will extend the GW window to low frequencies,
opening new investigations into dynamical processes involving these
massive black hole binaries. MBHB are also the primary
multimessenger astrophysics sources. The GW events will be accompanied
by electromagnetic (EM) counterparts and, since information carried
electromagnetically is complementary to that carried gravitationally,
a great deal can be learnt about an event and its environment (binary
AGN and its host galaxy and beyond) across cosmic history as it
becomes possible to measure both forms of radiation in concert.

 Devising observing strategies for the new multimessenger astrophysics
LISA opens up will significantly benefit from predictions of EM counterparts of binary AGN and SMBH host galaxies. It will
require 'full-physics' hydrodynamical
cosmological simulations with sufficient resolution and volume.
One of the major goal objectives of  
LISA  is to trace the origin, growth
and merger history of massive black holes across cosmic ages.

\begin{figure}
\hbox{
    \includegraphics[width=0.49\textwidth]{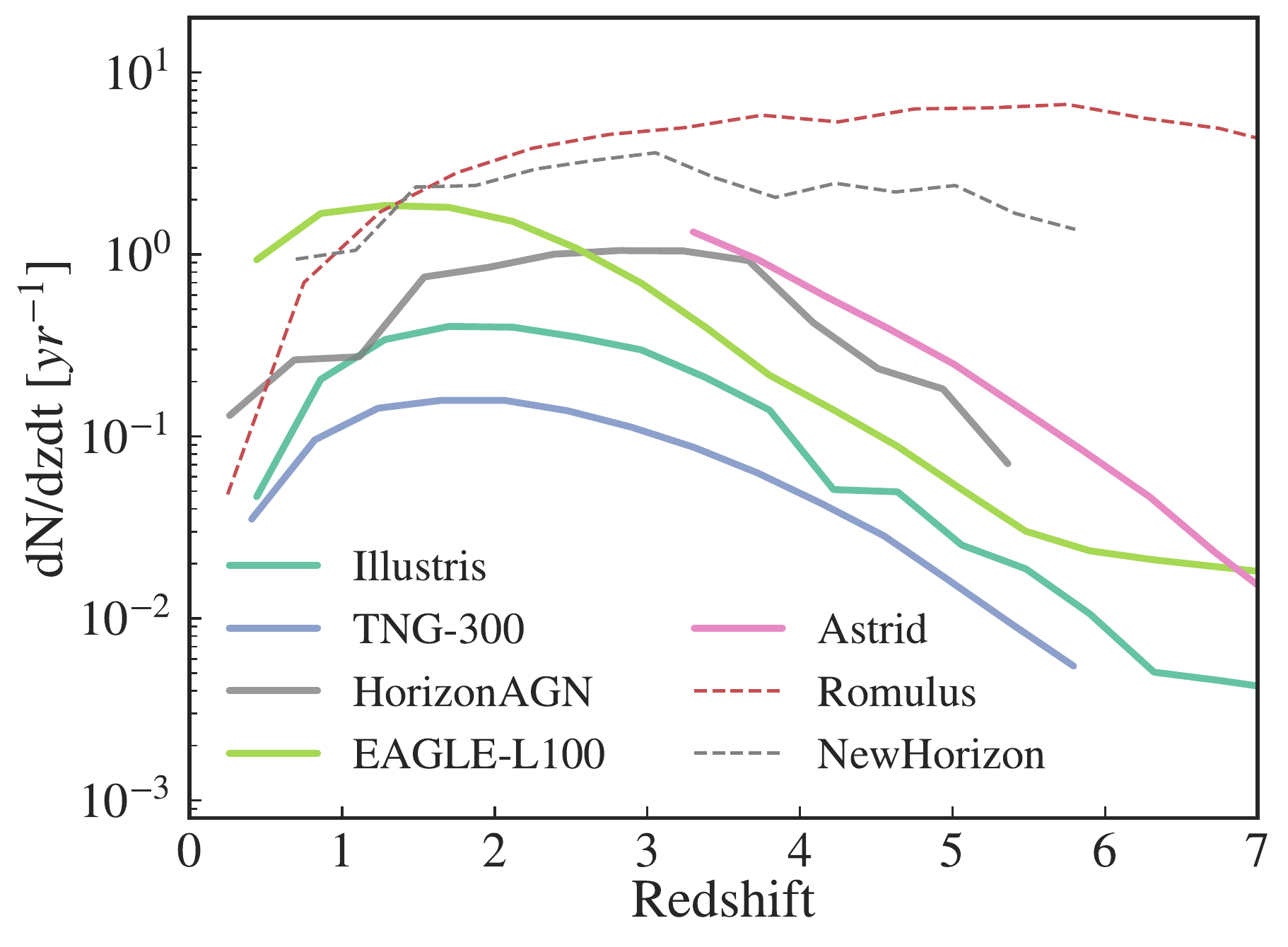}
    \includegraphics[width=0.51\textwidth]{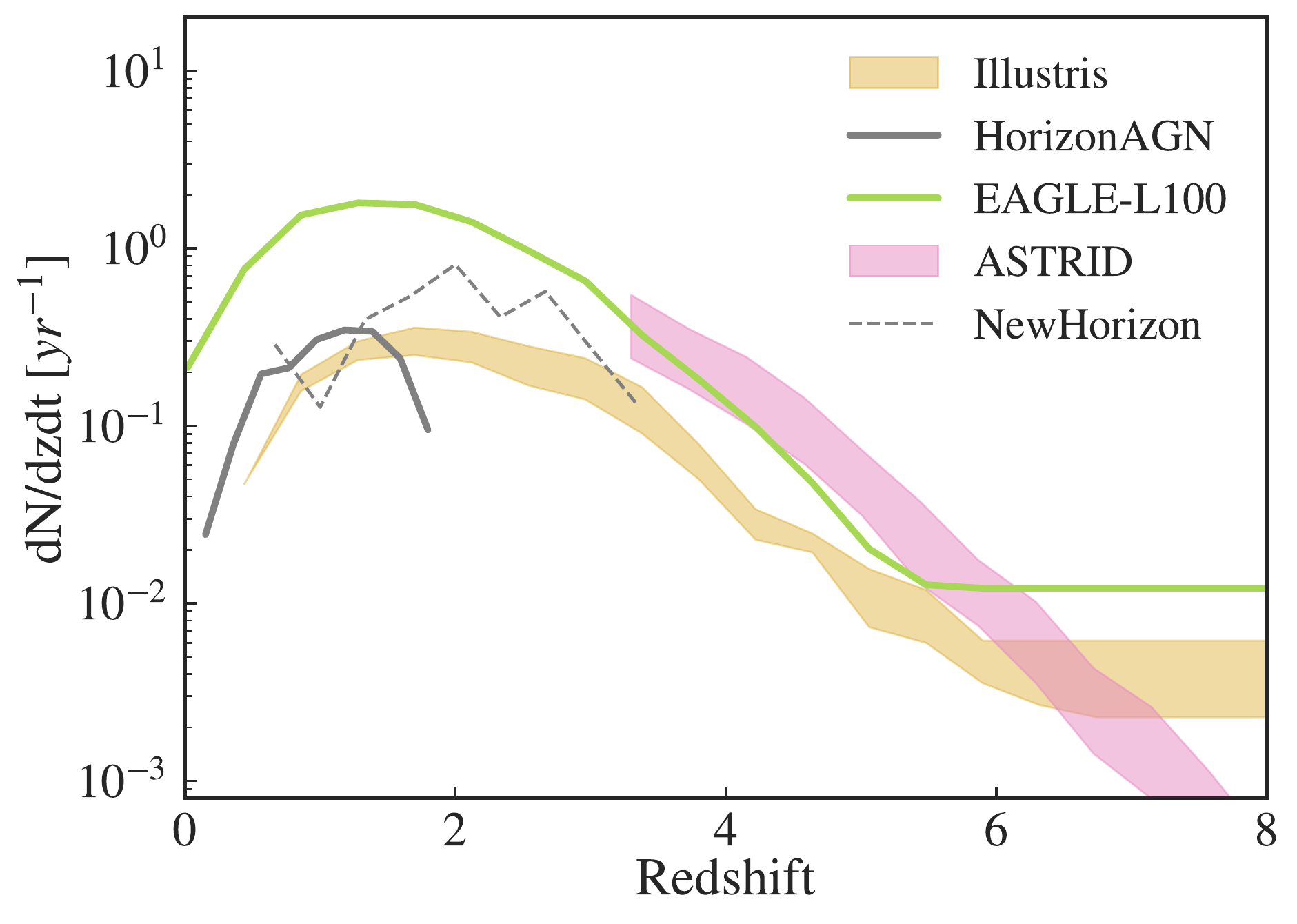}
    }
    \caption{The merger rates for all binaries in a suite of recent hydrodynamical simulations with different levels of delays. Without considering any post-processing delays  we expect a total of $\sim 2$ mergers per year. 
    The rate when considering only DF and hardening decreases  the merger rates significanlty at high redshifts. Credit for these figure: Nianyi Chen.}
    \label{fig:BHmergers}
\end{figure}

The state-of-the art multi-scale hydrodynamical
simulations that we have been describing  include different implementations for BH growth and
associated feedback. 
They can be used used to predict
SMBH mergers rates.
and perform accurate studies of the
predictions for SMBH merger rates (Figure~\ref{fig:BHmergers}).
Excitingly they can also 
provide corresponding EM counterparts and host galaxies for the MBHBs that can be observed in future and
upcoming facilities. Host galaxy identification of MBHB provides
unique information on galaxy-BH coevolution (and precise determination
of the distance-redshift relation). The first LISA detections of
massive black hole mergers will mobilize global astronomical resources
and be an astronomical event of enormous excitement. The mock
catalogs and synthetic observations that
one can obtain should be able bring
traditional astronomers into the LISA community and begin LISA
science with MBHM even before LISA is launched.

\begin{figure}
\begin{center}
\includegraphics[width=\textwidth]{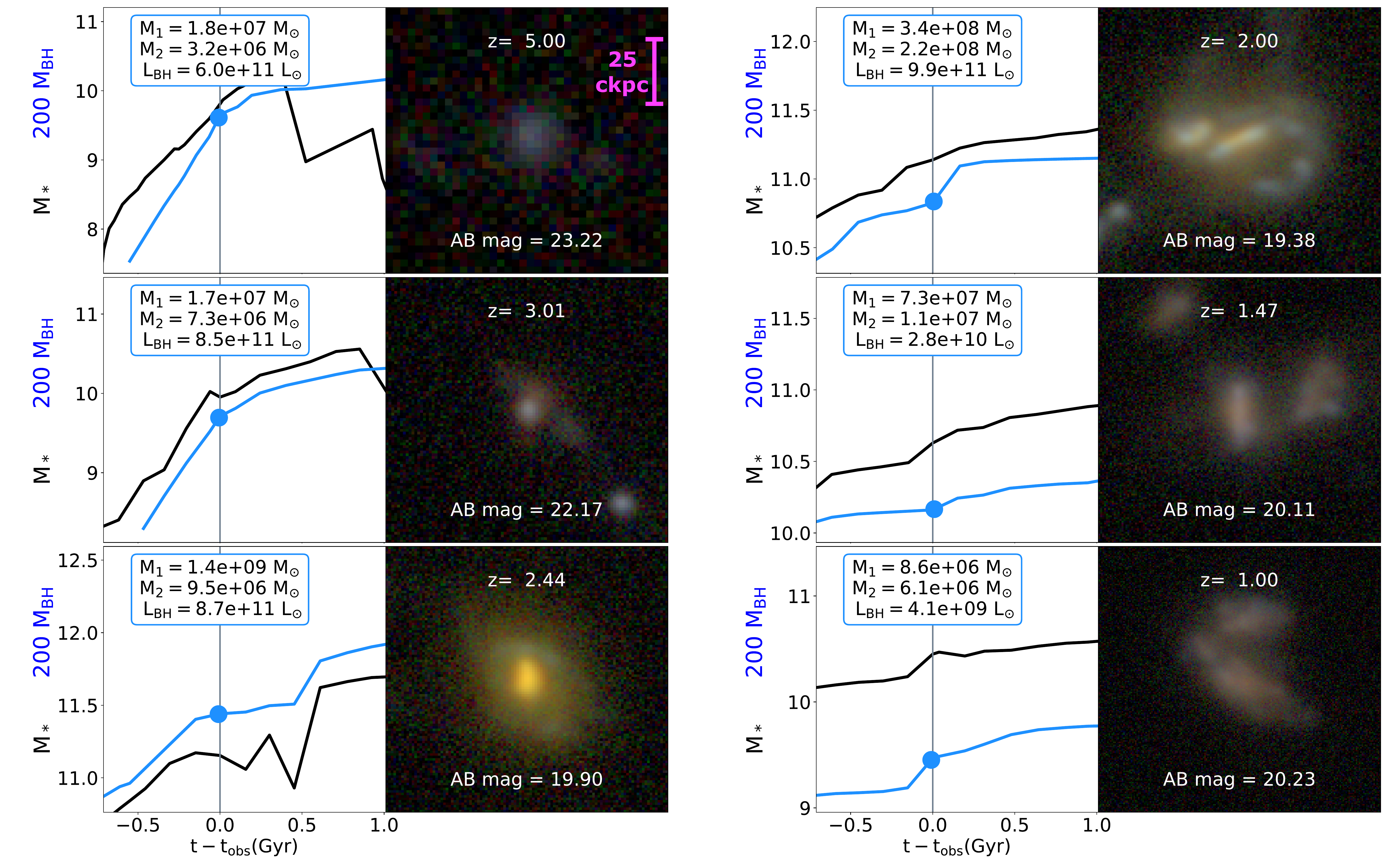}
\caption{Six example potential LISA sources from a catalog of MBHB
  mergers in Illustris, spanning mass, mass ratio, and
  redshift. These plots represent a small window on the $M_*$ and
  $M_{\rm BH}$ history of the main progenitor host in the left panel.
  The right panel shows simulated JWST-Nircam images of the
  hosts\cite{Snyder2015}. Many of these sources are associated with obvious galaxy
  mergers, and there is substantial diversity in the host morphology,
  luminosity, and color. \label{f:hosts}}
\end{center}
\end{figure}
\begin{figure}
    \centering
    \hbox{
    \includegraphics[width=0.5\textwidth]{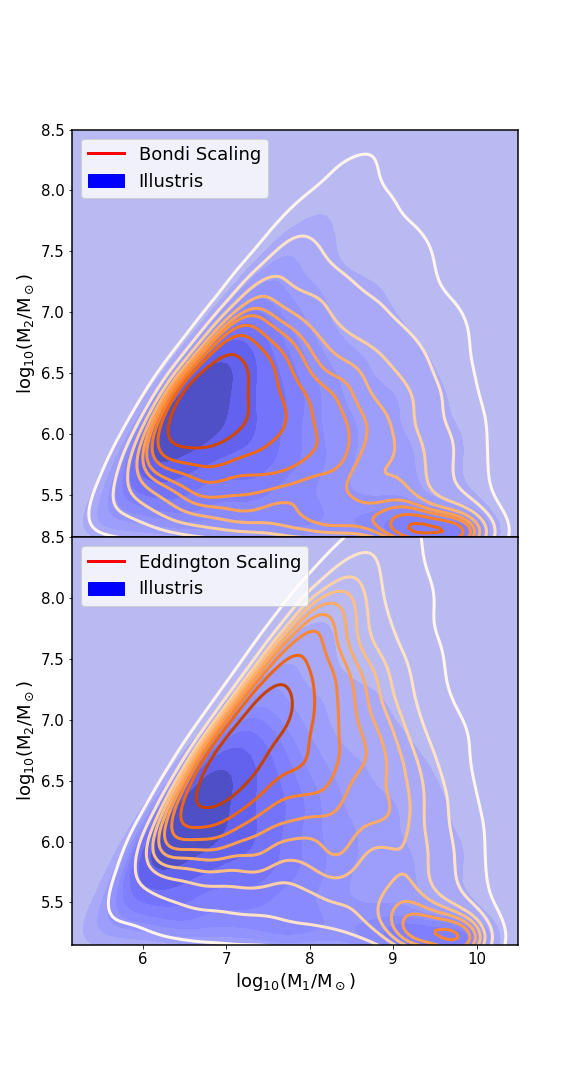}
    \includegraphics[width=0.5\textwidth]{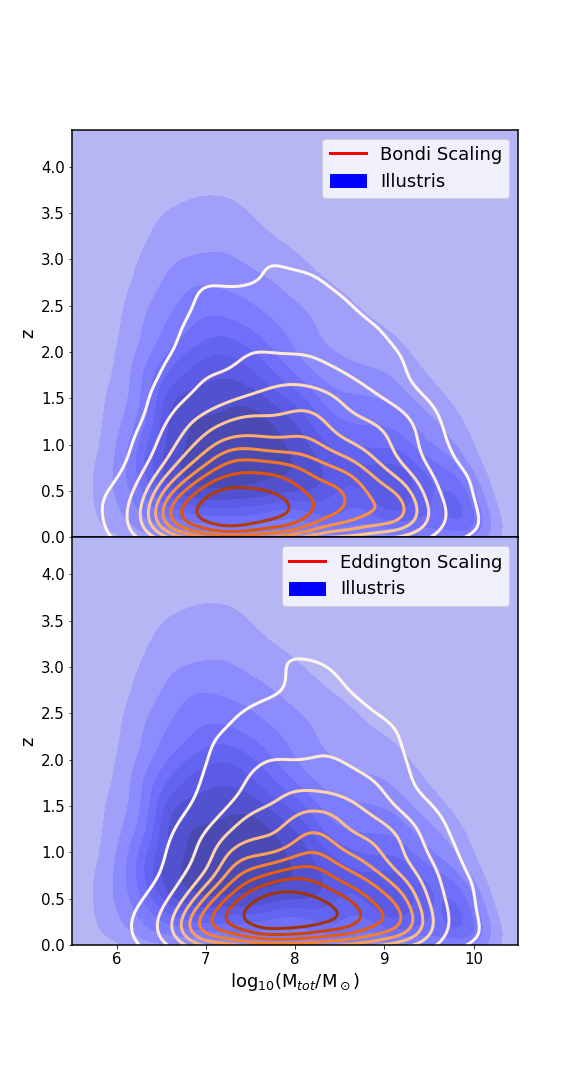}}
    \caption{\textit{Top:} Distribution of merger masses ($M_1$ and $M_2$), from the original Illustris simulation (blue) and after mass growth during the merger delay time, assuming growth follows a Bondi scaling (orange contours). \textit{Bottom:} Same as top, but assuming growth follows an Eddington scaling.
    Distribution of mergers across redshift and total merger mass ($M_{\rm{tot}} = M_1+M_2$) from the original Illustris simulation (blue), and after imposing a dynamical friction time delay, assuming growth follows a Bondi scaling.  \textit{Bottom:} Same as top, but assuming black hole growth follows an Eddington scaling.}
    \label{fig:M1M2}
    \end{figure}

To illustrate some of the data products that we have available in the
simulations we show some preliminary analyses. For each SMBH merger
that takes place in the simulations we store the mass of both SMBHs,
$M1$ and $M2$, and the redshift $z$ at which the merger event takes
place. Figure~\ref{fig:M1M2} we show the 2D histogram of the mass of
each BH member for all the mergers the Illustris
simulations considered here. The total number of BH mergers in each
simulation model is indicated in the figure.

After compiling a database of massive binary BH candidates from the
large cosmological simulations, we can
construct and organize
predictions for their host galaxy morphologies and AGN signatures.
The goal is to enable multi-messager astronomy with LISA
  sources via detailed comparisons between putative LISA events and
  telescope data that would illuminate properties of the EM
  counterparts and histories of their host galaxies.
 
\begin{figure}
\begin{center}
\includegraphics[width=\textwidth]{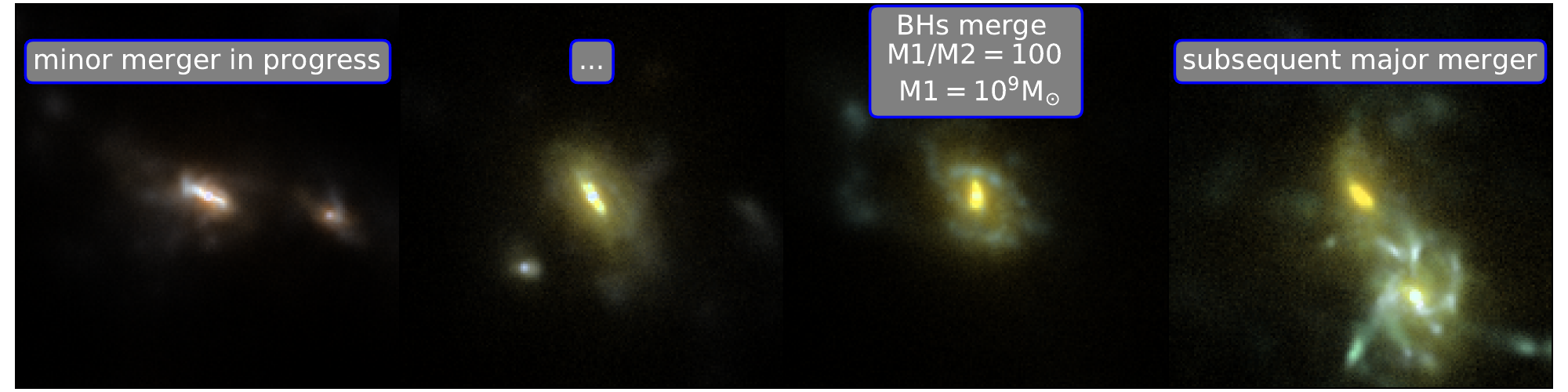}
\caption{Roughly $500 Myr$ of evolution of the host galaxy for the
  $z=2.44$ source in Figure~\ref{f:hosts} (3rd panel).  Labels
  describe the visible assembly processes acting on the galaxy and
  MBHB. The primary MBH has AGN emission visible as a bright blue
  point source in the first 3 panels.  Each panel has a fixed 50 
physical kpc field of view \cite{Snyder2015a}.  \label{f:sequence}}
\end{center}
\end{figure}

Following the dynamics of galaxy mergers with their black holes we
can  to estimate the incidence of {\it dual AGN} (at least
down to typical separation of a few kpc, at which we still can resolve
dynamical friction directly) and the detectability of these binary
systems.
Correspondingly we have detailed properties of the stellar
distribution of the host galaxies, with age, metallicities, star
formation rates and associated morphologies. As illustrated in
Fig.~\ref{f:hosts} we are able to statistically {\em characterize
  the type of galaxy for given minor/major BH merger events across
  redshift}.  Even with LISA more than a decade away, our aim would
be to predict which future facilities ( Webb, Luvoir, OST) will be
needed to to observe the type of galaxy which will be MBHB hosts out to high
redshifts. For example, the $z=5$ example shown in Fig.~\ref{f:hosts}
is the highest redshift host galaxy of a MBHB event in Illustris
that will be detactable by JWST (as at these redshifts a MBHB typically
involves lower masses) \cite{Snyder2015a}.

In Figure~\ref{f:hosts}, we show mock JWST images and mass assembly
histories of several galaxies hosting MBHB sources in Illustris. We
selected these sources to span a range of redshifts, masses, and mass
ratios  Such products can
be used to characterize galaxy morphology and AGN activity which may
indicate recent (bulges) or ongoing (companions, tails, etc) merging
activity and therefore link (in simulations) the population of LISA
sources to the story of how galaxy populations
assembled. Figure~\ref{f:sequence} shows the time evolution of a
single such source over $\sim 500 Myr$. In this evolution, a minor
galaxy merger delivers a MBH with $M2/M1 \sim 100$ to the primary
host ($M1 \sim 10^9$), and these BHs merge near the time shown in the
3rd panel. We can see that this MBHB merger occurs during a period of
rapid galaxy assembly in the host.

 \begin{figure*}
    \centering
    \includegraphics[width=0.9\textwidth]{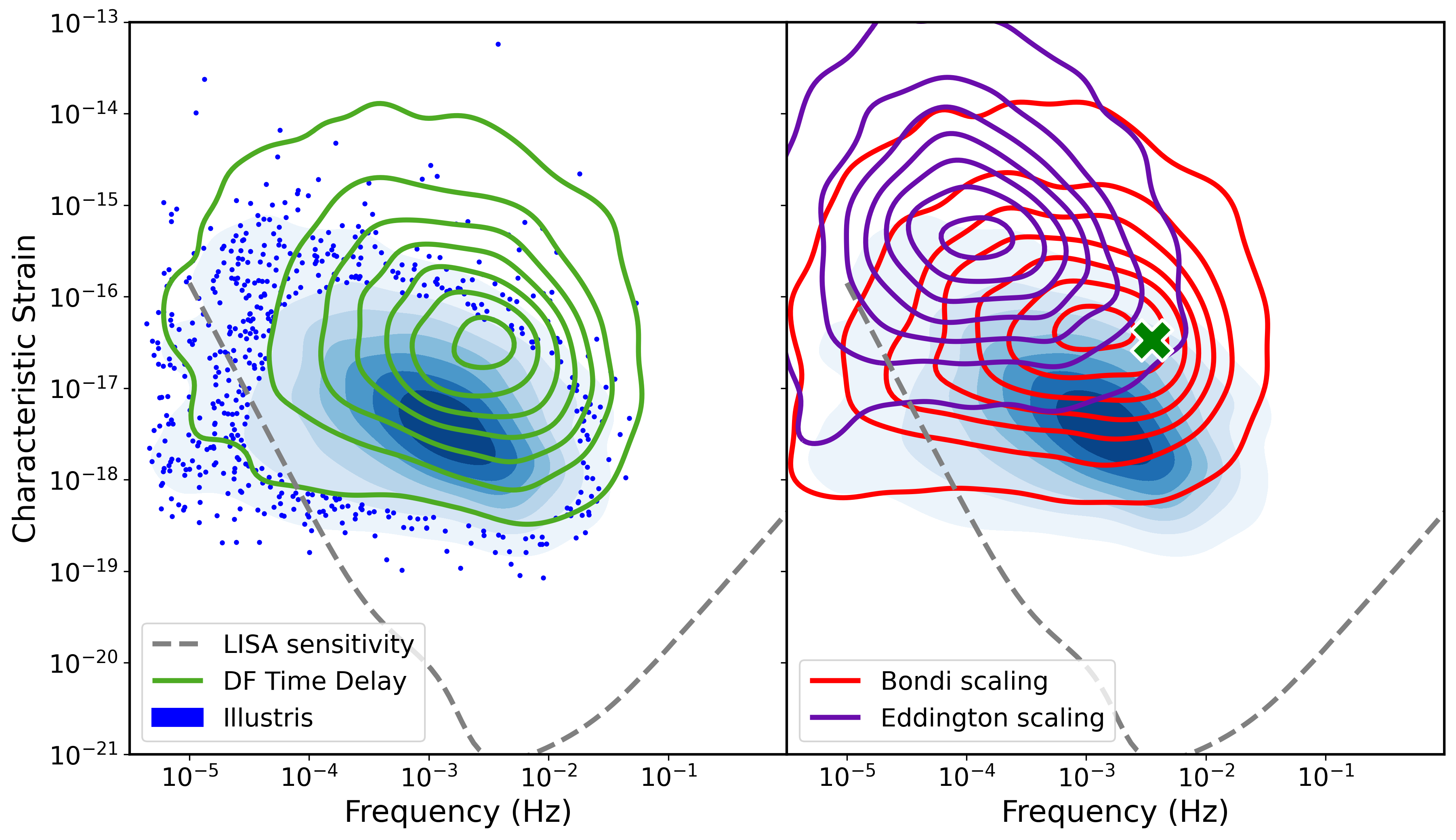}
    \caption{\textit{Left panel:} Illustration of the distribution of GW signals in frequency-strain space for BH bianries in Illustris simulation (blue), and after incorporating a dynamical friction time delay (green). In black the LISA sensitivity curve. \textit{Right panel:} Distribution of GW signals after incorporating black hole mass growth according to Bondi scaling (red) or Eddington scaling (purple).  The green cross shows the peak of the distribution without any mass growth (i.e. peak of the green contours from left panel).  Incorporating a time delay primarily increases the strain, with a minor increase in frequency.  Incorporating mass growth decreases the frequency and further increases the strain, especially for more efficient growth (i.e. assuming Eddington scaling) \cite{Banks2022}.}  
    \label{fig:freq-strain}
\end{figure*}

Currently a few 
teams have been able to carry out impressive MHD simulations of
circumbinary disks \cite{Bowen2018}, \cite{Lousto2017}, \cite{Farris2015}
around relativistic binary BHs which are now starting to produce
detailed EM counterpart signatures for these events. In the near future it will be possible to use large scale simulations
to provide reasonable 'initial conditions' of the gas environments
for these smaller scales around the BHs at the time of mergers that
the detailed simulations could use to derive  realistic {\em EM                                
signatures for a given mass ratio event in a given environment/galaxy                          
host}.
 
\begin{figure}
\begin{center}
\includegraphics[width=\textwidth]{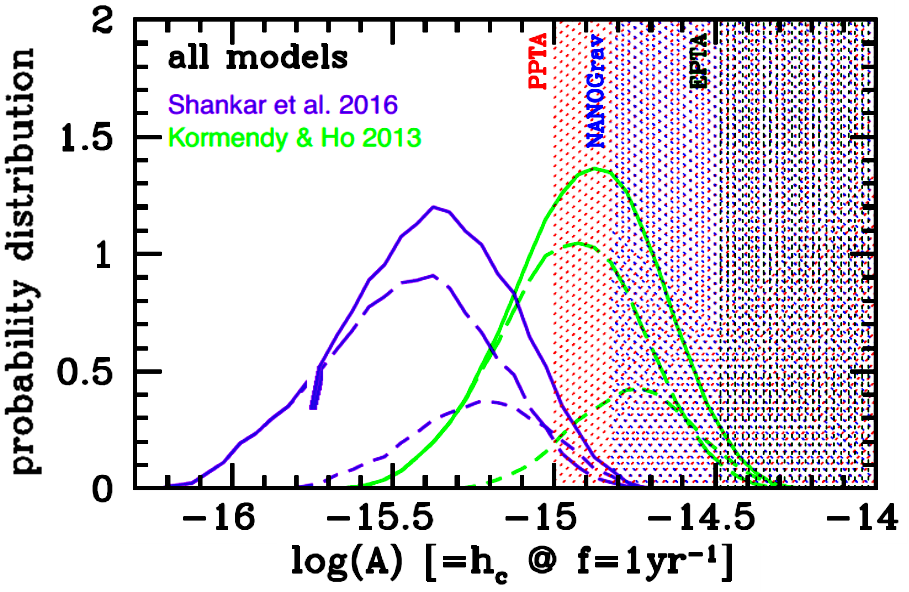}
\caption{Probability distribution \citep{Sesana16} of the signal amplitude $A$ assuming different scaling relations, as labelled. The two sets of lines for each scaling relation corresponds to two evolution patterns in the input stellar mass functions.  \label{fig:Sesana}}
\end{center}
\end{figure}

From an analytic modelling point of view, SMBH merger rates have been investigated by incorporating SMBHs in merging galaxies in a full cosmological context by assigning SMBHs to galaxies via different scaling relations \citep{Sesana16,Barausse20,Izqui22}. 
An example \citep{Sesana16} of this procedure is shown in Figure~\ref{fig:Sesana}, in which two scaling relations have been adopted, one characterizing the sample of local, dynamically measured SMBHs \citep{Kormendy2013} (green lines), and one that includes a possible correction for observational biases induced by the sphere of influence of the central SMBH \citep{Shankar16} (purple lines). It is found that the latter SMBH–host
galaxy relations imply a drop of a factor of $\sim 3$ in the signal amplitude $A$. This result by itself could help resolving any potential tension between recent PTA upper limits and theoretical predictions, without
invoking any additional physics related to the dynamics of SMBH
binaries, such as stalling, high eccentricity or strong coupling with
the surrounding stellar and gaseous environment. 

\section{Concluding remarks}

In this extensive review we have analysed in some detail several aspects of the modelling of SMBHs in a cosmological context, from full hydrodynamic simulations, to semi-analytic and semi-empirical models. A number of interesting conclusions can be drawn from this varied discussion:
\begin{itemize}
    \item Cosmological semi-analytic and hydrodynamic simulations grow SMBHs from seed BHs that range from $10^2$ to $10^6\, M_{\odot}$, roughly covering the full range of theoretical expectations. All recent both numerical and analytic models, predict a steep rise in the SMBH mass function at low masses \citep{Trinca22,Sicilia22}. Recent hydrodynamic simulations have revealed that such seed SMBHs can effectively grow into the giants rare quasars observed at $z>6$ thanks to the large-scale tidal fields destabilizing significant amounts of gas funnelling onto the seed BHs (see Figure~\ref{fig:firstqso}).  
    \item SMBHs grow primarily by (gas) accretion. Although other possibilities have been put forward in the literature \citep{Peirani08,Croker21}, as it stands, all current models have shown that, by making use of standard values of the radiative and kinetic efficiencies, gas accretion by itself could be a sufficient condition to reproduce the SMBH demography (in terms of both mass density and scaling relations) as calibrated in the local Universe (see, for example, Figures~\ref{fig:SMBHMFfromCE} and \ref{fig:LF}).
    \item Relatively frequent SMBH mergers are nevertheless predicted to occur in almost all hydrodynamic simulations, with a frequency that could be as high as a couple of mergers per year, but most probably much lower when including any sources of stalling (see Figure~\ref{fig:BHmergers}). LISA will be key to discern among the viable models of SMBH mergers (Figure~\ref{f:hosts}).
    \item SMBHs can feed back to the ISM significant amounts of energy and momentum via radiation pressure on free electrons and dust, as well as via large-scale mechanical winds and jets. Several cosmological models implement both a quasar-mode and radio-mode SMBH feedback, with the former regulating the initial growth of the SMBH and inner mass density, and the latter shown to be effective in producing a red galaxy sequence, in preventing overgrowth in massive galaxies, and in generating more realistic galactic and thermodynamic profiles.
    \item Strong degeneracies in the input parameters within a single model (among seeding, dynamics, accretion, and feedback), as well as degeneracies among the different implementations of SMBH feedback (e.g., Figure~\ref{fig:SimGallery}), still prevent to underpin the true underlying processes at play shaping SMBH and host galaxy growth. We note, however, that cosmological models implementing gravitational torque-based accretion predict the latter to be nearly independent of SMBH mass and with less need of self-regulation.
    \item Present cosmological models, both numerical and analytic ones, have been able not only to reproduce the number densities of SMBHs, both active and inactives ones, but also their spatial distribution as traced in the small and large-scale clustering properties of active SMBHs (see, e.g., Figure~\ref{fig:clustering}). Quasars hosts are predicted to be always related to host dark matter haloes of the order of $10^{12}-10^{13}\, M_{\odot}$, at all relevant redshifts, while the same models also suggest that excess small scale clustering is possibly due to quasar pairs in satellite galaxies.   
\end{itemize}

\bibliographystyle{ws-rv-van}
\bibliography{all_g}

\begin{thebibliography}{411}
\providecommand{\natexlab}[1]{#1}
\providecommand{\url}[1]{\texttt{#1}}
\expandafter\ifx\csname urlstyle\endcsname\relax
  \providecommand{\doi}[1]{doi: #1}\else
  \providecommand{\doi}{doi: \begingroup \urlstyle{rm}\Url}\fi

\bibitem{Spergel2006}
D.~N. {Spergel}, R.~{Bean}, O.~{Dore'} {\em et~al.}, {Wilkinson Microwave
  Anisotropy Probe (WMAP) Three Year Results: Implications for Cosmology},
  \emph{ArXiv Astrophysics e-prints}  (Mar., 2006).

\bibitem{Barnes1986}
J.~{Barnes} and P.~{Hut}, {A hierarchical O(N log N) force-calculation
  algorithm}, \emph{\nat}. {\bf 324}, \penalty0 446--449  (Dec., 1986).
\newblock \doi{10.1038/324446a0}.

\bibitem{springel05}
V.~{Springel}, {The cosmological simulation code GADGET-2}, \emph{\mnras}. {\bf
  364}\penalty0 (4), \penalty0 1105--1134  (Dec., 2005).
\newblock \doi{10.1111/j.1365-2966.2005.09655.x}.

\bibitem{Springel2010}
V.~{Springel}, {E pur si muove: Galilean-invariant cosmological hydrodynamical
  simulations on a moving mesh}, \emph{\mnras}. {\bf 401}, \penalty0 791--851
  (Jan., 2010).
\newblock \doi{10.1111/j.1365-2966.2009.15715.x}.

\bibitem{Hopkins2015_Gizmo}
P.~F. {Hopkins}, {A new class of accurate, mesh-free hydrodynamic simulation
  methods}, \emph{\mnras}. {\bf 450}, \penalty0 53--110  (June, 2015).
\newblock \doi{10.1093/mnras/stv195}.

\bibitem{Bauer2011}
A.~{Bauer} and V.~{Springel}, {Subsonic turbulence in smoothed particle
  hydrodynamics and moving-mesh simulations}, \emph{\mnras}. p. 3102  (May,
  2012).
\newblock \doi{10.1111/j.1365-2966.2012.21058.x}.

\bibitem{Sijacki2012}
D.~{Sijacki}, M.~{Vogelsberger}, D.~{Kere{\v s}} {\em et~al.}, {Moving mesh
  cosmology: the hydrodynamics of galaxy formation}, \emph{\mnras}. {\bf 424},
  \penalty0 2999--3027  (Aug., 2012).
\newblock \doi{10.1111/j.1365-2966.2012.21466.x}.

\bibitem{Vogelsberger2012}
M.~{Vogelsberger}, D.~{Sijacki}, D.~{Kere{\v s}} {\em et~al.}, {Moving mesh
  cosmology: numerical techniques and global statistics}, \emph{\mnras}. {\bf
  425}, \penalty0 3024--3057  (Oct., 2012).
\newblock \doi{10.1111/j.1365-2966.2012.21590.x}.

\bibitem{Torrey2012}
P.~{Torrey}, M.~{Vogelsberger}, D.~{Sijacki} {\em et~al.}, {Moving-mesh
  cosmology: properties of gas discs}, \emph{\mnras}. {\bf 427}, \penalty0
  2224--2238  (Dec., 2012).
\newblock \doi{10.1111/j.1365-2966.2012.22082.x}.

\bibitem{Genel2013}
S.~{Genel}, M.~{Vogelsberger}, D.~{Nelson} {\em et~al.}, {Following the flow:
  tracer particles in astrophysical fluid simulations}, \emph{\mnras}. {\bf
  435}, \penalty0 1426--1442  (Oct., 2013).
\newblock \doi{10.1093/mnras/stt1383}.

\bibitem{Nelson2013}
D.~{Nelson}, M.~{Vogelsberger}, S.~{Genel} {\em et~al.}, {Moving mesh
  cosmology: tracing cosmological gas accretion}, \emph{\mnras}. {\bf 429},
  \penalty0 3353--3370  (Mar., 2013).
\newblock \doi{10.1093/mnras/sts595}.

\bibitem{Hirschmann2014}
M.~{Hirschmann}, K.~{Dolag}, A.~{Saro} {\em et~al.}, {Cosmological simulations
  of black hole growth: AGN luminosities and downsizing}, \emph{\mnras}. {\bf
  442}\penalty0 (3), \penalty0 2304--2324  (Aug, 2014).
\newblock \doi{10.1093/mnras/stu1023}.

\bibitem{Steinborn2015}
L.~K. {Steinborn}, K.~{Dolag}, M.~{Hirschmann} {\em et~al.}, {A refined
  sub-grid model for black hole accretion and AGN feedback in large
  cosmological simulations}, \emph{\mnras}. {\bf 448}, \penalty0 1504--1525
  (Apr., 2015).
\newblock \doi{10.1093/mnras/stv072}.

\bibitem{Dubois2014}
Y.~{Dubois}, C.~{Pichon}, C.~{Welker} {\em et~al.}, {Dancing in the dark:
  galactic properties trace spin swings along the cosmic web}, \emph{\mnras}.
  {\bf 444}, \penalty0 1453--1468  (Oct., 2014).
\newblock \doi{10.1093/mnras/stu1227}.

\bibitem{Volonteri2016}
M.~{Volonteri}, Y.~{Dubois}, C.~{Pichon} {\em et~al.}, {The cosmic evolution of
  massive black holes in the Horizon-AGN simulation}, \emph{\mnras}. {\bf
  460}\penalty0 (3), \penalty0 2979--2996  (Aug, 2016).
\newblock \doi{10.1093/mnras/stw1123}.

\bibitem{Schaye2015}
J.~{Schaye}, R.~A. {Crain}, R.~G. {Bower} {\em et~al.}, {The EAGLE project:
  simulating the evolution and assembly of galaxies and their environments},
  \emph{\mnras}. {\bf 446}, \penalty0 521--554  (Jan., 2015).
\newblock \doi{10.1093/mnras/stu2058}.

\bibitem{Rosas-Guevara2015}
Y.~M. {Rosas-Guevara}, R.~G. {Bower}, J.~{Schaye} {\em et~al.}, {The impact of
  angular momentum on black hole accretion rates in simulations of galaxy
  formation}, \emph{\mnras}. {\bf 454}, \penalty0 1038--1057  (Nov., 2015).
\newblock \doi{10.1093/mnras/stv2056}.

\bibitem{Rosas-Guevara2016}
Y.~{Rosas-Guevara}, R.~G. {Bower}, J.~{Schaye} {\em et~al.}, {Supermassive
  black holes in the EAGLE Universe. Revealing the observables of their
  growth}, \emph{\mnras}. {\bf 462}\penalty0 (1), \penalty0 190--205  (Oct,
  2016).
\newblock \doi{10.1093/mnras/stw1679}.

\bibitem{Genel2014}
S.~{Genel}, M.~{Vogelsberger}, V.~{Springel} {\em et~al.}, {Introducing the
  Illustris project: the evolution of galaxy populations across cosmic time},
  \emph{\mnras}. {\bf 445}\penalty0 (1), \penalty0 175--200  (Nov, 2014).
\newblock \doi{10.1093/mnras/stu1654}.

\bibitem{Vogelsberger2014}
M.~{Vogelsberger}, S.~{Genel}, V.~{Springel} {\em et~al.}, {Introducing the
  Illustris Project: simulating the coevolution of dark and visible matter in
  the Universe}, \emph{\mnras}. {\bf 444}, \penalty0 1518--1547  (Oct., 2014).
\newblock \doi{10.1093/mnras/stu1536}.

\bibitem{Sijacki2015}
D.~{Sijacki}, M.~{Vogelsberger}, S.~{Genel} {\em et~al.}, {The Illustris
  simulation: the evolving population of black holes across cosmic time},
  \emph{\mnras}. {\bf 452}, \penalty0 575--596  (Sept., 2015).
\newblock \doi{10.1093/mnras/stv1340}.

\bibitem{Khandai2015_MassiveBlack2}
N.~{Khandai}, T.~{Di Matteo}, R.~{Croft} {\em et~al.}, {The MassiveBlack-II
  simulation: the evolution of haloes and galaxies to z {\ensuremath{\sim}} 0},
  \emph{\mnras}. {\bf 450}\penalty0 (2), \penalty0 1349--1374  (June, 2015).
\newblock \doi{10.1093/mnras/stv627}.

\bibitem{DeGraf2015}
C.~{DeGraf}, T.~{Di Matteo}, T.~{Treu} {\em et~al.}, {Scaling relations between
  black holes and their host galaxies: comparing theoretical and observational
  measurements, and the impact of selection effects}, \emph{\mnras}. {\bf 454},
  \penalty0 913--932  (Nov., 2015).
\newblock \doi{10.1093/mnras/stv2002}.

\bibitem{Feng2016}
Y.~{Feng}, T.~{Di-Matteo}, R.~A. {Croft} {\em et~al.}, {The BlueTides
  simulation: first galaxies and reionization}, \emph{\mnras}. {\bf 455},
  \penalty0 2778--2791  (Jan., 2016).
\newblock \doi{10.1093/mnras/stv2484}.

\bibitem{DiMatteo17}
T.~{Di Matteo}, R.~A.~C. {Croft}, Y.~{Feng} {\em et~al.}, {The origin of the
  most massive black holes at high-z: BlueTides and the next quasar frontier},
  \emph{\mnras}. {\bf 467}\penalty0 (4), \penalty0 4243--4251  (June, 2017).
\newblock \doi{10.1093/mnras/stx319}.

\bibitem{Huang2018}
K.-W. {Huang}, T.~{Di Matteo}, A.~K. {Bhowmick} {\em et~al.}, {BLUETIDES
  simulation: establishing black hole-galaxy relations at high redshift},
  \emph{\mnras}. {\bf 478}, \penalty0 5063--5073  (Aug., 2018).
\newblock \doi{10.1093/mnras/sty1329}.

\bibitem{Ni2020_QSOobscuration}
Y.~{Ni}, T.~{Di Matteo}, R.~{Gilli} {\em et~al.}, {QSO obscuration at high
  redshift (z {\ensuremath{\gtrsim}} 7): predictions from the BLUETIDES
  simulation}, \emph{\mnras}. {\bf 495}\penalty0 (2), \penalty0 2135--2151
  (June, 2020).
\newblock \doi{10.1093/mnras/staa1313}.

\bibitem{Tremmel2017_Romulus}
M.~{Tremmel}, M.~{Karcher}, F.~{Governato} {\em et~al.}, {The Romulus
  cosmological simulations: a physical approach to the formation, dynamics and
  accretion models of SMBHs}, \emph{\mnras}. {\bf 470}, \penalty0 1121--1139
  (Sept., 2017).
\newblock \doi{10.1093/mnras/stx1160}.

\bibitem{Ricarte2019}
A.~{Ricarte}, M.~{Tremmel}, P.~{Natarajan} {\em et~al.}, {Tracing black hole
  and galaxy co-evolution in the ROMULUS simulations}, \emph{\mnras}. {\bf
  489}\penalty0 (1), \penalty0 802--819  (Oct, 2019).
\newblock \doi{10.1093/mnras/stz2161}.

\bibitem{Sharma2020}
R.~S. {Sharma}, A.~M. {Brooks}, R.~S. {Somerville} {\em et~al.}, {Black Hole
  Growth and Feedback in Isolated ROMULUS25 Dwarf Galaxies}, \emph{\apj}.
  897\penalty0 (1):\penalty0 103  (July, 2020).
\newblock \doi{10.3847/1538-4357/ab960e}.

\bibitem{Weinberger2017}
R.~{Weinberger}, V.~{Springel}, L.~{Hernquist} {\em et~al.}, {Simulating galaxy
  formation with black hole driven thermal and kinetic feedback},
  \emph{\mnras}. {\bf 465}, \penalty0 3291--3308  (Mar., 2017).
\newblock \doi{10.1093/mnras/stw2944}.

\bibitem{Pillepich2018}
A.~{Pillepich}, V.~{Springel}, D.~{Nelson} {\em et~al.}, {Simulating galaxy
  formation with the IllustrisTNG model}, \emph{\mnras}. {\bf 473}\penalty0
  (3), \penalty0 4077--4106  (Jan, 2018).
\newblock \doi{10.1093/mnras/stx2656}.

\bibitem{Habouzit2019}
M.~{Habouzit}, S.~{Genel}, R.~S. {Somerville} {\em et~al.}, {Linking galaxy
  structural properties and star formation activity to black hole activity with
  IllustrisTNG}, \emph{\mnras}. {\bf 484}\penalty0 (4), \penalty0 4413--4443
  (Apr, 2019).
\newblock \doi{10.1093/mnras/stz102}.

\bibitem{Terrazas2020}
B.~A. {Terrazas}, E.~F. {Bell}, A.~{Pillepich} {\em et~al.}, {The relationship
  between black hole mass and galaxy properties: examining the black hole
  feedback model in IllustrisTNG}, \emph{\mnras}. {\bf 493}\penalty0 (2),
  \penalty0 1888--1906  (Apr., 2020).
\newblock \doi{10.1093/mnras/staa374}.

\bibitem{Dave2019_Simba}
R.~{Dav{\'e}}, D.~{Angl{\'e}s-Alc{\'a}zar}, D.~{Narayanan} {\em et~al.},
  {SIMBA: Cosmological simulations with black hole growth and feedback},
  \emph{\mnras}. {\bf 486}\penalty0 (2), \penalty0 2827--2849  (Jun, 2019).
\newblock \doi{10.1093/mnras/stz937}.

\bibitem{Thomas2019}
N.~{Thomas}, R.~{Dav{\'e}}, D.~{Angl{\'e}s-Alc{\'a}zar} {\em et~al.}, {Black
  hole - Galaxy correlations in SIMBA}, \emph{\mnras}. {\bf 487}\penalty0 (4),
  \penalty0 5764--5780  (Aug, 2019).
\newblock \doi{10.1093/mnras/stz1703}.

\bibitem{Thomas2021}
N.~{Thomas}, R.~{Dav{\'e}}, M.~J. {Jarvis} {\em et~al.}, {The Radio Galaxy
  Population in the SIMBA Simulations}, \emph{\mnras}  (Mar., 2021).
\newblock \doi{10.1093/mnras/stab654}.

\bibitem{Borrow2020}
J.~{Borrow}, D.~{Angl{\'e}s-Alc{\'a}zar} and R.~{Dav{\'e}}, {Cosmological
  baryon transfer in the SIMBA simulations}, \emph{\mnras}. {\bf 491}\penalty0
  (4), \penalty0 6102--6119  (Feb., 2020).
\newblock \doi{10.1093/mnras/stz3428}.

\bibitem{Chen2021_Astrid}
N.~{Chen}, Y.~{Ni}, A.~M. {Holgado} {\em et~al.}, {Massive Black Hole Mergers
  with Orbital Information: Predictions from the ASTRID Simulation},
  \emph{arXiv e-prints}. art. arXiv:2112.08555  (Dec., 2021).

\bibitem{Ni2021_astrid}
Y.~{Ni}, T.~{Di Matteo}, S.~{Bird} {\em et~al.}, {The ASTRID simulation: the
  evolution of Supermassive Black Holes}, \emph{arXiv e-prints}. art.
  arXiv:2110.14154  (Oct., 2021).

\bibitem{Bird2022_Astrid}
S.~{Bird}, Y.~{Ni}, T.~D. {Matteo} {\em et~al.}, {The ASTRID Simulation: Galaxy
  Formation and Reionization}, \emph{\mnras}  (Mar., 2022).
\newblock \doi{10.1093/mnras/stac648}.

\bibitem{Villaescusa-Navarro2021_CAMELS}
F.~{Villaescusa-Navarro}, D.~{Angl{\'e}s-Alc{\'a}zar}, S.~{Genel} {\em et~al.},
  {The CAMELS Project: Cosmology and Astrophysics with Machine-learning
  Simulations}, \emph{\apj}. 915\penalty0 (1):\penalty0 71  (July, 2021).
\newblock \doi{10.3847/1538-4357/abf7ba}.

\bibitem{Villaescusa-Navarro2022_CAMELSpublic}
F.~{Villaescusa-Navarro}, S.~{Genel}, D.~{Angl{\'e}s-Alc{\'a}zar} {\em et~al.},
  {The CAMELS project: public data release}, \emph{arXiv e-prints}. art.
  arXiv:2201.01300  (Jan., 2022).

\bibitem{Sawala2016}
T.~{Sawala}, C.~S. {Frenk}, A.~{Fattahi} {\em et~al.}, {The APOSTLE
  simulations: solutions to the Local Group's cosmic puzzles}, \emph{\mnras}.
  {\bf 457}\penalty0 (2), \penalty0 1931--1943  (Apr., 2016).
\newblock \doi{10.1093/mnras/stw145}.

\bibitem{Grand2017_Auriga}
R.~J.~J. {Grand}, F.~A. {G{\'o}mez}, F.~{Marinacci} {\em et~al.}, {The Auriga
  Project: the properties and formation mechanisms of disc galaxies across
  cosmic time}, \emph{\mnras}. {\bf 467}\penalty0 (1), \penalty0 179--207
  (May, 2017).
\newblock \doi{10.1093/mnras/stx071}.

\bibitem{Wang2015_nihao}
L.~{Wang}, A.~A. {Dutton}, G.~S. {Stinson} {\em et~al.}, {NIHAO project - I.
  Reproducing the inefficiency of galaxy formation across cosmic time with a
  large sample of cosmological hydrodynamical simulations}, \emph{\mnras}. {\bf
  454}\penalty0 (1), \penalty0 83--94  (Nov., 2015).
\newblock \doi{10.1093/mnras/stv1937}.

\bibitem{Blank2019}
M.~{Blank}, A.~V. {Macci{\`o}}, A.~A. {Dutton} {\em et~al.}, {NIHAO - XXII.
  Introducing black hole formation, accretion, and feedback into the NIHAO
  simulation suite}, \emph{\mnras}. {\bf 487}\penalty0 (4), \penalty0
  5476--5489  (Aug, 2019).
\newblock \doi{10.1093/mnras/stz1688}.

\bibitem{Bahe2017}
Y.~M. {Bah{\'e}}, D.~J. {Barnes}, C.~{Dalla Vecchia} {\em et~al.}, {The
  Hydrangea simulations: galaxy formation in and around massive clusters},
  \emph{\mnras}. {\bf 470}\penalty0 (4), \penalty0 4186--4208  (Oct., 2017).
\newblock \doi{10.1093/mnras/stx1403}.

\bibitem{Barnes2017}
D.~J. {Barnes}, S.~T. {Kay}, Y.~M. {Bah{\'e}} {\em et~al.}, {The Cluster-EAGLE
  project: global properties of simulated clusters with resolved galaxies},
  \emph{\mnras}. {\bf 471}\penalty0 (1), \penalty0 1088--1106  (Oct., 2017).
\newblock \doi{10.1093/mnras/stx1647}.

\bibitem{Weiguang2018_The300}
W.~{Cui}, A.~{Knebe}, G.~{Yepes} {\em et~al.}, {The Three Hundred project: a
  large catalogue of theoretically modelled galaxy clusters for cosmological
  and astrophysical applications}, \emph{\mnras}. {\bf 480}\penalty0 (3),
  \penalty0 2898--2915  (Nov., 2018).
\newblock \doi{10.1093/mnras/sty2111}.

\bibitem{Weiguang2022_The300}
W.~{Cui}, R.~{Dave}, A.~{Knebe} {\em et~al.}, {The Three Hundred project: The
  Gizmo-Simba runs}, \emph{arXiv e-prints}. art. arXiv:2202.14038  (Feb.,
  2022).

\bibitem{Choi2015_CosmoSim}
E.~{Choi}, J.~P. {Ostriker}, T.~{Naab} {\em et~al.}, {The impact of mechanical
  AGN feedback on the formation of massive early-type galaxies}, \emph{\mnras}.
  {\bf 449}, \penalty0 4105--4116  (June, 2015).
\newblock \doi{10.1093/mnras/stv575}.

\bibitem{Choi2018_AGNsizes}
E.~{Choi}, R.~S. {Somerville}, J.~P. {Ostriker} {\em et~al.}, {The Role of
  Black Hole Feedback on Size and Structural Evolution in Massive Galaxies},
  \emph{\apj}. 866\penalty0 (2):\penalty0 91  (Oct, 2018).
\newblock \doi{10.3847/1538-4357/aae076}.

\bibitem{Choi2020_AGNmetals}
E.~{Choi}, R.~{Brennan}, R.~S. {Somerville} {\em et~al.}, {The Impact of
  Outflows Driven by Active Galactic Nuclei on Metals in and around Galaxies},
  \emph{\apj}. 904\penalty0 (1):\penalty0 8  (Nov., 2020).
\newblock \doi{10.3847/1538-4357/abba7d}.

\bibitem{Costa2014_QSOenvironment}
T.~{Costa}, D.~{Sijacki}, M.~{Trenti} {\em et~al.}, {The environment of bright
  QSOs at z {\ensuremath{\sim}} 6: star-forming galaxies and X-ray emission},
  \emph{\mnras}. {\bf 439}\penalty0 (2), \penalty0 2146--2174  (Apr., 2014).
\newblock \doi{10.1093/mnras/stu101}.

\bibitem{Costa2015}
T.~{Costa}, D.~{Sijacki} and M.~G. {Haehnelt}, {Fast cold gas in hot AGN
  outflows.}, \emph{\mnras}. {\bf 448}, \penalty0 L30--L34  (Mar., 2015).
\newblock \doi{10.1093/mnrasl/slu193}.

\bibitem{Costa2018b}
T.~{Costa}, J.~{Rosdahl}, D.~{Sijacki} {\em et~al.}, {Quenching star formation
  with quasar outflows launched by trapped IR radiation}, \emph{\mnras}. {\bf
  479}\penalty0 (2), \penalty0 2079--2111  (Sept., 2018).
\newblock \doi{10.1093/mnras/sty1514}.

\bibitem{Bellovary2019}
J.~M. {Bellovary}, C.~E. {Cleary}, F.~{Munshi} {\em et~al.}, {Multimessenger
  signatures of massive black holes in dwarf galaxies}, \emph{\mnras}. {\bf
  482}\penalty0 (3), \penalty0 2913--2923  (Jan., 2019).
\newblock \doi{10.1093/mnras/sty2842}.

\bibitem{Bellovary2021}
J.~M. {Bellovary}, S.~{Hayoune}, K.~{Chafla} {\em et~al.}, {The origins of
  off-centre massive black holes in dwarf galaxies}, \emph{\mnras}. {\bf
  505}\penalty0 (4), \penalty0 5129--5141  (Aug., 2021).
\newblock \doi{10.1093/mnras/stab1665}.

\bibitem{Applebaum2021}
E.~{Applebaum}, A.~M. {Brooks}, C.~R. {Christensen} {\em et~al.}, {Ultrafaint
  Dwarfs in a Milky Way Context: Introducing the Mint Condition DC Justice
  League Simulations}, \emph{\apj}. 906\penalty0 (2):\penalty0 96  (Jan.,
  2021).
\newblock \doi{10.3847/1538-4357/abcafa}.

\bibitem{Volonteri2020}
M.~{Volonteri}, H.~{Pfister}, R.~S. {Beckmann} {\em et~al.}, {Black hole
  mergers from dwarf to massive galaxies with the NewHorizon and Horizon-AGN
  simulations}, \emph{\mnras}. {\bf 498}\penalty0 (2), \penalty0 2219--2238
  (Oct., 2020).
\newblock \doi{10.1093/mnras/staa2384}.

\bibitem{Dubois2021_newhorizon}
Y.~{Dubois}, R.~{Beckmann}, F.~{Bournaud} {\em et~al.}, {Introducing the
  NEWHORIZON simulation: Galaxy properties with resolved internal dynamics
  across cosmic time}, \emph{\aap}. 651:\penalty0 A109  (July, 2021).
\newblock \doi{10.1051/0004-6361/202039429}.

\bibitem{Angles-Alcazar2017_BHsOnFIRE}
D.~{Angl{\'e}s-Alc{\'a}zar}, C.-A. {Faucher-Gigu{\`e}re}, E.~{Quataert} {\em
  et~al.}, {Black holes on FIRE: stellar feedback limits early feeding of
  galactic nuclei}, \emph{\mnras}. {\bf 472}, \penalty0 L109--L114  (Nov.,
  2017).
\newblock \doi{10.1093/mnrasl/slx161}.

\bibitem{Hopkins2018_FIRE2methods}
P.~F. {Hopkins}, A.~{Wetzel}, D.~{Kere{\v{s}}} {\em et~al.}, {FIRE-2
  simulations: physics versus numerics in galaxy formation}, \emph{\mnras}.
  {\bf 480}, \penalty0 800--863  (Oct., 2018).
\newblock \doi{10.1093/mnras/sty1690}.

\bibitem{Angles-Alcazar2021}
D.~{Angl{\'e}s-Alc{\'a}zar}, E.~{Quataert}, P.~F. {Hopkins} {\em et~al.},
  {Cosmological Simulations of Quasar Fueling to Subparsec Scales Using
  Lagrangian Hyper-refinement}, \emph{\apj}. 917\penalty0 (2):\penalty0 53
  (Aug., 2021).
\newblock \doi{10.3847/1538-4357/ac09e8}.

\bibitem{Catmabacak2022}
O.~{{\c{C}}atmabacak}, R.~{Feldmann}, D.~{Angl{\'e}s-Alc{\'a}zar} {\em et~al.},
  {Black hole-galaxy scaling relations in FIRE: the importance of black hole
  location and mergers}, \emph{\mnras}. {\bf 511}\penalty0 (1), \penalty0
  506--535  (Mar., 2022).
\newblock \doi{10.1093/mnras/stac040}.

\bibitem{Hopkins2022_FIRE3}
P.~F. {Hopkins}, A.~{Wetzel}, C.~{Wheeler} {\em et~al.}, {FIRE-3: Updated
  Stellar Evolution Models, Yields, \& Microphysics and Fitting Functions for
  Applications in Galaxy Simulations}, \emph{arXiv e-prints}. art.
  arXiv:2203.00040  (Feb., 2022).

\bibitem{Wellons2022}
S.~{Wellons}, C.-A. {Faucher-Gigu{\`e}re}, P.~F. {Hopkins} {\em et~al.},
  {Exploring supermassive black hole physics and galaxy quenching across halo
  mass in FIRE cosmological zoom simulations}, \emph{arXiv e-prints}. art.
  arXiv:2203.06201  (Mar., 2022).

\bibitem{Volonteri2010}
M.~{Volonteri}, {Formation of supermassive black holes}, \emph{\aapr}. {\bf
  18}, \penalty0 279--315  (July, 2010).
\newblock \doi{10.1007/s00159-010-0029-x}.

\bibitem{Inayoshi2020_ARAA}
K.~{Inayoshi}, E.~{Visbal} and Z.~{Haiman}, {The Assembly of the First Massive
  Black Holes}, \emph{\araa}. {\bf 58}, \penalty0 27--97  (Aug., 2020).
\newblock \doi{10.1146/annurev-astro-120419-014455}.

\bibitem{Madau2001}
P.~{Madau} and M.~J. {Rees}, {Massive Black Holes as Population III Remnants},
  \emph{\apjl}. {\bf 551}, \penalty0 L27--L30  (Apr., 2001).
\newblock \doi{10.1086/319848}.

\bibitem{Volonteri2003}
M.~{Volonteri}, F.~{Haardt} and P.~{Madau}, {The Assembly and Merging History
  of Supermassive Black Holes in Hierarchical Models of Galaxy Formation},
  \emph{\apj}. {\bf 582}\penalty0 (2), \penalty0 559--573  (Jan., 2003).
\newblock \doi{10.1086/344675}.

\bibitem{Begelman2006}
M.~C. {Begelman}, M.~{Volonteri} and M.~J. {Rees}, {Formation of supermassive
  black holes by direct collapse in pre-galactic haloes}, \emph{\mnras}. {\bf
  370}, \penalty0 289--298  (July, 2006).
\newblock \doi{10.1111/j.1365-2966.2006.10467.x}.

\bibitem{LodatoNatarajan2006}
G.~{Lodato} and P.~{Natarajan}, {Supermassive black hole formation during the
  assembly of pre-galactic discs}, \emph{\mnras}. {\bf 371}\penalty0 (4),
  \penalty0 1813--1823  (Oct., 2006).
\newblock \doi{10.1111/j.1365-2966.2006.10801.x}.

\bibitem{LodatoNatarajan2007}
G.~{Lodato} and P.~{Natarajan}, {The mass function of high-redshift seed black
  holes}, \emph{\mnras}. {\bf 377}\penalty0 (1), \penalty0 L64--L68  (May,
  2007).
\newblock \doi{10.1111/j.1745-3933.2007.00304.x}.

\bibitem{DiMatteo2008}
T.~{Di Matteo}, J.~{Colberg}, V.~{Springel} {\em et~al.}, {Direct Cosmological
  Simulations of the Growth of Black Holes and Galaxies}, \emph{\apj}. {\bf
  676}, \penalty0 33--53  (Mar., 2008).
\newblock \doi{10.1086/524921}.

\bibitem{Booth2009}
C.~M. {Booth} and J.~{Schaye}, {Cosmological simulations of the growth of
  supermassive black holes and feedback from active galactic nuclei: method and
  tests}, \emph{\mnras}. {\bf 398}, \penalty0 53--74  (Sept., 2009).
\newblock \doi{10.1111/j.1365-2966.2009.15043.x}.

\bibitem{Angles-Alcazar2017_BHfeedback}
D.~{Angl{\'e}s-Alc{\'a}zar}, R.~{Dav{\'e}}, C.-A. {Faucher-Gigu{\`e}re} {\em
  et~al.}, {Gravitational torque-driven black hole growth and feedback in
  cosmological simulations}, \emph{\mnras}. {\bf 464}, \penalty0 2840--2853
  (Jan., 2017).
\newblock \doi{10.1093/mnras/stw2565}.

\bibitem{Bellovary2010}
J.~M. {Bellovary}, F.~{Governato}, T.~R. {Quinn} {\em et~al.}, {Wandering Black
  Holes in Bright Disk Galaxy Halos}, \emph{\apjl}. {\bf 721}, \penalty0
  L148--L152  (Oct., 2010).
\newblock \doi{10.1088/2041-8205/721/2/L148}.

\bibitem{Bellovary2011}
J.~{Bellovary}, M.~{Volonteri}, F.~{Governato} {\em et~al.}, {The First Massive
  Black Hole Seeds and Their Hosts}, \emph{\apj}. 742\penalty0 (1):\penalty0 13
   (Nov., 2011).
\newblock \doi{10.1088/0004-637X/742/1/13}.

\bibitem{TaylorKobayashi2014}
P.~{Taylor} and C.~{Kobayashi}, {Seeding black holes in cosmological
  simulations}, \emph{\mnras}. {\bf 442}\penalty0 (3), \penalty0 2751--2767
  (Aug., 2014).
\newblock \doi{10.1093/mnras/stu983}.

\bibitem{Habouzit2017}
M.~{Habouzit}, M.~{Volonteri} and Y.~{Dubois}, {Blossoms from black hole seeds:
  properties and early growth regulated by supernova feedback}, \emph{\mnras}.
  {\bf 468}, \penalty0 3935--3948  (July, 2017).
\newblock \doi{10.1093/mnras/stx666}.

\bibitem{MaLinhao2021}
L.~{Ma}, P.~F. {Hopkins}, X.~{Ma} {\em et~al.}, {Seeds Don't Sink: Even Massive
  Black Hole ``Seeds'' Cannot Migrate to Galaxy Centers Efficiently},
  \emph{arXiv e-prints}. art. arXiv:2101.02727  (Jan., 2021).

\bibitem{Springel2005_BHmodel}
V.~{Springel}, T.~{Di Matteo} and L.~{Hernquist}, {Modelling feedback from
  stars and black holes in galaxy mergers}, \emph{\mnras}. {\bf 361}, \penalty0
  776--794  (Aug., 2005).
\newblock \doi{10.1111/j.1365-2966.2005.09238.x}.

\bibitem{Woosley2002}
S.~E. {Woosley}, A.~{Heger} and T.~A. {Weaver}, {The evolution and explosion of
  massive stars}, \emph{Reviews of Modern Physics}. {\bf 74}\penalty0 (4),
  \penalty0 1015--1071  (Nov., 2002).
\newblock \doi{10.1103/RevModPhys.74.1015}.

\bibitem{Hirano2014}
S.~{Hirano}, T.~{Hosokawa}, N.~{Yoshida} {\em et~al.}, {One Hundred First
  Stars: Protostellar Evolution and the Final Masses}, \emph{\apj}.
  781\penalty0 (2):\penalty0 60  (Feb., 2014).
\newblock \doi{10.1088/0004-637X/781/2/60}.

\bibitem{VolonteriNatarajan2009}
M.~{Volonteri} and P.~{Natarajan}, {Journey to the
  M$_{BH}$-{\ensuremath{\sigma}} relation: the fate of low-mass black holes in
  the Universe}, \emph{\mnras}. {\bf 400}\penalty0 (4), \penalty0 1911--1918
  (Dec., 2009).
\newblock \doi{10.1111/j.1365-2966.2009.15577.x}.

\bibitem{vanWassenhove2010}
S.~{van Wassenhove}, M.~{Volonteri}, M.~G. {Walker} {\em et~al.}, {Massive
  black holes lurking in Milky Way satellites}, \emph{\mnras}. {\bf
  408}\penalty0 (2), \penalty0 1139--1146  (Oct., 2010).
\newblock \doi{10.1111/j.1365-2966.2010.17189.x}.

\bibitem{Bonoli2016}
S.~{Bonoli}, L.~{Mayer}, S.~{Kazantzidis} {\em et~al.}, {Black hole starvation
  and bulge evolution in a Milky Way-like galaxy}, \emph{\mnras}. {\bf 459},
  \penalty0 2603--2617  (July, 2016).
\newblock \doi{10.1093/mnras/stw694}.

\bibitem{Angles-Alcazar2013}
D.~{Angl{\'e}s-Alc{\'a}zar}, F.~{{\"O}zel} and R.~{Dav{\'e}}, {Black
  Hole-Galaxy Correlations without Self-regulation}, \emph{\apj}. 770:\penalty0
  5  (June, 2013).
\newblock \doi{10.1088/0004-637X/770/1/5}.

\bibitem{Angles-Alcazar2015}
D.~{Angl{\'e}s-Alc{\'a}zar}, F.~{{\"O}zel}, R.~{Dav{\'e}} {\em et~al.},
  {Torque-limited Growth of Massive Black Holes in Galaxies across Cosmic
  Time}, \emph{\apj}. 800:\penalty0 127  (Feb., 2015).
\newblock \doi{10.1088/0004-637X/800/2/127}.

\bibitem{Habouzit2016}
M.~{Habouzit}, M.~{Volonteri}, M.~{Latif} {\em et~al.}, {On the number density
  of `direct collapse' black hole seeds}, \emph{\mnras}. {\bf 463}\penalty0
  (1), \penalty0 529--540  (Nov., 2016).
\newblock \doi{10.1093/mnras/stw1924}.

\bibitem{DeGraf2020}
C.~{DeGraf} and D.~{Sijacki}, {Cosmological simulations of massive black hole
  seeds: predictions for next-generation electromagnetic and gravitational wave
  observations}, \emph{\mnras}. {\bf 491}\penalty0 (4), \penalty0 4973--4992
  (Feb., 2020).
\newblock \doi{10.1093/mnras/stz3309}.

\bibitem{Chandrasekhar1943}
S.~{Chandrasekhar}, {Dynamical Friction. I. General Considerations: the
  Coefficient of Dynamical Friction.}, \emph{\apj}. {\bf 97}, \penalty0 255
  (Mar., 1943).
\newblock \doi{10.1086/144517}.

\bibitem{Ostriker1999}
E.~C. {Ostriker}, {Dynamical Friction in a Gaseous Medium}, \emph{\apj}. {\bf
  513}\penalty0 (1), \penalty0 252--258  (Mar., 1999).
\newblock \doi{10.1086/306858}.

\bibitem{Beckmann2018}
R.~S. {Beckmann}, A.~{Slyz} and J.~{Devriendt}, {Bondi or not Bondi: the impact
  of resolution on accretion and drag force modelling for supermassive black
  holes}, \emph{\mnras}. {\bf 478}\penalty0 (1), \penalty0 995--1016  (Jul,
  2018).
\newblock \doi{10.1093/mnras/sty931}.

\bibitem{Park2017}
K.~{Park} and T.~{Bogdanovi{\'c}}, {Gaseous Dynamical Friction in Presence of
  Black Hole Radiative Feedback}, \emph{\apj}. 838\penalty0 (2):\penalty0 103
  (Apr., 2017).
\newblock \doi{10.3847/1538-4357/aa65ce}.

\bibitem{delValle2018}
L.~{del Valle} and M.~{Volonteri}, {The effect of AGN feedback on the migration
  time-scale of supermassive black holes binaries}, \emph{\mnras}. {\bf
  480}\penalty0 (1), \penalty0 439--450  (Oct., 2018).
\newblock \doi{10.1093/mnras/sty1815}.

\bibitem{Gruzinov2020}
A.~{Gruzinov}, Y.~{Levin} and C.~D. {Matzner}, {Negative dynamical friction on
  compact objects moving through dense gas}, \emph{\mnras}. {\bf 492}\penalty0
  (2), \penalty0 2755--2761  (Feb., 2020).
\newblock \doi{10.1093/mnras/staa013}.

\bibitem{Toyouchi2020}
D.~{Toyouchi}, T.~{Hosokawa}, K.~{Sugimura} {\em et~al.}, {Gaseous dynamical
  friction under radiative feedback: do intermediate-mass black holes speed up
  or down?}, \emph{\mnras}. {\bf 496}\penalty0 (2), \penalty0 1909--1921
  (Aug., 2020).
\newblock \doi{10.1093/mnras/staa1338}.

\bibitem{Bogdanovic2022_LRR}
T.~{Bogdanovi{\'c}}, M.~C. {Miller} and L.~{Blecha}, {Electromagnetic
  counterparts to massive black-hole mergers}, \emph{Living Reviews in
  Relativity}. 25\penalty0 (1):\penalty0 3  (Dec., 2022).
\newblock \doi{10.1007/s41114-022-00037-8}.

\bibitem{Pfister2017}
H.~{Pfister}, A.~{Lupi}, P.~R. {Capelo} {\em et~al.}, {The birth of a
  supermassive black hole binary}, \emph{\mnras}. {\bf 471}\penalty0 (3),
  \penalty0 3646--3656  (Nov., 2017).
\newblock \doi{10.1093/mnras/stx1853}.

\bibitem{Okamoto2008}
T.~{Okamoto}, R.~S. {Nemmen} and R.~G. {Bower}, {The impact of radio feedback
  from active galactic nuclei in cosmological simulations: formation of disc
  galaxies}, \emph{\mnras}. {\bf 385}\penalty0 (1), \penalty0 161--180  (Mar.,
  2008).
\newblock \doi{10.1111/j.1365-2966.2008.12883.x}.

\bibitem{Wurster2013}
J.~{Wurster} and R.~J. {Thacker}, {A comparative study of AGN feedback
  algorithms}, \emph{\mnras}. {\bf 431}, \penalty0 2513--2534  (May, 2013).
\newblock \doi{10.1093/mnras/stt346}.

\bibitem{Bahe2021}
Y.~M. {Bah{\'e}}, J.~{Schaye}, M.~{Schaller} {\em et~al.}, {The importance of
  black hole repositioning for galaxy formation simulations}, \emph{arXiv
  e-prints}. art. arXiv:2109.01489  (Sept., 2021).

\bibitem{Chen2022_dynfric}
N.~{Chen}, Y.~{Ni}, M.~{Tremmel} {\em et~al.}, {Dynamical friction modelling of
  massive black holes in cosmological simulations and effects on merger rate
  predictions}, \emph{\mnras}. {\bf 510}\penalty0 (1), \penalty0 531--550
  (Feb., 2022).
\newblock \doi{10.1093/mnras/stab3411}.

\bibitem{Biernacki2017}
P.~{Biernacki}, R.~{Teyssier} and A.~{Bleuler}, {On the dynamics of
  supermassive black holes in gas-rich, star-forming galaxies: the case for
  nuclear star cluster co-evolution}, \emph{\mnras}. {\bf 469}, \penalty0
  295--313  (July, 2017).
\newblock \doi{10.1093/mnras/stx845}.

\bibitem{Tremmel2015_DynFriction}
M.~{Tremmel}, F.~{Governato}, M.~{Volonteri} {\em et~al.}, {Off the beaten
  path: a new approach to realistically model the orbital decay of supermassive
  black holes in galaxy formation simulations}, \emph{\mnras}. {\bf 451},
  \penalty0 1868--1874  (Aug., 2015).
\newblock \doi{10.1093/mnras/stv1060}.

\bibitem{Pfister2019}
H.~{Pfister}, M.~{Volonteri}, Y.~{Dubois} {\em et~al.}, {The erratic dynamical
  life of black hole seeds in high-redshift galaxies}, \emph{\mnras}. {\bf
  486}\penalty0 (1), \penalty0 101--111  (June, 2019).
\newblock \doi{10.1093/mnras/stz822}.

\bibitem{Dubois2015_SNa}
Y.~{Dubois}, M.~{Volonteri}, J.~{Silk} {\em et~al.}, {Black hole evolution - I.
  Supernova-regulated black hole growth}, \emph{\mnras}. {\bf 452}, \penalty0
  1502--1518  (Sept., 2015).
\newblock \doi{10.1093/mnras/stv1416}.

\bibitem{Trebitsch2020}
M.~{Trebitsch}, M.~{Volonteri} and Y.~{Dubois}, {Modelling a bright z = 6
  galaxy at the faint end of the AGN luminosity function}, \emph{\mnras}. {\bf
  494}\penalty0 (3), \penalty0 3453--3463  (May, 2020).
\newblock \doi{10.1093/mnras/staa1012}.

\bibitem{Fiacconi2013}
D.~{Fiacconi}, L.~{Mayer}, R.~{Ro{\v s}kar} {\em et~al.}, {Massive Black Hole
  Pairs in Clumpy, Self-gravitating Circumnuclear Disks: Stochastic Orbital
  Decay}, \emph{\apjl}. 777:\penalty0 L14  (Nov., 2013).
\newblock \doi{10.1088/2041-8205/777/1/L14}.

\bibitem{Roskar2015}
R.~{Ro{\v{s}}kar}, D.~{Fiacconi}, L.~{Mayer} {\em et~al.}, {Orbital decay of
  supermassive black hole binaries in clumpy multiphase merger remnants},
  \emph{\mnras}. {\bf 449}\penalty0 (1), \penalty0 494--505  (May, 2015).
\newblock \doi{10.1093/mnras/stv312}.

\bibitem{Tamburello2017}
V.~{Tamburello}, P.~R. {Capelo}, L.~{Mayer} {\em et~al.}, {Supermassive black
  hole pairs in clumpy galaxies at high redshift: delayed binary formation and
  concurrent mass growth}, \emph{\mnras}. {\bf 464}\penalty0 (3), \penalty0
  2952--2962  (Jan., 2017).
\newblock \doi{10.1093/mnras/stw2561}.

\bibitem{Tamfal2018}
T.~{Tamfal}, P.~R. {Capelo}, S.~{Kazantzidis} {\em et~al.}, {Formation of LISA
  Black Hole Binaries in Merging Dwarf Galaxies: The Imprint of Dark Matter},
  \emph{\apjl}. 864\penalty0 (1):\penalty0 L19  (Sept., 2018).
\newblock \doi{10.3847/2041-8213/aada4b}.

\bibitem{Tremmel2018}
M.~{Tremmel}, F.~{Governato}, M.~{Volonteri} {\em et~al.}, {Dancing to CHANGA:
  a self-consistent prediction for close SMBH pair formation time-scales
  following galaxy mergers}, \emph{\mnras}. {\bf 475}\penalty0 (4), \penalty0
  4967--4977  (Apr., 2018).
\newblock \doi{10.1093/mnras/sty139}.

\bibitem{Reines2020}
A.~E. {Reines}, J.~J. {Condon}, J.~{Darling} {\em et~al.}, {A New Sample of
  (Wandering) Massive Black Holes in Dwarf Galaxies from High-resolution Radio
  Observations}, \emph{\apj}. 888\penalty0 (1):\penalty0 36  (Jan., 2020).
\newblock \doi{10.3847/1538-4357/ab4999}.

\bibitem{Mezcua2020}
M.~{Mezcua} and H.~{Dom{\'\i}nguez S{\'a}nchez}, {Hidden AGNs in Dwarf Galaxies
  Revealed by MaNGA: Light Echoes, Off-nuclear Wanderers, and a New Broad-line
  AGN}, \emph{\apjl}. 898\penalty0 (2):\penalty0 L30  (Aug., 2020).
\newblock \doi{10.3847/2041-8213/aba199}.

\bibitem{Ricarte2021}
A.~{Ricarte}, M.~{Tremmel}, P.~{Natarajan} {\em et~al.}, {Origins and
  demographics of wandering black holes}, \emph{\mnras}. {\bf 503}\penalty0
  (4), \penalty0 6098--6111  (June, 2021).
\newblock \doi{10.1093/mnras/stab866}.

\bibitem{Sharma2022}
R.~S. {Sharma}, A.~M. {Brooks}, M.~{Tremmel} {\em et~al.}, {A hidden population
  of massive black holes in simulated dwarf galaxies}, \emph{arXiv e-prints}.
  art. arXiv:2203.05580  (Mar., 2022).

\bibitem{Blecha2011}
L.~{Blecha}, T.~J. {Cox}, A.~{Loeb} {\em et~al.}, {Recoiling black holes in
  merging galaxies: relationship to active galactic nucleus lifetimes,
  starbursts and the M$_{BH}$-{$\sigma$} relation}, \emph{\mnras}. {\bf 412},
  \penalty0 2154--2182  (Apr., 2011).
\newblock \doi{10.1111/j.1365-2966.2010.18042.x}.

\bibitem{Sijacki2011}
D.~{Sijacki}, V.~{Springel} and M.~G. {Haehnelt}, {Gravitational recoils of
  supermassive black holes in hydrodynamical simulations of gas-rich galaxies},
  \emph{\mnras}. {\bf 414}, \penalty0 3656--3670  (July, 2011).
\newblock \doi{10.1111/j.1365-2966.2011.18666.x}.

\bibitem{Blecha2016}
L.~{Blecha}, D.~{Sijacki}, L.~Z. {Kelley} {\em et~al.}, {Recoiling black holes:
  prospects for detection and implications of spin alignment}, \emph{\mnras}.
  {\bf 456}, \penalty0 961--989  (Feb., 2016).
\newblock \doi{10.1093/mnras/stv2646}.

\bibitem{Bondi1944}
H.~{Bondi} and F.~{Hoyle}, {On the mechanism of accretion by stars},
  \emph{\mnras}. {\bf 104}, \penalty0 273  (1944).

\bibitem{Bondi1952}
H.~{Bondi}, {On spherically symmetrical accretion}, \emph{\mnras}. {\bf 112},
  \penalty0 195  (1952).

\bibitem{DiMatteo2005}
T.~{Di Matteo}, V.~{Springel} and L.~{Hernquist}, {Energy input from quasars
  regulates the growth and activity of black holes and their host galaxies},
  \emph{\nat}. {\bf 433}, \penalty0 604--607  (Feb., 2005).
\newblock \doi{10.1038/nature03335}.

\bibitem{Pelupessy2007}
F.~I. {Pelupessy}, T.~{Di Matteo} and B.~{Ciardi}, {How Rapidly Do Supermassive
  Black Hole ``Seeds'' Grow at Early Times?}, \emph{\apj}. {\bf 665}\penalty0
  (1), \penalty0 107--119  (Aug., 2007).
\newblock \doi{10.1086/519235}.

\bibitem{Jiang2014}
Y.-F. {Jiang}, J.~M. {Stone} and S.~W. {Davis}, {A Global Three-dimensional
  Radiation Magneto-hydrodynamic Simulation of Super-Eddington Accretion
  Disks}, \emph{\apj}. 796:\penalty0 106  (Dec., 2014).
\newblock \doi{10.1088/0004-637X/796/2/106}.

\bibitem{Jiang2019}
Y.-F. {Jiang}, J.~M. {Stone} and S.~W. {Davis}, {Super-Eddington Accretion
  Disks around Supermassive Black Holes}, \emph{\apj}. 880\penalty0
  (2):\penalty0 67  (Aug., 2019).
\newblock \doi{10.3847/1538-4357/ab29ff}.

\bibitem{Lupi2016}
A.~{Lupi}, F.~{Haardt}, M.~{Dotti} {\em et~al.}, {Growing massive black holes
  through supercritical accretion of stellar-mass seeds}, \emph{\mnras}. {\bf
  456}\penalty0 (3), \penalty0 2993--3003  (Mar., 2016).
\newblock \doi{10.1093/mnras/stv2877}.

\bibitem{YuTremaine2002}
Q.~{Yu} and S.~{Tremaine}, {Observational constraints on growth of massive
  black holes}, \emph{\mnras}. {\bf 335}, \penalty0 965--976  (Oct., 2002).
\newblock \doi{10.1046/j.1365-8711.2002.05532.x}.

\bibitem{Kormendy2013}
J.~{Kormendy} and L.~C. {Ho}, {Coevolution (Or Not) of Supermassive Black Holes
  and Host Galaxies}, \emph{\araa}. {\bf 51}\penalty0 (1), \penalty0 511--653
  (Aug., 2013).
\newblock \doi{10.1146/annurev-astro-082708-101811}.

\bibitem{Graham2016}
A.~W. {Graham}, {Galaxy Bulges and Their Massive Black Holes: A Review},
  \emph{Galactic Bulges}. {\bf 418}, \penalty0 263  (2016).
\newblock \doi{10.1007/978-3-319-19378-6_11}.

\bibitem{Hopkins2010_MultiScale}
P.~F. {Hopkins} and E.~{Quataert}, {How do massive black holes get their gas?},
  \emph{\mnras}. {\bf 407}, \penalty0 1529--1564  (Sept., 2010).
\newblock \doi{10.1111/j.1365-2966.2010.17064.x}.

\bibitem{Hopkins2011_Analytic}
P.~F. {Hopkins} and E.~{Quataert}, {An analytic model of angular momentum
  transport by gravitational torques: from galaxies to massive black holes},
  \emph{\mnras}. {\bf 415}, \penalty0 1027--1050  (Aug., 2011).
\newblock \doi{10.1111/j.1365-2966.2011.18542.x}.

\bibitem{Hobbs2012}
A.~{Hobbs}, C.~{Power}, S.~{Nayakshin} {\em et~al.}, {Modelling supermassive
  black hole growth: towards an improved sub-grid prescription}, \emph{\mnras}.
  {\bf 421}\penalty0 (4), \penalty0 3443--3449  (Apr, 2012).
\newblock \doi{10.1111/j.1365-2966.2012.20563.x}.

\bibitem{Gaspari2013}
M.~{Gaspari}, M.~{Ruszkowski} and S.~P. {Oh}, {Chaotic cold accretion on to
  black holes}, \emph{\mnras}. {\bf 432}\penalty0 (4), \penalty0 3401--3422
  (July, 2013).
\newblock \doi{10.1093/mnras/stt692}.

\bibitem{Hopkins2016_NuclearSims}
P.~F. {Hopkins}, P.~{Torrey}, C.-A. {Faucher-Gigu{\`e}re} {\em et~al.},
  {Stellar and quasar feedback in concert: effects on AGN accretion,
  obscuration, and outflows}, \emph{\mnras}. {\bf 458}, \penalty0 816--831
  (May, 2016).
\newblock \doi{10.1093/mnras/stw289}.

\bibitem{Negri2017}
A.~{Negri} and M.~{Volonteri}, {Black hole feeding and feedback: the physics
  inside the `sub-grid'}, \emph{\mnras}. {\bf 467}, \penalty0 3475--3492  (May,
  2017).
\newblock \doi{10.1093/mnras/stx362}.

\bibitem{Habouzit2021_Mbh}
M.~{Habouzit}, Y.~{Li}, R.~S. {Somerville} {\em et~al.}, {Supermassive black
  holes in cosmological simulations I: M$_{BH}$ - M$_{{\ensuremath{\star}}}$
  relation and black hole mass function}, \emph{\mnras}. {\bf 503}\penalty0
  (2), \penalty0 1940--1975  (May, 2021).
\newblock \doi{10.1093/mnras/stab496}.

\bibitem{Beckmann2019}
R.~S. {Beckmann}, J.~{Devriendt} and A.~{Slyz}, {Zooming in on supermassive
  black holes: how resolving their gas cloud host renders their accretion
  episodic}, \emph{\mnras}. {\bf 483}\penalty0 (3), \penalty0 3488--3509  (Mar,
  2019).
\newblock \doi{10.1093/mnras/sty2890}.

\bibitem{Catmabacak2020}
O.~{{\c{C}}atmabacak}, R.~{Feldmann}, D.~{Angl{\'e}s-Alc{\'a}zar} {\em et~al.},
  {Black hole -- galaxy co-evolution in FIRE: the importance of black hole
  location and mergers}, \emph{arXiv e-prints}. art. arXiv:2007.12185  (July,
  2020).

\bibitem{Bournaud2011}
F.~{Bournaud}, A.~{Dekel}, R.~{Teyssier} {\em et~al.}, {Black Hole Growth and
  Active Galactic Nuclei Obscuration by Instability-driven Inflows in
  High-redshift Disk Galaxies Fed by Cold Streams}, \emph{\apjl}. 741:\penalty0
  L33  (Nov., 2011).
\newblock \doi{10.1088/2041-8205/741/2/L33}.

\bibitem{Hobbs2011}
A.~{Hobbs}, S.~{Nayakshin}, C.~{Power} {\em et~al.}, {Feeding supermassive
  black holes through supersonic turbulence and ballistic accretion},
  \emph{\mnras}. {\bf 413}\penalty0 (4), \penalty0 2633--2650  (June, 2011).
\newblock \doi{10.1111/j.1365-2966.2011.18333.x}.

\bibitem{Gabor2013}
J.~M. {Gabor} and F.~{Bournaud}, {Simulations of supermassive black hole growth
  in high-redshift disc galaxies}, \emph{\mnras}. {\bf 434}\penalty0 (1),
  \penalty0 606--620  (Sep, 2013).
\newblock \doi{10.1093/mnras/stt1046}.

\bibitem{Power2011}
C.~{Power}, S.~{Nayakshin} and A.~{King}, {The accretion disc particle method
  for simulations of black hole feeding and feedback}, \emph{\mnras}. {\bf
  412}\penalty0 (1), \penalty0 269--276  (Mar, 2011).
\newblock \doi{10.1111/j.1365-2966.2010.17901.x}.

\bibitem{Fiacconi2018}
D.~{Fiacconi}, D.~{Sijacki} and J.~E. {Pringle}, {Galactic nuclei evolution
  with spinning black holes: method and implementation}, \emph{\mnras}. {\bf
  477}\penalty0 (3), \penalty0 3807--3835  (Jul, 2018).
\newblock \doi{10.1093/mnras/sty893}.

\bibitem{Talbot2021}
R.~Y. {Talbot}, M.~A. {Bourne} and D.~{Sijacki}, {Blandford-Znajek jets in
  galaxy formation simulations: method and implementation}, \emph{\mnras}. {\bf
  504}\penalty0 (3), \penalty0 3619--3650  (July, 2021).
\newblock \doi{10.1093/mnras/stab804}.

\bibitem{Dubois2014_spin}
Y.~{Dubois}, M.~{Volonteri} and J.~{Silk}, {Black hole evolution - III.
  Statistical properties of mass growth and spin evolution using large-scale
  hydrodynamical cosmological simulations}, \emph{\mnras}. {\bf 440}\penalty0
  (2), \penalty0 1590--1606  (May, 2014).
\newblock \doi{10.1093/mnras/stu373}.

\bibitem{Bustamante2019}
S.~{Bustamante} and V.~{Springel}, {Spin evolution and feedback of supermassive
  black holes in cosmological simulations}, \emph{\mnras}. {\bf 490}\penalty0
  (3), \penalty0 4133--4153  (Dec., 2019).
\newblock \doi{10.1093/mnras/stz2836}.

\bibitem{Bardeen72}
J.~M. {Bardeen}, W.~H. {Press} and S.~A. {Teukolsky}, {Rotating Black Holes:
  Locally Nonrotating Frames, Energy Extraction, and Scalar Synchrotron
  Radiation}, \emph{\apj}. {\bf 178}, \penalty0 347--370  (Dec., 1972).
\newblock \doi{10.1086/151796}.

\bibitem{Fabian2012}
A.~C. {Fabian}, {Observational Evidence of Active Galactic Nuclei Feedback},
  \emph{\araa}. {\bf 50}, \penalty0 455--489  (Sept., 2012).
\newblock \doi{10.1146/annurev-astro-081811-125521}.

\bibitem{Silk1998}
J.~{Silk} and M.~J. {Rees}, {Quasars and galaxy formation}, \emph{\aap}. {\bf
  331}, \penalty0 L1--L4  (Mar., 1998).

\bibitem{Murray2005}
N.~{Murray}, E.~{Quataert} and T.~A. {Thompson}, {On the Maximum Luminosity of
  Galaxies and Their Central Black Holes: Feedback from Momentum-driven Winds},
  \emph{\apj}. {\bf 618}, \penalty0 569--585  (Jan., 2005).
\newblock \doi{10.1086/426067}.

\bibitem{Tombesi2013}
F.~{Tombesi}, M.~{Cappi}, J.~N. {Reeves} {\em et~al.}, {Unification of X-ray
  winds in Seyfert galaxies: from ultra-fast outflows to warm absorbers},
  \emph{\mnras}. {\bf 430}, \penalty0 1102--1117  (Apr., 2013).
\newblock \doi{10.1093/mnras/sts692}.

\bibitem{Nardini2015}
E.~{Nardini}, J.~N. {Reeves}, J.~{Gofford} {\em et~al.}, {Black hole feedback
  in the luminous quasar PDS 456}, \emph{Science}. {\bf 347}, \penalty0
  860--863  (Feb., 2015).
\newblock \doi{10.1126/science.1259202}.

\bibitem{Greene2012_QSOoutflow}
J.~E. {Greene}, N.~L. {Zakamska} and P.~S. {Smith}, {A Spectacular Outflow in
  an Obscured Quasar}, \emph{\apj}. 746:\penalty0 86  (Feb., 2012).
\newblock \doi{10.1088/0004-637X/746/1/86}.

\bibitem{Cicone2014}
C.~{Cicone}, R.~{Maiolino}, E.~{Sturm} {\em et~al.}, {Massive molecular
  outflows and evidence for AGN feedback from CO observations}, \emph{\aap}.
  562:\penalty0 A21  (Feb., 2014).
\newblock \doi{10.1051/0004-6361/201322464}.

\bibitem{Zakamska2014}
N.~L. {Zakamska} and J.~E. {Greene}, {Quasar feedback and the origin of radio
  emission in radio-quiet quasars}, \emph{\mnras}. {\bf 442}, \penalty0
  784--804  (July, 2014).
\newblock \doi{10.1093/mnras/stu842}.

\bibitem{Wylezalek2020}
D.~{Wylezalek}, A.~M. {Flores}, N.~L. {Zakamska} {\em et~al.}, {Ionized gas
  outflow signatures in SDSS-IV MaNGA active galactic nuclei}, \emph{\mnras}.
  {\bf 492}\penalty0 (4), \penalty0 4680--4696  (Mar., 2020).
\newblock \doi{10.1093/mnras/staa062}.

\bibitem{Hlavacek-Larrondo2012}
J.~{Hlavacek-Larrondo}, A.~C. {Fabian}, A.~C. {Edge} {\em et~al.}, {Extreme AGN
  feedback in the MAssive Cluster Survey: a detailed study of X-ray cavities at
  z\&gt;0.3}, \emph{\mnras}. {\bf 421}\penalty0 (2), \penalty0 1360--1384
  (Apr., 2012).
\newblock \doi{10.1111/j.1365-2966.2011.20405.x}.

\bibitem{Trainor2013}
R.~{Trainor} and C.~C. {Steidel}, {Constraints on Hyperluminous QSO Lifetimes
  via Fluorescent Ly{\ensuremath{\alpha}} Emitters at Z
  \raisebox{-0.5ex}\textasciitilde= 2.7}, \emph{\apjl}. 775\penalty0
  (1):\penalty0 L3  (Sept., 2013).
\newblock \doi{10.1088/2041-8205/775/1/L3}.

\bibitem{Eilers2017}
A.-C. {Eilers}, F.~B. {Davies}, J.~F. {Hennawi} {\em et~al.}, {Implications of
  z {\textasciitilde} 6 Quasar Proximity Zones for the Epoch of Reionization
  and Quasar Lifetimes}, \emph{\apj}. 840\penalty0 (1):\penalty0 24  (May,
  2017).
\newblock \doi{10.3847/1538-4357/aa6c60}.

\bibitem{Baldry2012}
I.~K. {Baldry}, S.~P. {Driver}, J.~{Loveday} {\em et~al.}, {Galaxy And Mass
  Assembly (GAMA): the galaxy stellar mass function at z < 0.06},
  \emph{\mnras}. {\bf 421}\penalty0 (1), \penalty0 621--634  (Mar., 2012).
\newblock \doi{10.1111/j.1365-2966.2012.20340.x}.

\bibitem{Bernardi2017}
M.~{Bernardi}, A.~{Meert}, R.~K. {Sheth} {\em et~al.}, {The high mass end of
  the stellar mass function: Dependence on stellar population models and
  agreement between fits to the light profile}, \emph{\mnras}. {\bf
  467}\penalty0 (2), \penalty0 2217--2233  (May, 2017).
\newblock \doi{10.1093/mnras/stx176}.

\bibitem{Baldry2006}
I.~K. {Baldry}, M.~L. {Balogh}, R.~G. {Bower} {\em et~al.}, {Galaxy bimodality
  versus stellar mass and environment}, \emph{\mnras}. {\bf 373}\penalty0 (2),
  \penalty0 469--483  (Dec., 2006).
\newblock \doi{10.1111/j.1365-2966.2006.11081.x}.

\bibitem{Brammer2009}
G.~B. {Brammer}, K.~E. {Whitaker}, P.~G. {van Dokkum} {\em et~al.}, {The Dead
  Sequence: A Clear Bimodality in Galaxy Colors from z = 0 to z = 2.5},
  \emph{\apjl}. {\bf 706}\penalty0 (1), \penalty0 L173--L177  (Nov., 2009).
\newblock \doi{10.1088/0004-637X/706/1/L173}.

\bibitem{Debuhr2010}
J.~{Debuhr}, E.~{Quataert}, C.-P. {Ma} {\em et~al.}, {Self-regulated black hole
  growth via momentum deposition in galaxy merger simulations}, \emph{\mnras}.
  {\bf 406}\penalty0 (1), \penalty0 L55--L59  (July, 2010).
\newblock \doi{10.1111/j.1745-3933.2010.00881.x}.

\bibitem{Debuhr2012}
J.~{Debuhr}, E.~{Quataert} and C.-P. {Ma}, {Galaxy-scale outflows driven by
  active galactic nuclei}, \emph{\mnras}. {\bf 420}, \penalty0 2221--2231
  (Mar., 2012).
\newblock \doi{10.1111/j.1365-2966.2011.20187.x}.

\bibitem{Costa2018a}
T.~{Costa}, J.~{Rosdahl}, D.~{Sijacki} {\em et~al.}, {Driving gas shells with
  radiation pressure on dust in radiation-hydrodynamic simulations},
  \emph{\mnras}. {\bf 473}\penalty0 (3), \penalty0 4197--4219  (Jan., 2018).
\newblock \doi{10.1093/mnras/stx2598}.

\bibitem{Choi2012_BHmodel}
E.~{Choi}, J.~P. {Ostriker}, T.~{Naab} {\em et~al.}, {Radiative and
  Momentum-based Mechanical Active Galactic Nucleus Feedback in a
  Three-dimensional Galaxy Evolution Code}, \emph{\apj}. 754:\penalty0 125
  (Aug., 2012).
\newblock \doi{10.1088/0004-637X/754/2/125}.

\bibitem{Richings2018_MolecularOutflow}
A.~J. {Richings} and C.-A. {Faucher-Gigu{\`e}re}, {The origin of fast molecular
  outflows in quasars: molecule formation in AGN-driven galactic winds},
  \emph{\mnras}. {\bf 474}\penalty0 (3), \penalty0 3673--3699  (Mar, 2018).
\newblock \doi{10.1093/mnras/stx3014}.

\bibitem{Torrey2020}
P.~{Torrey}, P.~F. {Hopkins}, C.-A. {Faucher-Gigu{\`e}re} {\em et~al.}, {The
  impact of AGN wind feedback in simulations of isolated galaxies with a
  multiphase ISM}, \emph{\mnras}. {\bf 497}\penalty0 (4), \penalty0 5292--5308
  (Oct., 2020).
\newblock \doi{10.1093/mnras/staa2222}.

\bibitem{Richings2021}
A.~J. {Richings}, C.-A. {Faucher-Gigu{\`e}re} and J.~{Stern}, {Unravelling the
  physics of multiphase AGN winds through emission line tracers},
  \emph{\mnras}. {\bf 503}\penalty0 (2), \penalty0 1568--1585  (May, 2021).
\newblock \doi{10.1093/mnras/stab556}.

\bibitem{Kim2011}
J.-h. {Kim}, J.~H. {Wise}, M.~A. {Alvarez} {\em et~al.}, {Galaxy Formation with
  Self-consistently Modeled Stars and Massive Black Holes. I.
  Feedback-regulated Star Formation and Black Hole Growth}, \emph{\apj}.
  738\penalty0 (1):\penalty0 54  (Sept., 2011).
\newblock \doi{10.1088/0004-637X/738/1/54}.

\bibitem{Gibson2009}
R.~R. {Gibson}, L.~{Jiang}, W.~N. {Brandt} {\em et~al.}, {A Catalog of Broad
  Absorption Line Quasars in Sloan Digital Sky Survey Data Release 5},
  \emph{\apj}. {\bf 692}\penalty0 (1), \penalty0 758--777  (Feb., 2009).
\newblock \doi{10.1088/0004-637X/692/1/758}.

\bibitem{Gofford2015}
J.~{Gofford}, J.~N. {Reeves}, D.~E. {McLaughlin} {\em et~al.}, {The Suzaku view
  of highly ionized outflows in AGN - II. Location, energetics and scalings
  with bolometric luminosity}, \emph{\mnras}. {\bf 451}\penalty0 (4), \penalty0
  4169--4182  (Aug., 2015).
\newblock \doi{10.1093/mnras/stv1207}.

\bibitem{Costa2020}
T.~{Costa}, R.~{Pakmor} and V.~{Springel}, {Powering galactic superwinds with
  small-scale AGN winds}, \emph{\mnras}. {\bf 497}\penalty0 (4), \penalty0
  5229--5255  (Oct., 2020).
\newblock \doi{10.1093/mnras/staa2321}.

\bibitem{Faucher-Giguere2012_WindModel}
C.-A. {Faucher-Gigu{\`e}re} and E.~{Quataert}, {The physics of galactic winds
  driven by active galactic nuclei}, \emph{\mnras}. {\bf 425}, \penalty0
  605--622  (Sept., 2012).
\newblock \doi{10.1111/j.1365-2966.2012.21512.x}.

\bibitem{Zubovas2012}
K.~{Zubovas} and A.~{King}, {Clearing Out a Galaxy}, \emph{\apjl}. 745\penalty0
  (2):\penalty0 L34  (Feb., 2012).
\newblock \doi{10.1088/2041-8205/745/2/L34}.

\bibitem{Heckman2014}
T.~M. {Heckman} and P.~N. {Best}, {The Coevolution of Galaxies and Supermassive
  Black Holes: Insights from Surveys of the Contemporary Universe},
  \emph{\araa}. {\bf 52}, \penalty0 589--660  (Aug., 2014).
\newblock \doi{10.1146/annurev-astro-081913-035722}.

\bibitem{Sijacki2007}
D.~{Sijacki}, V.~{Springel}, T.~{Di Matteo} {\em et~al.}, {A unified model for
  AGN feedback in cosmological simulations of structure formation},
  \emph{\mnras}. {\bf 380}, \penalty0 877--900  (Sept., 2007).
\newblock \doi{10.1111/j.1365-2966.2007.12153.x}.

\bibitem{Bourne2017}
M.~A. {Bourne} and D.~{Sijacki}, {AGN jet feedback on a moving mesh: cocoon
  inflation, gas flows and turbulence}, \emph{\mnras}. {\bf 472}\penalty0 (4),
  \penalty0 4707--4735  (Dec, 2017).
\newblock \doi{10.1093/mnras/stx2269}.

\bibitem{Weinberger2017_jet}
R.~{Weinberger}, K.~{Ehlert}, C.~{Pfrommer} {\em et~al.}, {Simulating the
  interaction of jets with the intracluster medium}, \emph{\mnras}. {\bf
  470}\penalty0 (4), \penalty0 4530--4546  (Oct., 2017).
\newblock \doi{10.1093/mnras/stx1409}.

\bibitem{Su2021_Jets}
K.-Y. {Su}, P.~F. {Hopkins}, G.~L. {Bryan} {\em et~al.}, {Which AGN jets quench
  star formation in massive galaxies?}, \emph{\mnras}. {\bf 507}\penalty0 (1),
  \penalty0 175--204  (Oct., 2021).
\newblock \doi{10.1093/mnras/stab2021}.

\bibitem{Cole00}
S.~{Cole}, C.~G. {Lacey}, C.~M. {Baugh} {\em et~al.}, {Hierarchical galaxy
  formation}, \emph{\mnras}. {\bf 319}\penalty0 (1), \penalty0 168--204  (Nov.,
  2000).
\newblock \doi{10.1046/j.1365-8711.2000.03879.x}.

\bibitem{Benson03}
A.~J. {Benson}, R.~G. {Bower}, C.~S. {Frenk} {\em et~al.}, {What Shapes the
  Luminosity Function of Galaxies?}, \emph{\apj}. {\bf 599}\penalty0 (1),
  \penalty0 38--49  (Dec., 2003).
\newblock \doi{10.1086/379160}.

\bibitem{Granato2004}
G.~L. {Granato}, G.~{De Zotti}, L.~{Silva} {\em et~al.}, {A Physical Model for
  the Coevolution of QSOs and Their Spheroidal Hosts}, \emph{\apj}. {\bf 600},
  \penalty0 580--594  (Jan., 2004).

\bibitem{Baugh06}
C.~M. {Baugh}, {A primer on hierarchical galaxy formation: the semi-analytical
  approach}, \emph{Reports on Progress in Physics}. {\bf 69}\penalty0 (12),
  \penalty0 3101--3156  (Dec., 2006).
\newblock \doi{10.1088/0034-4885/69/12/R02}.

\bibitem{Bower06}
R.~G. {Bower}, A.~J. {Benson}, R.~{Malbon} {\em et~al.}, {Breaking the
  hierarchy of galaxy formation}, \emph{\mnras}. {\bf 370}\penalty0 (2),
  \penalty0 645--655  (Aug., 2006).
\newblock \doi{10.1111/j.1365-2966.2006.10519.x}.

\bibitem{Cattaneo06}
A.~{Cattaneo}, A.~{Dekel}, J.~{Devriendt} {\em et~al.}, {Modelling the galaxy
  bimodality: shutdown above a critical halo mass}, \emph{\mnras}. {\bf
  370}\penalty0 (4), \penalty0 1651--1665  (Aug., 2006).
\newblock \doi{10.1111/j.1365-2966.2006.10608.x}.

\bibitem{Croton2006}
D.~J. {Croton}, V.~{Springel}, S.~D.~M. {White} {\em et~al.}, {The many lives
  of active galactic nuclei: cooling flows, black holes and the luminosities
  and colours of galaxies}, \emph{\mnras}. {\bf 365}, \penalty0 11--28  (Jan.,
  2006).
\newblock \doi{10.1111/j.1365-2966.2005.09675.x}.

\bibitem{DeLucia2006}
G.~{De Lucia}, V.~{Springel}, S.~D.~M. {White} {\em et~al.}, {The formation
  history of elliptical galaxies}, \emph{\mnras}. {\bf 366}, \penalty0 499--509
   (Feb., 2006).
\newblock \doi{10.1111/j.1365-2966.2005.09879.x}.

\bibitem{Menci06}
N.~{Menci}, A.~{Fontana}, E.~{Giallongo} {\em et~al.}, {The Abundance of
  Distant and Extremely Red Galaxies: The Role of AGN Feedback in Hierarchical
  Models}, \emph{\apj}. {\bf 647}\penalty0 (2), \penalty0 753--762  (Aug.,
  2006).
\newblock \doi{10.1086/505528}.

\bibitem{Monaco07}
P.~{Monaco}, F.~{Fontanot} and G.~{Taffoni}, {The MORGANA model for the rise of
  galaxies and active nuclei}, \emph{\mnras}. {\bf 375}\penalty0 (4), \penalty0
  1189--1219  (Mar., 2007).
\newblock \doi{10.1111/j.1365-2966.2006.11253.x}.

\bibitem{Somerville99}
R.~S. {Somerville} and J.~R. {Primack}, {Semi-analytic modelling of galaxy
  formation: the local Universe}, \emph{\mnras}. {\bf 310}\penalty0 (4),
  \penalty0 1087--1110  (Dec., 1999).
\newblock \doi{10.1046/j.1365-8711.1999.03032.x}.

\bibitem{Somerville2008}
R.~S. {Somerville}, P.~F. {Hopkins}, T.~J. {Cox} {\em et~al.}, {A semi-analytic
  model for the co-evolution of galaxies, black holes and active galactic
  nuclei}, \emph{\mnras}. {\bf 391}\penalty0 (2), \penalty0 481--506  (Dec.,
  2008).
\newblock \doi{10.1111/j.1365-2966.2008.13805.x}.

\bibitem{Guo11}
Q.~{Guo}, S.~{White}, M.~{Boylan-Kolchin} {\em et~al.}, {From dwarf spheroidals
  to cD galaxies: simulating the galaxy population in a
  {\ensuremath{\Lambda}}CDM cosmology}, \emph{\mnras}. {\bf 413}\penalty0 (1),
  \penalty0 101--131  (May, 2011).
\newblock \doi{10.1111/j.1365-2966.2010.18114.x}.

\bibitem{Benson12}
A.~J. {Benson}, {G ALACTICUS: A semi-analytic model of galaxy formation},
  \emph{\na}. {\bf 17}\penalty0 (2), \penalty0 175--197  (Feb., 2012).
\newblock \doi{10.1016/j.newast.2011.07.004}.

\bibitem{Hirschmann12}
M.~{Hirschmann}, R.~S. {Somerville}, T.~{Naab} {\em et~al.}, {Origin of the
  antihierarchical growth of black holes}, \emph{\mnras}. {\bf 426}\penalty0
  (1), \penalty0 237--257  (Oct., 2012).
\newblock \doi{10.1111/j.1365-2966.2012.21626.x}.

\bibitem{Henri13}
B.~M.~B. {Henriques}, S.~D.~M. {White}, P.~A. {Thomas} {\em et~al.},
  {Simulations of the galaxy population constrained by observations from z = 3
  to the present day: implications for galactic winds and the fate of their
  ejecta}, \emph{\mnras}. {\bf 431}\penalty0 (4), \penalty0 3373--3395  (June,
  2013).
\newblock \doi{10.1093/mnras/stt415}.

\bibitem{Croton16}
D.~J. {Croton}, A.~R.~H. {Stevens}, C.~{Tonini} {\em et~al.}, {Semi-Analytic
  Galaxy Evolution (SAGE): Model Calibration and Basic Results}, \emph{\apjs}.
  222\penalty0 (2):\penalty0 22  (Feb., 2016).
\newblock \doi{10.3847/0067-0049/222/2/22}.

\bibitem{Lacey16}
C.~G. {Lacey}, C.~M. {Baugh}, C.~S. {Frenk} {\em et~al.}, {A unified
  multiwavelength model of galaxy formation}, \emph{\mnras}. {\bf 462}\penalty0
  (4), \penalty0 3854--3911  (Nov., 2016).
\newblock \doi{10.1093/mnras/stw1888}.

\bibitem{Lagos18}
C.~d.~P. {Lagos}, R.~J. {Tobar}, A.~S.~G. {Robotham} {\em et~al.}, {Shark:
  introducing an open source, free, and flexible semi-analytic model of galaxy
  formation}, \emph{\mnras}. {\bf 481}\penalty0 (3), \penalty0 3573--3603
  (Dec., 2018).
\newblock \doi{10.1093/mnras/sty2440}.

\bibitem{Cattaneo20}
A.~{Cattaneo}, I.~{Koutsouridou}, E.~{Tollet} {\em et~al.}, {GalICS 2.1: a new
  semianalytic model for cold accretion, cooling, feedback, and their roles in
  galaxy formation}, \emph{\mnras}. {\bf 497}\penalty0 (1), \penalty0 279--301
  (Sept., 2020).
\newblock \doi{10.1093/mnras/staa1832}.

\bibitem{Fontanot20}
F.~{Fontanot}, G.~{De Lucia}, M.~{Hirschmann} {\em et~al.}, {The rise of active
  galactic nuclei in the galaxy evolution and assembly semi-analytic model},
  \emph{\mnras}. {\bf 496}\penalty0 (3), \penalty0 3943--3960  (Aug., 2020).
\newblock \doi{10.1093/mnras/staa1716}.

\bibitem{Marulli2008}
F.~{Marulli}, S.~{Bonoli}, E.~{Branchini} {\em et~al.}, {Modelling the
  cosmological co-evolution of supermassive black holes and galaxies - I. BH
  scaling relations and the AGN luminosity function}, \emph{\mnras}. {\bf 385},
  \penalty0 1846--1858  (Apr., 2008).
\newblock \doi{10.1111/j.1365-2966.2008.12988.x}.

\bibitem{Menci08}
N.~{Menci}, F.~{Fiore}, S.~{Puccetti} {\em et~al.}, {The Blast Wave Model for
  AGN Feedback: Effects on AGN Obscuration}, \emph{\apj}. {\bf 686}\penalty0
  (1), \penalty0 219--229  (Oct., 2008).
\newblock \doi{10.1086/591438}.

\bibitem{Bonoli2009}
S.~{Bonoli}, F.~{Marulli}, V.~{Springel} {\em et~al.}, {Modelling the
  cosmological co-evolution of supermassive black holes and galaxies - II. The
  clustering of quasars and their dark environment}, \emph{\mnras}. p. 606
  (May, 2009).
\newblock \doi{10.1111/j.1365-2966.2009.14701.x}.

\bibitem{Fanidakis12}
N.~{Fanidakis}, C.~M. {Baugh}, A.~J. {Benson} {\em et~al.}, {The evolution of
  active galactic nuclei across cosmic time: what is downsizing?},
  \emph{\mnras}. {\bf 419}\penalty0 (4), \penalty0 2797--2820  (Feb., 2012).
\newblock \doi{10.1111/j.1365-2966.2011.19931.x}.

\bibitem{KauffHaehnelt}
G.~{Kauffmann} and M.~{Haehnelt}, {A unified model for the evolution of
  galaxies and quasars}, \emph{\mnras}. {\bf 311}\penalty0 (3), \penalty0
  576--588  (Jan., 2000).
\newblock \doi{10.1046/j.1365-8711.2000.03077.x}.

\bibitem{Umemura1993}
M.~{Umemura}, A.~{Loeb} and E.~L. {Turner}, {Early Cosmic Formation of Massive
  Black Holes}, \emph{\apj}. {\bf 419}, \penalty0 459--+  (Dec., 1993).
\newblock \doi{10.1086/173499}.

\bibitem{Lapi06}
A.~{Lapi}, F.~{Shankar}, J.~{Mao} {\em et~al.}, {Quasar Luminosity Functions
  from Joint Evolution of Black Holes and Host Galaxies}, \emph{\apj}. {\bf
  650}\penalty0 (1), \penalty0 42--56  (Oct., 2006).
\newblock \doi{10.1086/507122}.

\bibitem{Hopkins06LF}
P.~F. {Hopkins}, L.~{Hernquist}, T.~J. {Cox} {\em et~al.}, {The Evolution in
  the Faint-End Slope of the Quasar Luminosity Function}, \emph{\apj}. {\bf
  639}\penalty0 (2), \penalty0 700--709  (Mar., 2006).
\newblock \doi{10.1086/499351}.

\bibitem{Malbon07}
R.~K. {Malbon}, C.~M. {Baugh}, C.~S. {Frenk} {\em et~al.}, {Black hole growth
  in hierarchical galaxy formation}, \emph{\mnras}. {\bf 382}\penalty0 (4),
  \penalty0 1394--1414  (Dec., 2007).
\newblock \doi{10.1111/j.1365-2966.2007.12317.x}.

\bibitem{Lidz06}
A.~{Lidz}, P.~F. {Hopkins}, T.~J. {Cox} {\em et~al.}, {The Luminosity
  Dependence of Quasar Clustering}, \emph{\apj}. {\bf 641}\penalty0 (1),
  \penalty0 41--49  (Apr., 2006).
\newblock \doi{10.1086/500444}.

\bibitem{SilkRees1998}
J.~{Silk} and M.~J. {Rees}, {Quasars and galaxy formation}, \emph{\aap}. {\bf
  331}, \penalty0 L1--L4  (Mar., 1998).

\bibitem{Soltan82}
A.~{Soltan}, {Masses of quasars.}, \emph{\mnras}. {\bf 200}, \penalty0 115--122
   (July, 1982).
\newblock \doi{10.1093/mnras/200.1.115}.

\bibitem{HaringRix}
N.~{H{\"a}ring} and H.-W. {Rix}, {On the Black Hole Mass-Bulge Mass Relation},
  \emph{\apjl}. {\bf 604}\penalty0 (2), \penalty0 L89--L92  (Apr., 2004).
\newblock \doi{10.1086/383567}.

\bibitem{GrahamSigma}
A.~W. {Graham}, C.~A. {Onken}, E.~{Athanassoula} {\em et~al.}, {An expanded
  M$_{bh}$-{\ensuremath{\sigma}} diagram, and a new calibration of active
  galactic nuclei masses}, \emph{\mnras}. {\bf 412}\penalty0 (4), \penalty0
  2211--2228  (Apr., 2011).
\newblock \doi{10.1111/j.1365-2966.2010.18045.x}.

\bibitem{GrahamLight}
A.~W. {Graham}, P.~{Erwin}, N.~{Caon} {\em et~al.}, {A Correlation between
  Galaxy Light Concentration and Supermassive Black Hole Mass}, \emph{\apjl}.
  {\bf 563}\penalty0 (1), \penalty0 L11--L14  (Dec., 2001).
\newblock \doi{10.1086/338500}.

\bibitem{Bogdan}
{\'A}.~{Bogd{\'a}n} and A.~D. {Goulding}, {Connecting Dark Matter Halos with
  the Galaxy Center and the Supermassive Black Hole}, \emph{\apj}. 800\penalty0
  (2):\penalty0 124  (Feb., 2015).
\newblock \doi{10.1088/0004-637X/800/2/124}.

\bibitem{Marconi04}
A.~{Marconi}, G.~{Risaliti}, R.~{Gilli} {\em et~al.}, {Local supermassive black
  holes, relics of active galactic nuclei and the X-ray background},
  \emph{\mnras}. {\bf 351}\penalty0 (1), \penalty0 169--185  (June, 2004).
\newblock \doi{10.1111/j.1365-2966.2004.07765.x}.

\bibitem{Shankar13}
F.~{Shankar}, D.~H. {Weinberg} and J.~{Miralda-Escud{\'e}}, {Accretion-driven
  evolution of black holes: Eddington ratios, duty cycles and active galaxy
  fractions}, \emph{\mnras}. {\bf 428}\penalty0 (1), \penalty0 421--446  (Jan.,
  2013).
\newblock \doi{10.1093/mnras/sts026}.

\bibitem{Ueda14}
Y.~{Ueda}, M.~{Akiyama}, G.~{Hasinger} {\em et~al.}, {Toward the Standard
  Population Synthesis Model of the X-Ray Background: Evolution of X-Ray
  Luminosity and Absorption Functions of Active Galactic Nuclei Including
  Compton-thick Populations}, \emph{\apj}. 786\penalty0 (2):\penalty0 104
  (May, 2014).
\newblock \doi{10.1088/0004-637X/786/2/104}.

\bibitem{Shankar20}
F.~{Shankar}, D.~H. {Weinberg}, C.~{Marsden} {\em et~al.}, {Probing black hole
  accretion tracks, scaling relations, and radiative efficiencies from stacked
  X-ray active galactic nuclei}, \emph{\mnras}. {\bf 493}\penalty0 (1),
  \penalty0 1500--1511  (Mar., 2020).
\newblock \doi{10.1093/mnras/stz3522}.

\bibitem{Ananna20}
T.~T. {Ananna}, C.~M. {Urry}, E.~{Treister} {\em et~al.}, {Accretion History of
  AGNs. III. Radiative Efficiency and AGN Contribution to Reionization},
  \emph{\apj}. 903\penalty0 (2):\penalty0 85  (Nov., 2020).
\newblock \doi{10.3847/1538-4357/abb815}.

\bibitem{Duras20}
F.~{Duras}, A.~{Bongiorno}, F.~{Ricci} {\em et~al.}, {Universal bolometric
  corrections for active galactic nuclei over seven luminosity decades},
  \emph{\aap}. 636:\penalty0 A73  (Apr., 2020).
\newblock \doi{10.1051/0004-6361/201936817}.

\bibitem{Shankar16}
F.~{Shankar}, M.~{Bernardi}, R.~K. {Sheth} {\em et~al.}, {Selection bias in
  dynamically measured supermassive black hole samples: its consequences and
  the quest for the most fundamental relation}, \emph{\mnras}. {\bf
  460}\penalty0 (3), \penalty0 3119--3142  (Aug., 2016).
\newblock \doi{10.1093/mnras/stw678}.

\bibitem{SmallBlandford}
T.~A. {Small} and R.~D. {Blandford}, {Quasar evolution and the growth of black
  holes.}, \emph{\mnras}. {\bf 259}, \penalty0 725--737  (Dec., 1992).
\newblock \doi{10.1093/mnras/259.4.725}.

\bibitem{Shankar09}
F.~{Shankar}, D.~H. {Weinberg} and J.~{Miralda-Escud{\'e}}, {Self-Consistent
  Models of the AGN and Black Hole Populations: Duty Cycles, Accretion Rates,
  and the Mean Radiative Efficiency}, \emph{\apj}. {\bf 690}\penalty0 (1),
  \penalty0 20--41  (Jan., 2009).
\newblock \doi{10.1088/0004-637X/690/1/20}.

\bibitem{Aversa15}
R.~{Aversa}, A.~{Lapi}, G.~{de Zotti} {\em et~al.}, {Black Hole and Galaxy
  Coevolution from Continuity Equation and Abundance Matching}, \emph{\apj}.
  810\penalty0 (1):\penalty0 74  (Sept., 2015).
\newblock \doi{10.1088/0004-637X/810/1/74}.

\bibitem{TucciVolonteri}
M.~{Tucci} and M.~{Volonteri}, {Constraining supermassive black hole evolution
  through the continuity equation}, \emph{\aap}. 600:\penalty0 A64  (Apr.,
  2017).
\newblock \doi{10.1051/0004-6361/201628419}.

\bibitem{ShankarNat}
F.~{Shankar}, V.~{Allevato}, M.~{Bernardi} {\em et~al.}, {Constraining black
  hole-galaxy scaling relations and radiative efficiency from galaxy
  clustering}, \emph{Nature Astronomy}. {\bf 4}, \penalty0 282--291  (Jan.,
  2020).
\newblock \doi{10.1038/s41550-019-0949-y}.

\bibitem{DavisLaor}
S.~W. {Davis} and A.~{Laor}, {The Radiative Efficiency of Accretion Flows in
  Individual Active Galactic Nuclei}, \emph{\apj}. 728\penalty0 (2):\penalty0
  98  (Feb., 2011).
\newblock \doi{10.1088/0004-637X/728/2/98}.

\bibitem{Cao10}
X.~{Cao}, {Cosmological Evolution of Massive Black Holes: Effects of Eddington
  Ratio Distribution and Quasar Lifetime}, \emph{\apj}. {\bf 725}\penalty0 (1),
  \penalty0 388--393  (Dec., 2010).
\newblock \doi{10.1088/0004-637X/725/1/388}.

\bibitem{Shankar10}
F.~{Shankar}, M.~{Crocce}, J.~{Miralda-Escud{\'e}} {\em et~al.}, {On the
  Radiative Efficiencies, Eddington Ratios, and Duty Cycles of Luminous
  High-redshift Quasars}, \emph{\apj}. {\bf 718}\penalty0 (1), \penalty0
  231--250  (July, 2010).
\newblock \doi{10.1088/0004-637X/718/1/231}.

\bibitem{Hopkins_SEM}
P.~F. {Hopkins}, J.~D. {Younger}, C.~C. {Hayward} {\em et~al.}, {Mergers,
  active galactic nuclei and `normal' galaxies: contributions to the
  distribution of star formation rates and infrared luminosity functions},
  \emph{\mnras}. {\bf 402}\penalty0 (3), \penalty0 1693--1713  (Mar., 2010).
\newblock \doi{10.1111/j.1365-2966.2009.15990.x}.

\bibitem{Yang17}
G.~{Yang}, C.~T.~J. {Chen}, F.~{Vito} {\em et~al.}, {Black Hole Growth Is
  Mainly Linked to Host-galaxy Stellar Mass Rather Than Star Formation Rate},
  \emph{\apj}. 842\penalty0 (2):\penalty0 72  (June, 2017).
\newblock \doi{10.3847/1538-4357/aa7564}.

\bibitem{Trinity}
H.~{Zhang}, P.~{Behroozi}, M.~{Volonteri} {\em et~al.}, {Trinity I:
  Self-Consistently Modeling the Dark Matter Halo-Galaxy-Supermassive Black
  Hole Connection from $z=0-10$}, \emph{arXiv e-prints}. art. arXiv:2105.10474
  (May, 2021).

\bibitem{Georgakakis19}
A.~{Georgakakis}, J.~{Comparat}, A.~{Merloni} {\em et~al.}, {Exploring the halo
  occupation of AGN using dark-matter cosmological simulations}, \emph{\mnras}.
  {\bf 487}\penalty0 (1), \penalty0 275--295  (July, 2019).
\newblock \doi{10.1093/mnras/sty3454}.

\bibitem{AirdCoil}
J.~{Aird} and A.~L. {Coil}, {The AGN-galaxy-halo connection: the distribution
  of AGN host halo masses to z = 2.5}, \emph{\mnras}. {\bf 502}\penalty0 (4),
  \penalty0 5962--5980  (Apr., 2021).
\newblock \doi{10.1093/mnras/stab312}.

\bibitem{Allevato21}
V.~{Allevato}, F.~{Shankar}, C.~{Marsden} {\em et~al.}, {Building Robust Active
  Galactic Nuclei Mock Catalogs to Unveil Black Hole Evolution and for Survey
  Planning}, \emph{\apj}. 916\penalty0 (1):\penalty0 34  (July, 2021).
\newblock \doi{10.3847/1538-4357/abfe59}.

\bibitem{Viitanen21}
A.~{Viitanen}, V.~{Allevato}, A.~{Finoguenov} {\em et~al.}, {The role of
  scatter and satellites in shaping the large-scale clustering of X-ray AGN as
  a function of host galaxy stellar mass}, \emph{\mnras}. {\bf 507}\penalty0
  (4), \penalty0 6148--6160  (Nov., 2021).
\newblock \doi{10.1093/mnras/stab2538}.

\bibitem{Carraro22}
R.~{Carraro}, F.~{Shankar}, V.~{Allevato} {\em et~al.}, {An Eddington
  ratio-driven origin for the L$_{X}$ - M$_{*}$ relation in quiescent and star
  forming active galaxies}, \emph{\mnras}  (Feb., 2022).
\newblock \doi{10.1093/mnras/stac441}.

\bibitem{ConroyWhite}
C.~{Conroy} and M.~{White}, {A Simple Model for Quasar Demographics},
  \emph{\apj}. 762\penalty0 (2):\penalty0 70  (Jan., 2013).
\newblock \doi{10.1088/0004-637X/762/2/70}.

\bibitem{Euclid}
R.~{Laureijs}, J.~{Amiaux}, S.~{Arduini} {\em et~al.}, {Euclid Definition Study
  Report}, \emph{arXiv e-prints}. art. arXiv:1110.3193  (Oct., 2011).

\bibitem{LSST}
J.~{Yao}, M.~{Ishak}, M.~A. {Troxel} {\em et~al.}
\newblock {Effect of Self-Calibration of Intrinsic Alignment on the
  Cosmological Parameter Constraints for LSST}.
\newblock In \emph{American Astronomical Society Meeting Abstracts \#229}, vol.
  229, \emph{American Astronomical Society Meeting Abstracts}, p. 125.07  (Jan,
  2017).

\bibitem{Kravtsov04}
A.~V. {Kravtsov}, A.~A. {Berlind}, R.~H. {Wechsler} {\em et~al.}, {The Dark
  Side of the Halo Occupation Distribution}, \emph{\apj}. {\bf 609}\penalty0
  (1), \penalty0 35--49  (July, 2004).
\newblock \doi{10.1086/420959}.

\bibitem{Vale04}
A.~{Vale} and J.~P. {Ostriker}, {Linking halo mass to galaxy luminosity},
  \emph{\mnras}. {\bf 353}\penalty0 (1), \penalty0 189--200  (Sept., 2004).
\newblock \doi{10.1111/j.1365-2966.2004.08059.x}.

\bibitem{Shankar06}
F.~{Shankar}, A.~{Lapi}, P.~{Salucci} {\em et~al.}, {New Relationships between
  Galaxy Properties and Host Halo Mass, and the Role of Feedbacks in Galaxy
  Formation}, \emph{\apj}. {\bf 643}\penalty0 (1), \penalty0 14--25  (May,
  2006).
\newblock \doi{10.1086/502794}.

\bibitem{Moster10}
B.~P. {Moster}, R.~S. {Somerville}, C.~{Maulbetsch} {\em et~al.}, {Constraints
  on the Relationship between Stellar Mass and Halo Mass at Low and High
  Redshift}, \emph{\apj}. {\bf 710}\penalty0 (2), \penalty0 903--923  (Feb.,
  2010).
\newblock \doi{10.1088/0004-637X/710/2/903}.

\bibitem{Moster13}
B.~P. {Moster}, T.~{Naab} and S.~D.~M. {White}, {Galactic star formation and
  accretion histories from matching galaxies to dark matter haloes},
  \emph{\mnras}. {\bf 428}\penalty0 (4), \penalty0 3121--3138  (Feb., 2013).
\newblock \doi{10.1093/mnras/sts261}.

\bibitem{Behroozi13}
P.~S. {Behroozi}, R.~H. {Wechsler} and C.~{Conroy}, {The Average Star Formation
  Histories of Galaxies in Dark Matter Halos from z = 0-8}, \emph{\apj}.
  770\penalty0 (1):\penalty0 57  (June, 2013).
\newblock \doi{10.1088/0004-637X/770/1/57}.

\bibitem{Kravtsov18}
A.~V. {Kravtsov}, A.~A. {Vikhlinin} and A.~V. {Meshcheryakov}, {Stellar
  Mass{\textemdash}Halo Mass Relation and Star Formation Efficiency in
  High-Mass Halos}, \emph{Astronomy Letters}. {\bf 44}\penalty0 (1), \penalty0
  8--34  (Jan., 2018).
\newblock \doi{10.1134/S1063773717120015}.

\bibitem{Vika09}
M.~{Vika}, S.~P. {Driver}, A.~W. {Graham} {\em et~al.}, {The Millennium Galaxy
  Catalogue: the M$_{bh}$-L$_{spheroid}$ derived supermassive black hole mass
  function}, \emph{\mnras}. {\bf 400}\penalty0 (3), \penalty0 1451--1460
  (Dec., 2009).
\newblock \doi{10.1111/j.1365-2966.2009.15544.x}.

\bibitem{Hopkins09}
P.~F. {Hopkins}, L.~{Hernquist}, T.~J. {Cox} {\em et~al.}, {Dissipation and
  Extra Light in Galactic Nuclei. IV. Evolution in the Scaling Relations of
  Spheroids}, \emph{\apj}. {\bf 691}\penalty0 (2), \penalty0 1424--1458  (Feb.,
  2009).
\newblock \doi{10.1088/0004-637X/691/2/1424}.

\bibitem{Zavala12}
J.~{Zavala}, V.~{Avila-Reese}, C.~{Firmani} {\em et~al.}, {The growth of
  galactic bulges through mergers in {\ensuremath{\Lambda}} CDM haloes
  revisited - I. Present-day properties}, \emph{\mnras}. {\bf 427}\penalty0
  (2), \penalty0 1503--1516  (Dec., 2012).
\newblock \doi{10.1111/j.1365-2966.2012.22100.x}.

\bibitem{Shankar14}
F.~{Shankar}, S.~{Mei}, M.~{Huertas-Company} {\em et~al.}, {Environmental
  dependence of bulge-dominated galaxy sizes in hierarchical models of galaxy
  formation. Comparison with the local Universe}, \emph{\mnras}. {\bf
  439}\penalty0 (4), \penalty0 3189--3212  (Apr., 2014).
\newblock \doi{10.1093/mnras/stt2470}.

\bibitem{Cattaneo11}
A.~{Cattaneo}, G.~A. {Mamon}, K.~{Warnick} {\em et~al.}, {How do galaxies
  acquire their mass?}, \emph{\aap}. 533:\penalty0 A5  (Sept., 2011).
\newblock \doi{10.1051/0004-6361/201015780}.

\bibitem{BuchanShankar}
S.~{Buchan} and F.~{Shankar}, {Setting firmer constraints on the evolution of
  the most massive, central galaxies from their local abundances and ages},
  \emph{\mnras}. {\bf 462}\penalty0 (2), \penalty0 2001--2010  (Oct., 2016).
\newblock \doi{10.1093/mnras/stw1771}.

\bibitem{Grylls19}
P.~J. {Grylls}, F.~{Shankar}, L.~{Zanisi} {\em et~al.}, {A statistical
  semi-empirical model: satellite galaxies in groups and clusters},
  \emph{\mnras}. {\bf 483}\penalty0 (2), \penalty0 2506--2523  (Feb., 2019).
\newblock \doi{10.1093/mnras/sty3281}.

\bibitem{Grylls20SFH}
P.~J. {Grylls}, F.~{Shankar}, J.~{Leja} {\em et~al.}, {Predicting fully
  self-consistent satellite richness, galaxy growth, and star formation rates
  from the STatistical sEmi-Empirical modeL STEEL}, \emph{\mnras}. {\bf
  491}\penalty0 (1), \penalty0 634--654  (Jan., 2020).
\newblock \doi{10.1093/mnras/stz2956}.

\bibitem{Grylls20Mergers}
P.~J. {Grylls}, F.~{Shankar} and C.~J. {Conselice}, {The significant effects of
  stellar mass estimation on galaxy pair fractions.}, \emph{\mnras}. {\bf
  499}\penalty0 (2), \penalty0 2265--2275  (Dec., 2020).
\newblock \doi{10.1093/mnras/staa2966}.

\bibitem{Oleary}
J.~A. {O'Leary}, B.~P. {Moster} and E.~{Kr{\"a}mer}, {EMERGE: constraining
  merging probabilities and time-scales of close galaxy pairs}, \emph{\mnras}.
  {\bf 503}\penalty0 (4), \penalty0 5646--5657  (June, 2021).
\newblock \doi{10.1093/mnras/stab889}.

\bibitem{Moster18}
B.~P. {Moster}, T.~{Naab} and S.~D.~M. {White}, {EMERGE - an empirical model
  for the formation of galaxies since z {\ensuremath{\sim}} 10}, \emph{\mnras}.
  {\bf 477}\penalty0 (2), \penalty0 1822--1852  (June, 2018).
\newblock \doi{10.1093/mnras/sty655}.

\bibitem{Behroozi19}
P.~{Behroozi}, R.~H. {Wechsler}, A.~P. {Hearin} {\em et~al.}, {UNIVERSEMACHINE:
  The correlation between galaxy growth and dark matter halo assembly from z =
  0-10}, \emph{\mnras}. {\bf 488}\penalty0 (3), \penalty0 3143--3194  (Sept.,
  2019).
\newblock \doi{10.1093/mnras/stz1182}.

\bibitem{WyitheLoeb}
J.~S.~B. {Wyithe} and A.~{Loeb}, {Self-regulated Growth of Supermassive Black
  Holes in Galaxies as the Origin of the Optical and X-Ray Luminosity Functions
  of Quasars}, \emph{\apj}. {\bf 595}\penalty0 (2), \penalty0 614--623  (Oct.,
  2003).
\newblock \doi{10.1086/377475}.

\bibitem{Hopkins05}
P.~F. {Hopkins}, L.~{Hernquist}, T.~J. {Cox} {\em et~al.},
  {Luminosity-dependent Quasar Lifetimes: A New Interpretation of the Quasar
  Luminosity Function}, \emph{\apj}. {\bf 630}\penalty0 (2), \penalty0 716--720
   (Sept., 2005).
\newblock \doi{10.1086/432463}.

\bibitem{Shen09}
Y.~{Shen}, {Supermassive Black Holes in the Hierarchical Universe: A General
  Framework and Observational Tests}, \emph{\apj}. {\bf 704}\penalty0 (1),
  \penalty0 89--108  (Oct., 2009).
\newblock \doi{10.1088/0004-637X/704/1/89}.

\bibitem{ShankarIAU}
F.~{Shankar}.
\newblock {Merger-Induced Quasars, Their Light Curves, and Their Host Halos}.
\newblock In eds. B.~M. {Peterson}, R.~S. {Somerville} and
  T.~{Storchi-Bergmann}, \emph{Co-Evolution of Central Black Holes and
  Galaxies}, vol. 267, pp. 248--253  (May, 2010).
\newblock \doi{10.1017/S1743921310006356}.

\bibitem{Haiman01}
Z.~{Haiman} and L.~{Hui}, {Constraining the Lifetime of Quasars from Their
  Spatial Clustering}, \emph{\apj}. {\bf 547}\penalty0 (1), \penalty0 27--38
  (Jan., 2001).
\newblock \doi{10.1086/318330}.

\bibitem{MartiniWeinberg}
P.~{Martini} and D.~H. {Weinberg}, {Quasar Clustering and the Lifetime of
  Quasars}, \emph{\apj}. {\bf 547}\penalty0 (1), \penalty0 12--26  (Jan.,
  2001).
\newblock \doi{10.1086/318331}.

\bibitem{Sesana16}
A.~{Sesana}, F.~{Shankar}, M.~{Bernardi} {\em et~al.}, {Selection bias in
  dynamically measured supermassive black hole samples: consequences for pulsar
  timing arrays}, \emph{\mnras}. {\bf 463}\penalty0 (1), \penalty0 L6--L11
  (Nov., 2016).
\newblock \doi{10.1093/mnrasl/slw139}.

\bibitem{Hopkins06UnifiedModel}
P.~F. {Hopkins}, L.~{Hernquist}, T.~J. {Cox} {\em et~al.}, {A Unified,
  Merger-driven Model of the Origin of Starbursts, Quasars, the Cosmic X-Ray
  Background, Supermassive Black Holes, and Galaxy Spheroids}, \emph{\apjs}.
  {\bf 163}\penalty0 (1), \penalty0 1--49  (Mar., 2006).
\newblock \doi{10.1086/499298}.

\bibitem{DiMatteo2017}
T.~{Di Matteo}, R.~A.~C. {Croft}, Y.~{Feng} {\em et~al.}, {The origin of the
  most massive black holes at high-z: BlueTides and the next quasar frontier},
  \emph{\mnras}. {\bf 467}, \penalty0 4243--4251  (June, 2017).
\newblock \doi{10.1093/mnras/stx319}.

\bibitem{Tenneti2019}
A.~{Tenneti}, S.~M. {Wilkins}, T.~{Di Matteo} {\em et~al.}, {A tiny host galaxy
  for the first giant black hole: z = 7.5 quasar in BlueTides}, \emph{\mnras}.
  {\bf 483}\penalty0 (1), \penalty0 1388--1399  (Feb., 2019).
\newblock \doi{10.1093/mnras/sty3161}.

\bibitem{Ni2022}
Y.~{Ni}, T.~{Di Matteo} and Y.~{Feng}, {Not all peaks are created equal: the
  early growth of supermassive black holes}, \emph{\mnras}. {\bf 509}\penalty0
  (2), \penalty0 3043--3064  (Jan., 2022).
\newblock \doi{10.1093/mnras/stab3162}.

\bibitem{huang20}
K.-W. {Huang}, Y.~{Ni}, Y.~{Feng} {\em et~al.}, {The early growth of
  supermassive black holes in cosmological hydrodynamic simulations with
  constrained Gaussian realizations}, \emph{\mnras}. {\bf 496}\penalty0 (1),
  \penalty0 1--12  (June, 2020).
\newblock \doi{10.1093/mnras/staa1515}.

\bibitem{Bertschinger1987}
E.~{Bertschinger}, {Path Integral Methods for Primordial Density Perturbations:
  Sampling of Constrained Gaussian Random Fields}, \emph{\apjl}. {\bf 323},
  \penalty0 L103  (Dec., 1987).
\newblock \doi{10.1086/185066}.

\bibitem{Binney1991}
J.~{Binney} and T.~{Quinn}, {Gaussian random fields in spherical coordinates},
  \emph{\mnras}. {\bf 249}, \penalty0 678--683  (Apr., 1991).
\newblock \doi{10.1093/mnras/249.4.678}.

\bibitem{vandeWeygaert1996}
R.~{van de Weygaert} and E.~{Bertschinger}, {Peak and gravity constraints in
  Gaussian primordial density fields: An application of the Hoffman-Ribak
  method}, \emph{\mnras}. {\bf 281}, \penalty0 84  (July, 1996).
\newblock \doi{10.1093/mnras/281.1.84}.

\bibitem{Latif}
M.~A. {Latif} and A.~{Ferrara}, {Formation of Supermassive Black Hole Seeds},
  \emph{\pasa}. 33:\penalty0 e051  (Oct., 2016).
\newblock \doi{10.1017/pasa.2016.41}.

\bibitem{Madau14}
P.~{Madau}, F.~{Haardt} and M.~{Dotti}, {Super-critical Growth of Massive Black
  Holes from Stellar-mass Seeds}, \emph{\apjl}. 784\penalty0 (2):\penalty0 L38
  (Apr., 2014).
\newblock \doi{10.1088/2041-8205/784/2/L38}.

\bibitem{Volonteri15}
M.~{Volonteri}, J.~{Silk} and G.~{Dubus}, {The Case for Supercritical Accretion
  onto Massive Black Holes at High Redshift}, \emph{\apj}. 804\penalty0
  (2):\penalty0 148  (May, 2015).
\newblock \doi{10.1088/0004-637X/804/2/148}.

\bibitem{Lupi16}
A.~{Lupi}, F.~{Haardt}, M.~{Dotti} {\em et~al.}, {Growing massive black holes
  through supercritical accretion of stellar-mass seeds}, \emph{\mnras}. {\bf
  456}\penalty0 (3), \penalty0 2993--3003  (Mar., 2016).
\newblock \doi{10.1093/mnras/stv2877}.

\bibitem{Regan19}
J.~A. {Regan}, T.~P. {Downes}, M.~{Volonteri} {\em et~al.}, {Super-Eddington
  accretion and feedback from the first massive seed black holes},
  \emph{\mnras}. {\bf 486}\penalty0 (3), \penalty0 3892--3906  (July, 2019).
\newblock \doi{10.1093/mnras/stz1045}.

\bibitem{DiMatteo2012}
T.~{Di Matteo}, N.~{Khandai}, C.~{DeGraf} {\em et~al.}, {Cold Flows and the
  First Quasars}, \emph{\apjl}. 745:\penalty0 L29  (Feb., 2012).
\newblock \doi{10.1088/2041-8205/745/2/L29}.

\bibitem{MayerBonoli}
L.~{Mayer} and S.~{Bonoli}, {The route to massive black hole formation via
  merger-driven direct collapse: a review}, \emph{Reports on Progress in
  Physics}. 82\penalty0 (1):\penalty0 016901  (Jan., 2019).
\newblock \doi{10.1088/1361-6633/aad6a5}.

\bibitem{Boco20}
L.~{Boco}, A.~{Lapi} and L.~{Danese}, {Growth of Supermassive Black Hole Seeds
  in ETG Star-forming Progenitors: Multiple Merging of Stellar Compact Remnants
  via Gaseous Dynamical Friction and Gravitational-wave Emission}, \emph{\apj}.
  891\penalty0 (1):\penalty0 94  (Mar., 2020).
\newblock \doi{10.3847/1538-4357/ab7446}.

\bibitem{Tagawa2020}
H.~{Tagawa}, Z.~{Haiman} and B.~{Kocsis}, {Making a Supermassive Star by
  Stellar Bombardment}, \emph{\apj}. 892\penalty0 (1):\penalty0 36  (Mar.,
  2020).
\newblock \doi{10.3847/1538-4357/ab7922}.

\bibitem{Trinca22}
A.~{Trinca}, R.~{Schneider}, R.~{Valiante} {\em et~al.}, {The low-end of the
  black hole mass function at cosmic dawn}, \emph{\mnras}. {\bf 511}\penalty0
  (1), \penalty0 616--640  (Mar., 2022).
\newblock \doi{10.1093/mnras/stac062}.

\bibitem{McConnell2013}
N.~J. {McConnell} and C.-P. {Ma}, {Revisiting the Scaling Relations of Black
  Hole Masses and Host Galaxy Properties}, \emph{\apj}. 764:\penalty0 184
  (Feb., 2013).
\newblock \doi{10.1088/0004-637X/764/2/184}.

\bibitem{Reines2015}
A.~E. {Reines} and M.~{Volonteri}, {Relations between Central Black Hole Mass
  and Total Galaxy Stellar Mass in the Local Universe}, \emph{\apj}.
  813:\penalty0 82  (Nov., 2015).
\newblock \doi{10.1088/0004-637X/813/2/82}.

\bibitem{Trakhtenbrot2010}
B.~{Trakhtenbrot} and H.~{Netzer}, {The evolution of M$_{*}$/M$_{BH}$ between z
  = 2 and z = 0}, \emph{\mnras}. {\bf 406}, \penalty0 L35--L39  (July, 2010).
\newblock \doi{10.1111/j.1745-3933.2010.00876.x}.

\bibitem{Bongiorno2014}
A.~{Bongiorno}, R.~{Maiolino}, M.~{Brusa} {\em et~al.}, {The M$_{BH}$-M$_{*}$
  relation for X-ray-obscured, red QSOs at 1.2 $\lt$ z $\lt$ 2.6},
  \emph{\mnras}. {\bf 443}, \penalty0 2077--2091  (Sept., 2014).
\newblock \doi{10.1093/mnras/stu1248}.

\bibitem{Schulze2014}
A.~{Schulze} and L.~{Wisotzki}, {Accounting for selection effects in the
  BH-bulge relations: no evidence for cosmological evolution}, \emph{\mnras}.
  {\bf 438}, \penalty0 3422--3433  (Mar., 2014).
\newblock \doi{10.1093/mnras/stt2457}.

\bibitem{Shen2015}
Y.~{Shen}, J.~E. {Greene}, L.~C. {Ho} {\em et~al.}, {The Sloan Digital Sky
  Survey Reverberation Mapping Project: No Evidence for Evolution in the
  $M_{\sigma} $ Relation to z ~1 }, \emph{\apj}. 805:\penalty0 96  (June,
  2015).
\newblock \doi{10.1088/0004-637X/805/2/96}.

\bibitem{Sun2015}
M.~{Sun}, J.~R. {Trump}, W.~N. {Brandt} {\em et~al.}, {Evolution in the Black
  Hole Galaxy Scaling Relations and the Duty Cycle of Nuclear Activity in
  Star-forming Galaxies}, \emph{\apj}. 802:\penalty0 14  (Mar., 2015).
\newblock \doi{10.1088/0004-637X/802/1/14}.

\bibitem{Willott2015}
C.~J. {Willott}, J.~{Bergeron} and A.~{Omont}, {Star Formation Rate and
  Dynamical Mass of 10$^{8}$ Solar Mass Black Hole Host Galaxies At Redshift
  6}, \emph{\apj}. 801:\penalty0 123  (Mar., 2015).
\newblock \doi{10.1088/0004-637X/801/2/123}.

\bibitem{King2003}
A.~{King}, {Black Holes, Galaxy Formation, and the M$_{BH}$-{$\sigma$}
  Relation}, \emph{\apjl}. {\bf 596}, \penalty0 L27--L29  (Oct., 2003).
\newblock \doi{10.1086/379143}.

\bibitem{Hopkins2007_BHplane}
P.~F. {Hopkins}, L.~{Hernquist}, T.~J. {Cox} {\em et~al.}, {A Theoretical
  Interpretation of the Black Hole Fundamental Plane}, \emph{\apj}. {\bf 669},
  \penalty0 45--66  (Nov., 2007).
\newblock \doi{10.1086/521590}.

\bibitem{Peng2007}
C.~Y. {Peng}, {How Mergers May Affect the Mass Scaling Relation between
  Gravitationally Bound Systems}, \emph{\apj}. {\bf 671}, \penalty0 1098--1107
  (Dec., 2007).
\newblock \doi{10.1086/522774}.

\bibitem{Hirschmann2010}
M.~{Hirschmann}, S.~{Khochfar}, A.~{Burkert} {\em et~al.}, {On the evolution of
  the intrinsic scatter in black hole versus galaxy mass relations},
  \emph{\mnras}. {\bf 407}, \penalty0 1016--1032  (Sept., 2010).
\newblock \doi{10.1111/j.1365-2966.2010.17006.x}.

\bibitem{Jahnke2011}
K.~{Jahnke} and A.~V. {Macci{\`o}}, {The Non-causal Origin of the
  Black-hole-galaxy Scaling Relations}, \emph{\apj}. 734:\penalty0 92  (June,
  2011).
\newblock \doi{10.1088/0004-637X/734/2/92}.

\bibitem{Kauffmann2009}
G.~{Kauffmann} and T.~M. {Heckman}, {Feast and Famine: regulation of black hole
  growth in low-redshift galaxies}, \emph{\mnras}. {\bf 397}, \penalty0
  135--147  (July, 2009).
\newblock \doi{10.1111/j.1365-2966.2009.14960.x}.

\bibitem{Chen2013}
C.-T.~J. {Chen}, R.~C. {Hickox}, S.~{Alberts} {\em et~al.}, {A Correlation
  between Star Formation Rate and Average Black Hole Accretion in Star-forming
  Galaxies}, \emph{\apj}. 773:\penalty0 3  (Aug., 2013).
\newblock \doi{10.1088/0004-637X/773/1/3}.

\bibitem{Li2020_TNGbhs}
Y.~{Li}, M.~{Habouzit}, S.~{Genel} {\em et~al.}, {Correlations between Black
  Holes and Host Galaxies in the Illustris and IllustrisTNG Simulations},
  \emph{\apj}. 895\penalty0 (2):\penalty0 102  (June, 2020).
\newblock \doi{10.3847/1538-4357/ab8f8d}.

\bibitem{Crain2015}
R.~A. {Crain}, J.~{Schaye}, R.~G. {Bower} {\em et~al.}, {The EAGLE simulations
  of galaxy formation: calibration of subgrid physics and model variations},
  \emph{\mnras}. {\bf 450}, \penalty0 1937--1961  (June, 2015).
\newblock \doi{10.1093/mnras/stv725}.

\bibitem{Terrazas2017}
B.~A. {Terrazas}, E.~F. {Bell}, J.~{Woo} {\em et~al.}, {Supermassive Black
  Holes as the Regulators of Star Formation in Central Galaxies}, \emph{\apj}.
  844\penalty0 (2):\penalty0 170  (Aug., 2017).
\newblock \doi{10.3847/1538-4357/aa7d07}.

\bibitem{Bower2017}
R.~G. {Bower}, J.~{Schaye}, C.~S. {Frenk} {\em et~al.}, {The dark nemesis of
  galaxy formation: why hot haloes trigger black hole growth and bring star
  formation to an end}, \emph{\mnras}. {\bf 465}, \penalty0 32--44  (Feb.,
  2017).
\newblock \doi{10.1093/mnras/stw2735}.

\bibitem{Prieto2017}
J.~{Prieto}, A.~{Escala}, M.~{Volonteri} {\em et~al.}, {How AGN and SN Feedback
  Affect Mass Transport and Black Hole Growth in High-redshift Galaxies},
  \emph{\apj}. 836:\penalty0 216  (Feb., 2017).
\newblock \doi{10.3847/1538-4357/aa5be5}.

\bibitem{McAlpine2018}
S.~{McAlpine}, R.~G. {Bower}, D.~J. {Rosario} {\em et~al.}, {The rapid growth
  phase of supermassive black holes}, \emph{\mnras}. {\bf 481}, \penalty0
  3118--3128  (Dec., 2018).
\newblock \doi{10.1093/mnras/sty2489}.

\bibitem{Trebitsch2018}
M.~{Trebitsch}, M.~{Volonteri}, Y.~{Dubois} {\em et~al.}, {Escape of ionizing
  radiation from high-redshift dwarf galaxies: role of AGN feedback},
  \emph{\mnras}. {\bf 478}\penalty0 (4), \penalty0 5607--5625  (Aug, 2018).
\newblock \doi{10.1093/mnras/sty1406}.

\bibitem{Lupi2019}
A.~{Lupi}, M.~{Volonteri}, R.~{Decarli} {\em et~al.}, {High-redshift quasars
  and their host galaxies - I. Kinematical and dynamical properties and their
  tracers}, \emph{\mnras}. {\bf 488}\penalty0 (3), \penalty0 4004--4022  (Sep,
  2019).
\newblock \doi{10.1093/mnras/stz1959}.

\bibitem{Lapiner2021}
S.~{Lapiner}, A.~{Dekel} and Y.~{Dubois}, {Compaction-driven black hole
  growth}, \emph{\mnras}. {\bf 505}\penalty0 (1), \penalty0 172--190  (July,
  2021).
\newblock \doi{10.1093/mnras/stab1205}.

\bibitem{Tillman2022}
M.~T. {Tillman}, S.~{Wellons}, C.-A. {Faucher-Gigu{\`e}re} {\em et~al.},
  {Running late: testing delayed supermassive black hole growth models against
  the quasar luminosity function}, \emph{\mnras}. {\bf 511}\penalty0 (4),
  \penalty0 5756--5767  (Apr., 2022).
\newblock \doi{10.1093/mnras/stac398}.

\bibitem{Graham2013}
A.~W. {Graham} and N.~{Scott}, {The M $_{BH}$-L $_{spheroid}$ Relation at High
  and Low Masses, the Quadratic Growth of Black Holes, and Intermediate-mass
  Black Hole Candidates}, \emph{\apj}. 764:\penalty0 151  (Feb., 2013).
\newblock \doi{10.1088/0004-637X/764/2/151}.

\bibitem{Savorgnan2016}
G.~A.~D. {Savorgnan}, A.~W. {Graham}, A.~{Marconi} {\em et~al.}, {Supermassive
  Black Holes and Their Host Spheroids. II. The Red and Blue Sequence in the
  M$_{BH}$--M$_{*,sph}$ Diagram}, \emph{\apj}. 817:\penalty0 21  (Jan., 2016).
\newblock \doi{10.3847/0004-637X/817/1/21}.

\bibitem{DeGraf2012}
C.~{DeGraf}, T.~{Di Matteo}, N.~{Khandai} {\em et~al.}, {Early black holes in
  cosmological simulations: luminosity functions and clustering behaviour},
  \emph{\mnras}. {\bf 424}, \penalty0 1892--1898  (Aug., 2012).
\newblock \doi{10.1111/j.1365-2966.2012.21294.x}.

\bibitem{Haring2004}
N.~{H{\"a}ring} and H.-W. {Rix}, {On the Black Hole Mass-Bulge Mass Relation},
  \emph{\apjl}. {\bf 604}, \penalty0 L89--L92  (Apr., 2004).
\newblock \doi{10.1086/383567}.

\bibitem{Hirschmann10}
M.~{Hirschmann}, S.~{Khochfar}, A.~{Burkert} {\em et~al.}, {On the evolution of
  the intrinsic scatter in black hole versus galaxy mass relations},
  \emph{\mnras}. {\bf 407}\penalty0 (2), \penalty0 1016--1032  (Sept., 2010).
\newblock \doi{10.1111/j.1365-2966.2010.17006.x}.

\bibitem{Carraro20}
R.~{Carraro}, G.~{Rodighiero}, P.~{Cassata} {\em et~al.}, {Coevolution of black
  hole accretion and star formation in galaxies up to z = 3.5}, \emph{\aap}.
  642:\penalty0 A65  (Oct., 2020).
\newblock \doi{10.1051/0004-6361/201936649}.

\bibitem{Shen15}
Y.~{Shen}, J.~E. {Greene}, L.~C. {Ho} {\em et~al.}, {The Sloan Digital Sky
  Survey Reverberation Mapping Project: No Evidence for Evolution in the
  M{\textbullet} -{\ensuremath{\sigma}}$_{*}$ Relation to z{\ensuremath{\sim}}
  1}, \emph{\apj}. 805\penalty0 (2):\penalty0 96  (June, 2015).
\newblock \doi{10.1088/0004-637X/805/2/96}.

\bibitem{Suh20}
H.~{Suh}, F.~{Civano}, B.~{Trakhtenbrot} {\em et~al.}, {No Significant
  Evolution of Relations between Black Hole Mass and Galaxy Total Stellar Mass
  Up to z {\ensuremath{\sim}} 2.5}, \emph{\apj}. 889\penalty0 (1):\penalty0 32
  (Jan., 2020).
\newblock \doi{10.3847/1538-4357/ab5f5f}.

\bibitem{Li21}
Y.~{Li}, Y.~{Ni}, R.~A.~C. {Croft} {\em et~al.}, {AI-assisted superresolution
  cosmological simulations}, \emph{Proceedings of the National Academy of
  Science}. 118\penalty0 (19):\penalty0 2022038118  (May, 2021).
\newblock \doi{10.1073/pnas.2022038118}.

\bibitem{ShankarMbhSigma}
F.~{Shankar}, M.~{Bernardi}, K.~{Richardson} {\em et~al.}, {Black hole scaling
  relations of active and quiescent galaxies: Addressing selection effects and
  constraining virial factors}, \emph{\mnras}. {\bf 485}\penalty0 (1),
  \penalty0 1278--1292  (May, 2019).
\newblock \doi{10.1093/mnras/stz376}.

\bibitem{Marsden22}
C.~{Marsden}, F.~{Shankar}, M.~{Bernardi} {\em et~al.}, {The weak dependence of
  velocity dispersion on disc fractions, mass-to-light ratio, and redshift:
  implications for galaxy and black hole evolution}, \emph{\mnras}. {\bf
  510}\penalty0 (4), \penalty0 5639--5660  (Mar., 2022).
\newblock \doi{10.1093/mnras/stab3705}.

\bibitem{Sahu19}
N.~{Sahu}, A.~W. {Graham} and B.~L. {Davis}, {Black Hole Mass Scaling Relations
  for Early-type Galaxies. I. M $_{BH}$-M $_{*,}$ $_{sph}$ and M $_{BH}$-M
  $_{*,gal}$}, \emph{\apj}. 876\penalty0 (2):\penalty0 155  (May, 2019).
\newblock \doi{10.3847/1538-4357/ab0f32}.

\bibitem{ReinesVolonteri}
A.~E. {Reines} and M.~{Volonteri}, {Relations between Central Black Hole Mass
  and Total Galaxy Stellar Mass in the Local Universe}, \emph{\apj}.
  813\penalty0 (2):\penalty0 82  (Nov., 2015).
\newblock \doi{10.1088/0004-637X/813/2/82}.

\bibitem{Giallongo2015}
E.~{Giallongo}, A.~{Grazian}, F.~{Fiore} {\em et~al.}, {Faint AGNs at z $\gt$ 4
  in the CANDELS GOODS-S field: looking for contributors to the reionization of
  the Universe}, \emph{\aap}. 578:\penalty0 A83  (June, 2015).
\newblock \doi{10.1051/0004-6361/201425334}.

\bibitem{Fan2019}
X.~{Fan}, A.~{Barth}, E.~{Banados} {\em et~al.}, {The First Luminous Quasars
  and Their Host Galaxies}, \emph{\baas}. 51\penalty0 (3):\penalty0 121  (May,
  2019).

\bibitem{Habouzit2022_AGN}
M.~{Habouzit}, R.~S. {Somerville}, Y.~{Li} {\em et~al.}, {Supermassive black
  holes in cosmological simulations - II: the AGN population and predictions
  for upcoming X-ray missions}, \emph{\mnras}. {\bf 509}\penalty0 (2),
  \penalty0 3015--3042  (Jan., 2022).
\newblock \doi{10.1093/mnras/stab3147}.

\bibitem{kollmeier2006}
J.~A. {Kollmeier}, C.~A. {Onken}, C.~S. {Kochanek} {\em et~al.}, {Black Hole
  Masses and Eddington Ratios at 0.3 \&lt; z \&lt; 4}, \emph{\apj}. {\bf
  648}\penalty0 (1), \penalty0 128--139  (Sept., 2006).
\newblock \doi{10.1086/505646}.

\bibitem{Shankar2013}
F.~{Shankar}, D.~H. {Weinberg} and J.~{Miralda-Escud{\'e}}, {Accretion-driven
  evolution of black holes: Eddington ratios, duty cycles and active galaxy
  fractions}, \emph{\mnras}. {\bf 428}\penalty0 (1), \penalty0 421--446  (Jan.,
  2013).
\newblock \doi{10.1093/mnras/sts026}.

\bibitem{Madau2014}
P.~{Madau} and M.~{Dickinson}, {Cosmic Star-Formation History}, \emph{\araa}.
  {\bf 52}, \penalty0 415--486  (Aug., 2014).
\newblock \doi{10.1146/annurev-astro-081811-125615}.

\bibitem{Aird2018}
J.~{Aird}, A.~L. {Coil} and A.~{Georgakakis}, {X-rays across the galaxy
  population - II. The distribution of AGN accretion rates as a function of
  stellar mass and redshift.}, \emph{\mnras}. {\bf 474}, \penalty0 1225--1249
  (Jan., 2018).
\newblock \doi{10.1093/mnras/stx2700}.

\bibitem{Ueda2014}
Y.~{Ueda}, M.~{Akiyama}, G.~{Hasinger} {\em et~al.}, {Toward the Standard
  Population Synthesis Model of the X-Ray Background: Evolution of X-Ray
  Luminosity and Absorption Functions of Active Galactic Nuclei Including
  Compton-thick Populations}, \emph{\apj}. 786\penalty0 (2):\penalty0 104
  (May, 2014).
\newblock \doi{10.1088/0004-637X/786/2/104}.

\bibitem{Aird2015}
J.~{Aird}, A.~L. {Coil}, A.~{Georgakakis} {\em et~al.}, {The evolution of the
  X-ray luminosity functions of unabsorbed and absorbed AGNs out to
  z{\ensuremath{\sim}} 5}, \emph{\mnras}. {\bf 451}\penalty0 (2), \penalty0
  1892--1927  (Aug., 2015).
\newblock \doi{10.1093/mnras/stv1062}.

\bibitem{MerloniHeinz2008}
A.~{Merloni} and S.~{Heinz}, {A synthesis model for AGN evolution: supermassive
  black holes growth and feedback modes}, \emph{\mnras}. {\bf 388}\penalty0
  (3), \penalty0 1011--1030  (Aug., 2008).
\newblock \doi{10.1111/j.1365-2966.2008.13472.x}.

\bibitem{ShenKelly2012}
Y.~{Shen} and B.~C. {Kelly}, {The Demographics of Broad-line Quasars in the
  Mass-Luminosity Plane. I. Testing FWHM-based Virial Black Hole Masses},
  \emph{\apj}. 746\penalty0 (2):\penalty0 169  (Feb., 2012).
\newblock \doi{10.1088/0004-637X/746/2/169}.

\bibitem{Buchner2015}
J.~{Buchner}, A.~{Georgakakis}, K.~{Nandra} {\em et~al.},
  {Obscuration-dependent Evolution of Active Galactic Nuclei}, \emph{\apj}.
  802\penalty0 (2):\penalty0 89  (Apr., 2015).
\newblock \doi{10.1088/0004-637X/802/2/89}.

\bibitem{McLureDunlop04}
R.~J. {McLure} and J.~S. {Dunlop}, {The cosmological evolution of quasar black
  hole masses}, \emph{\mnras}. {\bf 352}\penalty0 (4), \penalty0 1390--1404
  (Aug., 2004).
\newblock \doi{10.1111/j.1365-2966.2004.08034.x}.

\bibitem{Merloni04}
A.~{Merloni}, {The anti-hierarchical growth of supermassive black holes},
  \emph{\mnras}. {\bf 353}\penalty0 (4), \penalty0 1035--1047  (Oct., 2004).
\newblock \doi{10.1111/j.1365-2966.2004.08147.x}.

\bibitem{Shankar04}
F.~{Shankar}, P.~{Salucci}, G.~L. {Granato} {\em et~al.}, {Supermassive black
  hole demography: the match between the local and accreted mass functions},
  \emph{\mnras}. {\bf 354}\penalty0 (4), \penalty0 1020--1030  (Nov., 2004).
\newblock \doi{10.1111/j.1365-2966.2004.08261.x}.

\bibitem{Kollmeier06}
J.~A. {Kollmeier}, C.~A. {Onken}, C.~S. {Kochanek} {\em et~al.}, {Black Hole
  Masses and Eddington Ratios at 0.3 < z < 4}, \emph{\apj}. {\bf 648}\penalty0
  (1), \penalty0 128--139  (Sept., 2006).
\newblock \doi{10.1086/505646}.

\bibitem{Cao08}
X.~{Cao} and F.~{Li}, {Rapidly spinning massive black holes in active galactic
  nuclei: evidence from the black hole mass function}, \emph{\mnras}. {\bf
  390}\penalty0 (2), \penalty0 561--566  (Oct., 2008).
\newblock \doi{10.1111/j.1365-2966.2008.13800.x}.

\bibitem{Silverman08}
J.~D. {Silverman}, P.~J. {Green}, W.~A. {Barkhouse} {\em et~al.}, {The
  Luminosity Function of X-Ray-selected Active Galactic Nuclei: Evolution of
  Supermassive Black Holes at High Redshift}, \emph{\apj}. {\bf 679}\penalty0
  (1), \penalty0 118--139  (May, 2008).
\newblock \doi{10.1086/529572}.

\bibitem{DeGraf2011}
C.~{Degraf}, T.~{Di Matteo} and V.~{Springel}, {Black hole clustering in
  cosmological hydrodynamic simulations: evidence for mergers}, \emph{\mnras}.
  {\bf 413}, \penalty0 1383--1394  (May, 2011).
\newblock \doi{10.1111/j.1365-2966.2011.18221.x}.

\bibitem{oogi2016}
T.~{Oogi}, M.~{Enoki}, T.~{Ishiyama} {\em et~al.}, {Quasar clustering in a
  galaxy and quasar formation model based on ultra high-resolution N-body
  simulations}, \emph{\mnras}. {\bf 456}\penalty0 (1), \penalty0 L30--L34
  (Feb., 2016).
\newblock \doi{10.1093/mnrasl/slv169}.

\bibitem{Bhowmick2018}
A.~K. {Bhowmick}, T.~{Di Matteo}, Y.~{Feng} {\em et~al.}, {The clustering of z
  $>$ 7 galaxies: predictions from the BLUETIDES simulation}, \emph{\mnras}.
  {\bf 474}, \penalty0 5393--5405  (Mar., 2018).
\newblock \doi{10.1093/mnras/stx3149}.

\bibitem{HaimanHui2001}
Z.~{Haiman} and L.~{Hui}, {Constraining the Lifetime of Quasars from Their
  Spatial Clustering}, \emph{\apj}. {\bf 547}, \penalty0 27--38  (Jan., 2001).
\newblock \doi{10.1086/318330}.

\bibitem{MartiniWeinberg2001}
P.~{Martini} and D.~H. {Weinberg}, {Quasar Clustering and the Lifetime of
  Quasars}, \emph{\apj}. {\bf 547}, \penalty0 12--26  (Jan., 2001).
\newblock \doi{10.1086/318331}.

\bibitem{Bhowmick2017}
A.~K. {Bhowmick}, T.~{Di Matteo}, Y.~{Feng} {\em et~al.}, {The clustering of z
  $\gt$ 7 galaxies: predictions from the BLUETIDES simulation}, \emph{\mnras}.
  {\bf 474}, \penalty0 5393--5405  (Mar., 2018).
\newblock \doi{10.1093/mnras/stx3149}.

\bibitem{Farina2013}
E.~P. {Farina}, C.~{Montuori}, R.~{Decarli} {\em et~al.}, {Caught in the act:
  discovery of a physical quasar triplet}, \emph{\mnras}. {\bf 431}\penalty0
  (2), \penalty0 1019--1025  (May, 2013).
\newblock \doi{10.1093/mnras/stt209}.

\bibitem{Djorgovski2007}
S.~G. {Djorgovski}, F.~{Courbin}, G.~{Meylan} {\em et~al.}, {Discovery of a
  Probable Physical Triple Quasar}, \emph{\apjl}. {\bf 662}\penalty0 (1),
  \penalty0 L1--L5  (June, 2007).
\newblock \doi{10.1086/519162}.

\bibitem{SomervilleDave2015}
R.~S. {Somerville} and R.~{Dav{\'e}}, {Physical Models of Galaxy Formation in a
  Cosmological Framework}, \emph{\araa}. {\bf 53}, \penalty0 51--113  (Aug.,
  2015).
\newblock \doi{10.1146/annurev-astro-082812-140951}.

\bibitem{DESI2016}
{DESI Collaboration}, A.~{Aghamousa}, J.~{Aguilar} {\em et~al.}, {The DESI
  Experiment Part I: Science,Targeting, and Survey Design}, \emph{arXiv
  e-prints}. art. arXiv:1611.00036  (Oct., 2016).

\bibitem{DES2018}
T.~M.~C. {Abbott}, F.~B. {Abdalla}, A.~{Alarcon} {\em et~al.}, {Dark Energy
  Survey year 1 results: Cosmological constraints from galaxy clustering and
  weak lensing}, \emph{\prd}. 98\penalty0 (4):\penalty0 043526  (Aug., 2018).
\newblock \doi{10.1103/PhysRevD.98.043526}.

\bibitem{Ivezic2019_LSST}
{\v{Z}}.~{Ivezi{\'c}}, S.~M. {Kahn}, J.~A. {Tyson} {\em et~al.}, {LSST: From
  Science Drivers to Reference Design and Anticipated Data Products},
  \emph{\apj}. 873\penalty0 (2):\penalty0 111  (Mar., 2019).
\newblock \doi{10.3847/1538-4357/ab042c}.

\bibitem{Spergel2013_WFIRST}
D.~{Spergel}, N.~{Gehrels}, J.~{Breckinridge} {\em et~al.}, {WFIRST-2.4: What
  Every Astronomer Should Know}, \emph{arXiv e-prints}. art. arXiv:1305.5425
  (May, 2013).

\bibitem{Laureijs2011_Euclid}
R.~{Laureijs}, J.~{Amiaux}, S.~{Arduini} {\em et~al.}, {Euclid Definition Study
  Report}, \emph{arXiv e-prints}. art. arXiv:1110.3193  (Oct., 2011).

\bibitem{Chisari2019_review}
N.~E. {Chisari}, A.~J. {Mead}, S.~{Joudaki} {\em et~al.}, {Modelling baryonic
  feedback for survey cosmology}, \emph{The Open Journal of Astrophysics}.
  2\penalty0 (1):\penalty0 4  (June, 2019).
\newblock \doi{10.21105/astro.1905.06082}.

\bibitem{vanDaalen2011}
M.~P. {van Daalen}, J.~{Schaye}, C.~M. {Booth} {\em et~al.}, {The effects of
  galaxy formation on the matter power spectrum: a challenge for precision
  cosmology}, \emph{\mnras}. {\bf 415}\penalty0 (4), \penalty0 3649--3665
  (Aug., 2011).
\newblock \doi{10.1111/j.1365-2966.2011.18981.x}.

\bibitem{Semboloni2011}
E.~{Semboloni}, H.~{Hoekstra}, J.~{Schaye} {\em et~al.}, {Quantifying the
  effect of baryon physics on weak lensing tomography}, \emph{\mnras}. {\bf
  417}, \penalty0 2020--2035  (Nov., 2011).
\newblock \doi{10.1111/j.1365-2966.2011.19385.x}.

\bibitem{Tenneti2015}
A.~{Tenneti}, R.~{Mandelbaum} and T.~{Di Matteo}, {Intrinsic alignments of disk
  and elliptical galaxies in the MassiveBlack-II and Illustris simulations},
  \emph{ArXiv e-prints}  (Oct., 2015).

\bibitem{Chisari2018}
N.~E. {Chisari}, M.~L.~A. {Richardson}, J.~{Devriendt} {\em et~al.}, {The
  impact of baryons on the matter power spectrum from the Horizon-AGN
  cosmological hydrodynamical simulation}, \emph{\mnras}. {\bf 480}\penalty0
  (3), \penalty0 3962--3977  (Nov., 2018).
\newblock \doi{10.1093/mnras/sty2093}.

\bibitem{Springel2018}
V.~{Springel}, R.~{Pakmor}, A.~{Pillepich} {\em et~al.}, {First results from
  the IllustrisTNG simulations: matter and galaxy clustering}, \emph{\mnras}.
  {\bf 475}, \penalty0 676--698  (Mar., 2018).
\newblock \doi{10.1093/mnras/stx3304}.

\bibitem{vanDaalen2020}
M.~P. {van Daalen}, I.~G. {McCarthy} and J.~{Schaye}, {Exploring the effects of
  galaxy formation on matter clustering through a library of simulation power
  spectra}, \emph{\mnras}. {\bf 491}\penalty0 (2), \penalty0 2424--2446  (Jan.,
  2020).
\newblock \doi{10.1093/mnras/stz3199}.

\bibitem{Dubois13}
Y.~{Dubois}, C.~{Pichon}, J.~{Devriendt} {\em et~al.}, {Blowing cold flows
  away: the impact of early AGN activity on the formation of a brightest
  cluster galaxy progenitor}, \emph{\mnras}. {\bf 428}\penalty0 (4), \penalty0
  2885--2900  (Feb., 2013).
\newblock \doi{10.1093/mnras/sts224}.

\bibitem{Menci19}
N.~{Menci}, F.~{Fiore}, C.~{Feruglio} {\em et~al.}, {Outflows in the Disks of
  Active Galaxies}, \emph{\apj}. 877\penalty0 (2):\penalty0 74  (June, 2019).
\newblock \doi{10.3847/1538-4357/ab1a3a}.

\bibitem{Menci14}
N.~{Menci}, M.~{Gatti}, F.~{Fiore} {\em et~al.}, {Triggering active galactic
  nuclei in hierarchical galaxy formation: disk instability vs. interactions},
  \emph{\aap}. 569:\penalty0 A37  (Sept., 2014).
\newblock \doi{10.1051/0004-6361/201424217}.

\bibitem{Fiore17}
F.~{Fiore}, C.~{Feruglio}, F.~{Shankar} {\em et~al.}, {AGN wind scaling
  relations and the co-evolution of black holes and galaxies}, \emph{\aap}.
  601:\penalty0 A143  (May, 2017).
\newblock \doi{10.1051/0004-6361/201629478}.

\bibitem{LISA2017arXiv170200786A}
P.~{Amaro-Seoane}, H.~{Audley}, S.~{Babak} {\em et~al.}, {Laser Interferometer
  Space Antenna}, \emph{arXiv e-prints}. art. arXiv:1702.00786  (Feb., 2017).

\bibitem{Mingarelli2017}
C.~M.~F. {Mingarelli}, T.~J.~W. {Lazio}, A.~{Sesana} {\em et~al.}, {The local
  nanohertz gravitational-wave landscape from supermassive black hole
  binaries}, \emph{Nature Astronomy}. {\bf 1}, \penalty0 886--892  (Nov.,
  2017).
\newblock \doi{10.1038/s41550-017-0299-6}.

\bibitem{Amaro-Seoane2012}
P.~{Amaro-Seoane}, S.~{Aoudia}, S.~{Babak} {\em et~al.}, {Low-frequency
  gravitational-wave science with eLISA/NGO}, \emph{Classical and Quantum
  Gravity}. 29\penalty0 (12):\penalty0 124016  (June, 2012).
\newblock \doi{10.1088/0264-9381/29/12/124016}.

\bibitem{Amaro2012}
P.~{Amaro-Seoane}, S.~{Aoudia}, S.~{Babak} {\em et~al.}, {Low-frequency
  gravitational-wave science with eLISA/NGO}, \emph{Classical and Quantum
  Gravity}. 29\penalty0 (12):\penalty0 124016  (June, 2012).
\newblock \doi{10.1088/0264-9381/29/12/124016}.

\bibitem{Amaro2013}
P.~{Amaro-Seoane}, S.~{Aoudia}, S.~{Babak} {\em et~al.}, {eLISA: Astrophysics
  and cosmology in the millihertz regime}, \emph{GW Notes, Vol.~6, p.~4-110}.
  {\bf 6}, \penalty0 4--110  (May, 2013).

\bibitem{WyitheLoeb2003}
J.~S.~B. {Wyithe} and A.~{Loeb}, {Low-Frequency Gravitational Waves from
  Massive Black Hole Binaries: Predictions for LISA and Pulsar Timing Arrays},
  \emph{\apj}. {\bf 590}, \penalty0 691--706  (June, 2003).
\newblock \doi{10.1086/375187}.

\bibitem{Enoki2004}
M.~{Enoki}, K.~T. {Inoue}, M.~{Nagashima} {\em et~al.}, {Gravitational Waves
  from Supermassive Black Hole Coalescence in a Hierarchical Galaxy Formation
  Model}, \emph{\apj}. {\bf 615}, \penalty0 19--28  (Nov., 2004).
\newblock \doi{10.1086/424475}.

\bibitem{Koushiappas2006}
S.~M. {Koushiappas} and A.~R. {Zentner}, {Testing Models of Supermassive Black
  Hole Seed Formation through Gravity Waves}, \emph{\apj}. {\bf 639}, \penalty0
  7--22  (Mar., 2006).
\newblock \doi{10.1086/499325}.

\bibitem{Micic2007}
M.~{Micic}, K.~{Holley-Bockelmann}, S.~{Sigurdsson} {\em et~al.}, {Supermassive
  black hole growth and merger rates from cosmological N-body simulations},
  \emph{\mnras}. {\bf 380}, \penalty0 1533--1540  (Oct., 2007).
\newblock \doi{10.1111/j.1365-2966.2007.12162.x}.

\bibitem{Sesana2009}
A.~{Sesana}, J.~{Gair}, I.~{Mandel} {\em et~al.}, {Observing Gravitational
  Waves from the First Generation of Black Holes}, \emph{\apjl}. {\bf 698},
  \penalty0 L129--L132  (June, 2009).
\newblock \doi{10.1088/0004-637X/698/2/L129}.

\bibitem{Klein2016}
A.~{Klein}, E.~{Barausse}, A.~{Sesana} {\em et~al.}, {Science with the
  space-based interferometer eLISA: Supermassive black hole binaries},
  \emph{\prd}. 93\penalty0 (2):\penalty0 024003  (Jan., 2016).
\newblock \doi{10.1103/PhysRevD.93.024003}.

\bibitem{ColpiDotti2011}
M.~{Colpi} and M.~{Dotti}, {Massive Binary Black Holes in the Cosmic
  Landscape}, \emph{Advanced Science Letters}. {\bf 4}, \penalty0 181--203
  (Feb., 2011).
\newblock \doi{10.1166/asl.2011.1205}.

\bibitem{Mayer2013}
L.~{Mayer}, {Massive black hole binaries in gas-rich galaxy mergers; multiple
  regimes of orbital decay and interplay with gas inflows}, \emph{Classical and
  Quantum Gravity}. 30\penalty0 (24):\penalty0 244008  (Dec., 2013).
\newblock \doi{10.1088/0264-9381/30/24/244008}.

\bibitem{Colpi2014}
M.~{Colpi}, {Massive Binary Black Holes in Galactic Nuclei and Their Path to
  Coalescence}, \emph{\ssr}. {\bf 183}, \penalty0 189--221  (Sept., 2014).
\newblock \doi{10.1007/s11214-014-0067-1}.

\bibitem{Quinlan1996}
G.~D. {Quinlan}, {The dynamical evolution of massive black hole binaries I.
  Hardening in a fixed stellar background}, \emph{\na}. {\bf 1}\penalty0 (1),
  \penalty0 35--56  (July, 1996).
\newblock \doi{10.1016/S1384-1076(96)00003-6}.

\bibitem{Berczik2006}
P.~{Berczik}, D.~{Merritt}, R.~{Spurzem} {\em et~al.}, {Efficient Merger of
  Binary Supermassive Black Holes in Nonaxisymmetric Galaxies}, \emph{\apjl}.
  {\bf 642}\penalty0 (1), \penalty0 L21--L24  (May, 2006).
\newblock \doi{10.1086/504426}.

\bibitem{Sesana2007b}
A.~{Sesana}, F.~{Haardt} and P.~{Madau}, {Interaction of Massive Black Hole
  Binaries with Their Stellar Environment. II. Loss Cone Depletion and Binary
  Orbital Decay}, \emph{\apj}. {\bf 660}\penalty0 (1), \penalty0 546--555
  (May, 2007).
\newblock \doi{10.1086/513016}.

\bibitem{Berentzen2009}
I.~{Berentzen}, M.~{Preto}, P.~{Berczik} {\em et~al.}, {Binary Black Hole
  Merger in Galactic Nuclei: Post-Newtonian Simulations}, \emph{\apj}. {\bf
  695}\penalty0 (1), \penalty0 455--468  (Apr., 2009).
\newblock \doi{10.1088/0004-637X/695/1/455}.

\bibitem{Khan2011}
F.~M. {Khan}, A.~{Just} and D.~{Merritt}, {Efficient Merger of Binary
  Supermassive Black Holes in Merging Galaxies}, \emph{\apj}. 732\penalty0
  (2):\penalty0 89  (May, 2011).
\newblock \doi{10.1088/0004-637X/732/2/89}.

\bibitem{Khan2013}
F.~M. {Khan}, K.~{Holley-Bockelmann}, P.~{Berczik} {\em et~al.}, {Supermassive
  Black Hole Binary Evolution in Axisymmetric Galaxies: The Final Parsec
  Problem is Not a Problem}, \emph{\apj}. 773\penalty0 (2):\penalty0 100
  (Aug., 2013).
\newblock \doi{10.1088/0004-637X/773/2/100}.

\bibitem{Vasiliev2015}
E.~{Vasiliev}, F.~{Antonini} and D.~{Merritt}, {The Final-parsec Problem in the
  Collisionless Limit}, \emph{\apj}. 810\penalty0 (1):\penalty0 49  (Sept.,
  2015).
\newblock \doi{10.1088/0004-637X/810/1/49}.

\bibitem{Haiman2009}
Z.~{Haiman}, B.~{Kocsis} and K.~{Menou}, {The Population of Viscosity- and
  Gravitational Wave-driven Supermassive Black Hole Binaries Among Luminous
  Active Galactic Nuclei}, \emph{\apj}. {\bf 700}\penalty0 (2), \penalty0
  1952--1969  (Aug., 2009).
\newblock \doi{10.1088/0004-637X/700/2/1952}.

\bibitem{Bonetti2018}
M.~{Bonetti}, F.~{Haardt}, A.~{Sesana} {\em et~al.}, {Post-Newtonian evolution
  of massive black hole triplets in galactic nuclei - II. Survey of the
  parameter space}, \emph{\mnras}. {\bf 477}\penalty0 (3), \penalty0 3910--3926
   (July, 2018).
\newblock \doi{10.1093/mnras/sty896}.

\bibitem{Tremmel2017}
M.~{Tremmel}, M.~{Karcher}, F.~{Governato} {\em et~al.}, {The Romulus
  cosmological simulations: a physical approach to the formation, dynamics and
  accretion models of SMBHs}, \emph{\mnras}. {\bf 470}\penalty0 (1), \penalty0
  1121--1139  (Sept., 2017).
\newblock \doi{10.1093/mnras/stx1160}.

\bibitem{Mannerkoski2021}
M.~{Mannerkoski}, P.~H. {Johansson}, A.~{Rantala} {\em et~al.}, {Signatures of
  the Many Supermassive Black Hole Mergers in a Cosmologically Forming Massive
  Early-Type Galaxy}, \emph{arXiv e-prints}. art. arXiv:2112.03576  (Dec.,
  2021).

\bibitem{Chen2021}
N.~{Chen}, Y.~{Ni}, A.~M. {Holgado} {\em et~al.}, {Massive Black Hole Mergers
  with Orbital Information: Predictions from the ASTRID Simulation},
  \emph{arXiv e-prints}. art. arXiv:2112.08555  (Dec., 2021).

\bibitem{Tremmel2015}
M.~{Tremmel}, F.~{Governato}, M.~{Volonteri} {\em et~al.}, {Off the beaten
  path: a new approach to realistically model the orbital decay of supermassive
  black holes in galaxy formation simulations}, \emph{\mnras}. {\bf
  451}\penalty0 (2), \penalty0 1868--1874  (Aug., 2015).
\newblock \doi{10.1093/mnras/stv1060}.

\bibitem{Snyder2015}
G.~F. Snyder, P.~Torrey, J.~M. Lotz {\em et~al.}, {Galaxy morphology and star
  formation in the Illustris Simulation at z\ = 0}, \emph{\mnras}. {\bf
  454}\penalty0 (2), \penalty0 1886--1908  (oct, 2015).
\newblock ISSN 0035-8711.
\newblock \doi{10.1093/mnras/stv2078}.
\newblock URL \url{http://adsabs.harvard.edu/abs/2015MNRAS.454.1886S}.

\bibitem{Snyder2015a}
G.~F. Snyder, J.~Lotz, C.~Moody {\em et~al.}, {Diverse structural evolution at
  z > 1 in cosmologically simulated gal\ axies}, \emph{\mnras}. {\bf
  451}\penalty0 (4), \penalty0 4290--4310  (jun, 2015).
\newblock ISSN 0035-8711.
\newblock \doi{10.1093/mnras/stv1231}.
\newblock URL \url{http://adsabs.harvard.edu/abs/2015MNRAS.451.4290S}.

\bibitem{Banks2022}
S.~{Banks}, K.~{Lee}, N.~{Azimi} {\em et~al.}, {On the detectability of massive
  black hole merger eventsby LISA}, \emph{arXiv e-prints}. art.
  arXiv:2107.09084  (July, 2021).

\bibitem{Bowen2018}
D.~B. {Bowen}, V.~{Mewes}, M.~{Campanelli} {\em et~al.}, {Quasi-periodic
  Behavior of Mini-disks in Binary Black Holes Approaching Merger},
  \emph{\apjl}. 853:\penalty0 L17  (Jan., 2018).
\newblock \doi{10.3847/2041-8213/aaa756}.

\bibitem{Lousto2017}
C.~O. {Lousto}, Y.~{Zlochower} and M.~{Campanelli}, {Modeling the Black Hole
  Merger of QSO 3C 186}, \emph{\apjl}. 841:\penalty0 L28  (June, 2017).
\newblock \doi{10.3847/2041-8213/aa733c}.

\bibitem{Farris2015}
B.~D. {Farris}, P.~{Duffell}, A.~I. {MacFadyen} {\em et~al.}, {Characteristic
  signatures in the thermal emission from accreting binary black holes},
  \emph{\mnras}. {\bf 446}, \penalty0 L36--L40  (Jan., 2015).
\newblock \doi{10.1093/mnrasl/slu160}.

\bibitem{Barausse20}
E.~{Barausse}, I.~{Dvorkin}, M.~{Tremmel} {\em et~al.}, {Massive Black Hole
  Merger Rates: The Effect of Kiloparsec Separation Wandering and Supernova
  Feedback}, \emph{\apj}. 904\penalty0 (1):\penalty0 16  (Nov., 2020).
\newblock \doi{10.3847/1538-4357/abba7f}.

\bibitem{Izqui22}
D.~{Izquierdo-Villalba}, A.~{Sesana}, S.~{Bonoli} {\em et~al.}, {Massive black
  hole evolution models confronting the n-Hz amplitude of the stochastic
  gravitational wave background}, \emph{\mnras}. {\bf 509}\penalty0 (3),
  \penalty0 3488--3503  (Jan., 2022).
\newblock \doi{10.1093/mnras/stab3239}.

\bibitem{Sicilia22}
A.~{Sicilia}, A.~{Lapi}, L.~{Boco} {\em et~al.}, {The Black Hole Mass Function
  across Cosmic Time. II. Heavy Seeds and (Super)Massive Black Holes},
  \emph{\apj}. 934\penalty0 (1):\penalty0 66  (July, 2022).
\newblock \doi{10.3847/1538-4357/ac7873}.

\bibitem{Peirani08}
S.~{Peirani} and J.~A. {de Freitas Pacheco}, {Dark matter accretion into
  supermassive black holes}, \emph{\prd}. 77\penalty0 (6):\penalty0 064023
  (Mar., 2008).
\newblock \doi{10.1103/PhysRevD.77.064023}.

\bibitem{Croker21}
K.~S. {Croker}, M.~{Zevin}, D.~{Farrah} {\em et~al.}, {Cosmologically Coupled
  Compact Objects: A Single-parameter Model for LIGO-Virgo Mass and Redshift
  Distributions}, \emph{\apjl}. 921\penalty0 (2):\penalty0 L22  (Nov., 2021).
\newblock \doi{10.3847/2041-8213/ac2fad}.

\end{thebibliography}

%

\end{document}